# Evaluation of the DiversiNews diversified news service

**Technical report**

**Daniele Pighin, Google Inc.**
**Enrique Alfonseca, Google Inc.**
**Felix Leif Keppmann, Karlsruhe Institute of Technology**
**Mitja Trampus, Jozef Stefan Institute**

# Abstract

In this report we present the outcome of an extensive evaluation of the DiversiNews platform [8, 10] for diversified browsing of news, developed in the scope of the RENDER project.

The evaluation was carried out along two main directions: a component evaluation, in which we assessed the maturity of the components underlying DiversiNews, and a user experience (UX) evaluation involving users of online news services. The results of the evaluation confirm the high value of DiversiNews as a novel paradigm for diversity-aware news browsing.

# Table of Contents

## List of Figures

# Abbreviations

| | |
|---|---|
| ICC | Inter-Class Correlation |
| UX | User Experience |

# 1 Introduction

This document presents a comprehensive evaluation of DiversiNews, a prototype web platform for browsing online news in a diversity-aware fashion, developed in scope of the RENDER project. For broader scope, refer to the project web site (http://render-project.eu/), documents describing DiversiNews [8, 10] and the online demo of DiversiNews (http://aidemo.ijs.si/diversinews).

We evaluated both the technological aspects of the platform and the reactions of internet users who have been exposed to the interface. The feedback that we collected from both evaluations has later been incorporated in newer versions of the DiversiNews platform, which now offers tighter integration among its components and a more polished and refined user experience.

The first aspect of evaluation, which we refer to as the "components" evaluation, is an assessment of all the components of the interface and their degree of integration. In particular, we measured the perceived quality of the generated summaries both in terms of fluency and informativeness, the extent to which the generated summaries are sensitive to the user specification and the quality of the related entities served by the related entities panel. For this evaluation, we employed expert annotators who were involved in several sessions of static data analysis. The methodology and the results of this evaluation are discussed in more detail in section 2.

The second aspect of evaluation, which we refer to as "User eXperience" (UX) evaluation, is an assessment of the ergonomics of the platform, its degree of self-documentation and the utility of the service provided, as perceived by potential end users. In this evaluation we involved both casual internet users and people who process news streams as part of their daily job. The subjects of the experiments were first asked to inspect and understand the interface visually, then to use it interactively and finally to provide their feedback, impressions and suggestions for improvement. The screenshots that were used are shown in Figure 13 (View 1: news search / cluster selection) and Figure 21 (View 2: diversified news browsing). More details about the UX evaluation are discussed in Section 3, while the complete text of the administered questionnaire along with the subjects' answers is listed in Annex A.

# 2 Components Evaluation

## 2.1 Objectives

The components evaluation of DiversiNews aims at quantifying the quality of the components of the platform and at assessing the level of maturity of the technologies employed. More precisely, this evaluation wants to answer the following questions:

- What summarization technology is more suitable for DiversiNews?
- Is state-of-the-art summarization technology ready for deployment in a user-facing platform on real news streams?
- Do the controls of DiversiNews interface have a measurable effect on the contents of the generated summaries? Or, turning the question around, are the employed summarization technologies sensitive enough to diversity-related parameters?

## 2.2 Methodology

For this evaluation we randomly selected 20 of the news clusters listed in the news search/cluster selection view of the DiversiNews demo. For each of the summaries, we generated 8 different summaries based on the combinations of the following dimensions:

- **Summarization technology:**

  DiversiNews users can choose between two different methods to generate summaries of news collections. The summarization algorithm can be toggled by acting on the summary panel (labelled (1) in Figure 21). The two summarizers are:

  - **TopicSum**, a state-of-the-art statistical summarizer described in D5.2.2 [8], DiversiNews' core summarization technology since the inception of the case study. In the interface and in the rest of this document we will refer to it as **Type1**.
  - The semantic-aware summarizer developed by JSI and described in D3.2.1 [7], which we will refer to as **Type2**.

- **Diversity controls settings:**

  In order to evaluate the effect of the diversity controls on the content and quality of the summaries, for each news collection and summarization engine we generated 4 summaries according to different configurations of the "relevant topics" and "polarity" controls (labelled (3) and (5) in Figure 21, respectively[1]. To evaluate each control in isolation, in each configuration we varied only one of the parameters, leaving the others in their neutral position:

  - **Neutral**: both controls are in the default position.
  - **Topic:** the "relevant topics" control is moved to overlap a specific target topic. The "polarity" control is in the default position.
  - **Positive:** the "polarity" control is moved to the rightmost position. The "relevant topics" control is in the default position.
  - **Negative:** the "polarity" control is moved to the leftmost position. The "relevant topics" control is in the default position.

Two expert annotators have annotated each of the generated summaries according to the following dimensions:

[1] We did not evaluate the effect of the "geographical source" widget as the panel controls a dimension pertaining an article's metadata (publisher location) as opposed to an emerging property of the article's content, as in the case of topic relevance or sentiment.

- **Fluency:** to what extent is the summary understandable and grammatical adequate? Each summary was rated on a 3-point Likert scale, with values:
    1. **Completely disconnected:** the summary is difficult to understand, non adequate for user consumption.
    2. **Local disfluencies:** the summary is understandable even though it contains local disfluencies. It is adequate for user consumption.
    3. **Fluent and understandable:** the summary is human grade, or nearly so.
- **Informativeness:** does the summary contain relevant information with respect to the selected news collection? How useful is the information in the summary? Each summary was rated on a 3-point Likert scale, with values:
    1. **Uninformative:** nothing relevant can be learned from the summary.
    2. **Partially informative:** the summary conveys useful information, but it's not exhaustive.
    3. **Informative:** the summary is informative and captures relevant aspect of at least one of the main topics in the news cluster.
- **Topicality:** Among the 8 summaries generated for each news collection, the raters have been asked to select up to 2 of them which are especially centered on a specific topic.
- **Polarity:** Among the 8 summaries generated for each news collection, the raters have been asked to select up to 2 of them which have an especially positive connotation, and 2 with an especially negative one.

As a quality metric for fluency and adequacy we considered the averaged value on the corresponding Likert scale. For topicality and polarity, we measured Precision (P), Recall (R) and F1-measure (F1) of the raters' decisions with respect to the configuration according to which the summaries were generated. As an example, a summary labeled as having negative polarity by a rater contributed a true positive for P, R and F1 calculation only if the summary had been generated with the polarity slider set to "negative". On the contrary, if the summary had been generated with any other setting of the polarity slider, then the example was considered as a false positive. Similarly, a false negative was emitted every time a rater wrongly marked a summary as neither positively nor negatively slanted.

Inter-annotator agreement on all dimensions was measured by means of Intra-Class Correlation (ICC) [3], calculated on 5 news collections (40 summaries) annotated by both raters. The raters reached a high agreement on Fluency (0.6) and Relatedness (0.71), while we observed a relatively low agreement on Polarity (0.44) and Informativeness (0.19). These figures are somehow expected, as while the former dimensions can be associated with somewhat objective criteria (i.e., passage grammaticality and the presence/absence of mentions of a topic in a summary), the latter are inherently more subjective. In particular, the annotation for informativeness is complicated by the fact that real-world news collections very often aggregate articles focusing of different events. In such cases, the annotation of this dimension requires the raters to implicitly take a number of arbitrary decisions, such as: deciding which of the reported events are the most relevant, deciding what information for each topic should be conveyed by the summary, deciding whether the content of the summary is adequate.

## 2.3 Main Findings

As shown in Figure 1 and Figure 2, according to raters **~65% of the summaries are at least adequately fluent, while ~70% are sufficiently informative**, with almost one third of the generated summaries being rated as human grade quality along both dimensions. These figures confirm the positive impressions that we gathered from the subjects of the UX evaluation (see Section 3), and we regard them as **extremely positive** in the light of the fact that for our evaluation we used real, unfiltered data in a completely open domain setting. Such an environment, with noisy, multi-topic news collections of very different sizes and documents of the most different lengths, is far more challenging than the data typically used for summarization benchmarking symposia such as DUC or TAC[2].

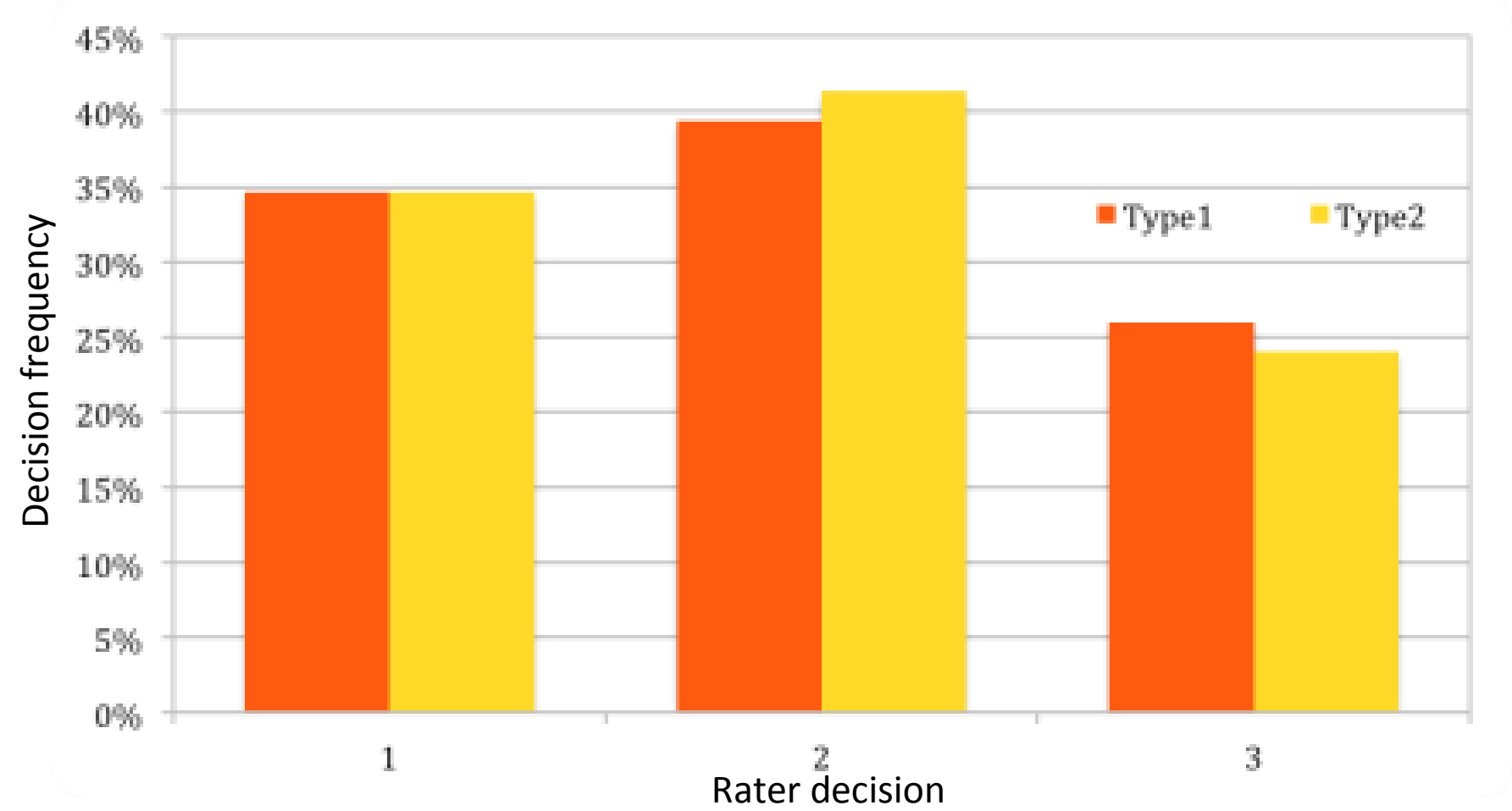

**Figure 1. Fluency assessment of the generated summaries.**

Concerning the comparison between the two summarizers, they are basically equivalent in terms of fluency, while Type1 is generally rated higher in terms of Informativeness. We speculate that this latest finding may be related to the fact that Type1 produces summaries that, on average, are 40% longer than those produced by Type2 (as shown Figure 3), and therefore they are more likely to contain relevant bits of information.

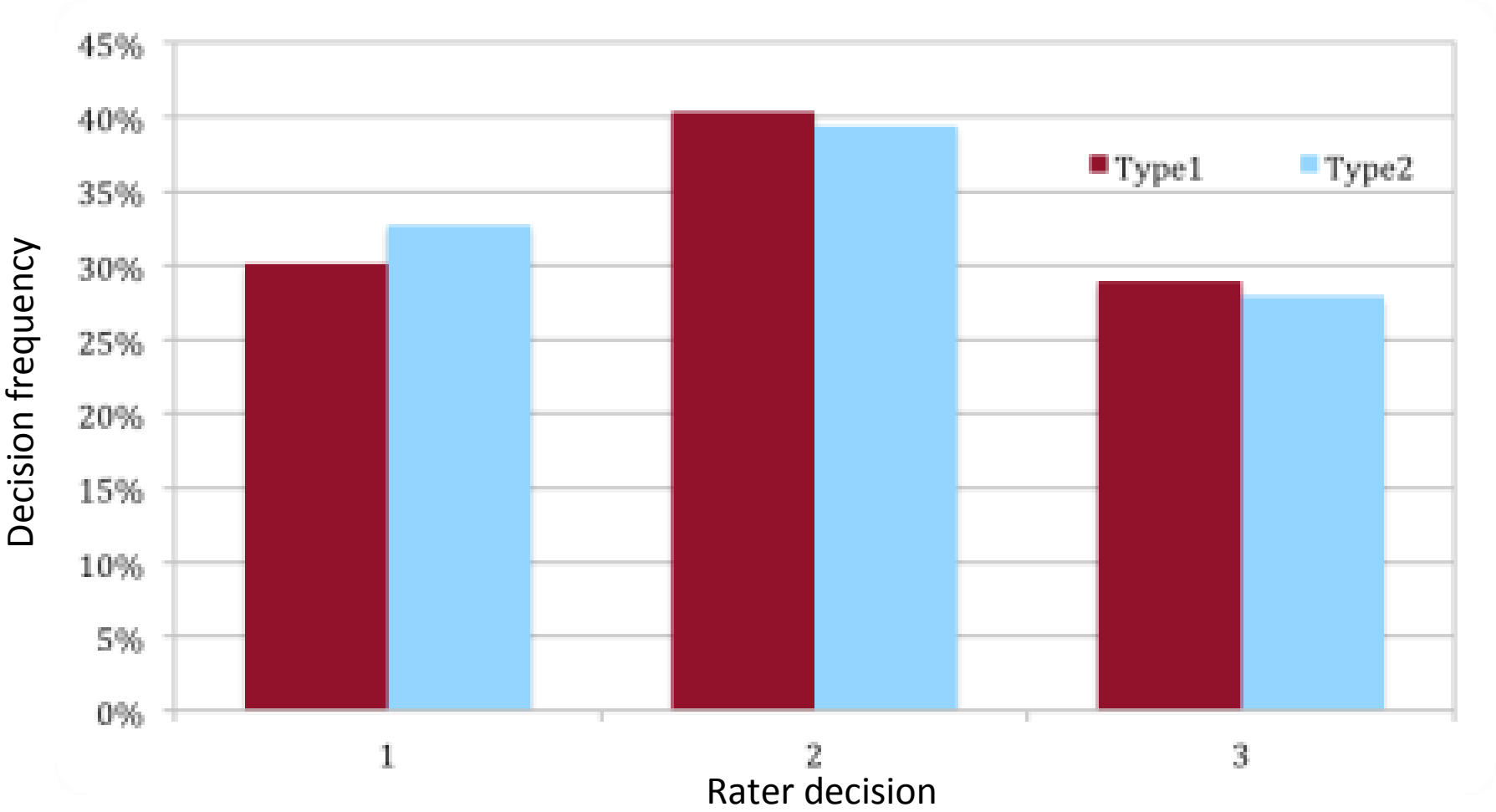

**Figure 2. Informativeness assessment of the generated summaries.**

According to raters, the fluency of the summaries rated as inadequate is mostly affected by the following factors:

1. Disconnected discourse markers, e.g.: "After all" or "In the light of this" appearing in the first sentence or in a subsequent sentence which is unrelated to the previous one.
2. Bad co-references.
3. Wrong sentence order.
4. Undefined terms of acronyms.
5. Interleaved topics within the same summary (A / B / A).
6. Too long sentences.

Similarly, we collected feedback on how to improve the summaries in terms of informativeness:

[2] http://www.nist.gov/tac/

7. Uneven coverage of topics (e.g., of three sentences making up a summary, two are relative to one topic while the third one covers a different one).
8. Uneven content of single sentences (e.g., a very long and detailed followed by a very short one).
9. Focus on not too relevant sub-topic (a common problem especially with noisy clusters).
10. Too many details for a summary.

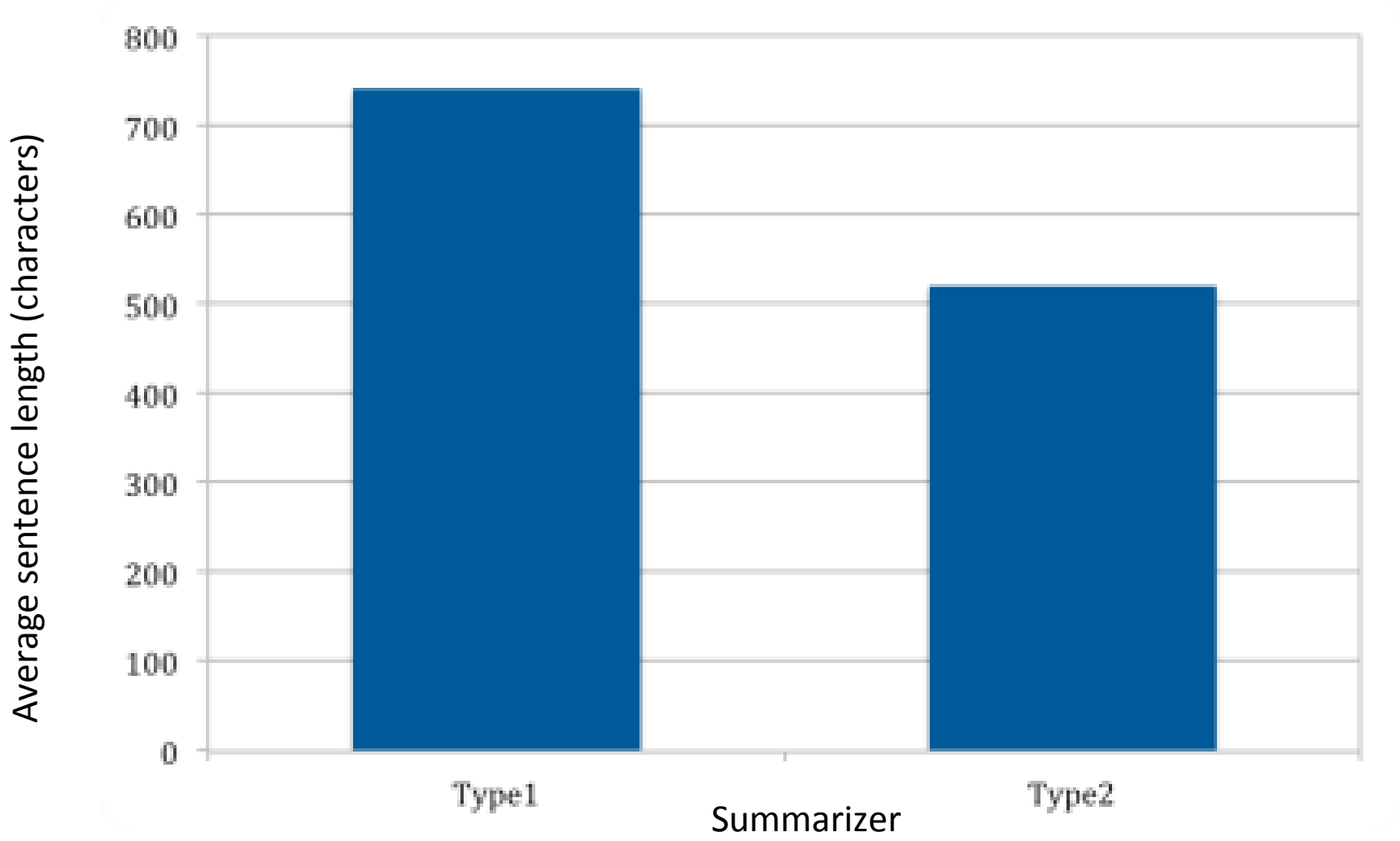

**Figure 3. Average length of the evaluated summaries (characters).**

All these comments are related to open research problems within multi-document automatic text summarization, such as discourse modelling (1, 2, 3, 4, 5) and the combination of extractive and abstractive technology to modulate the information content of each sentence (6 and 8). Both research lines are actively being explored at Google, and the latter is already showing promising results [4][5][6].

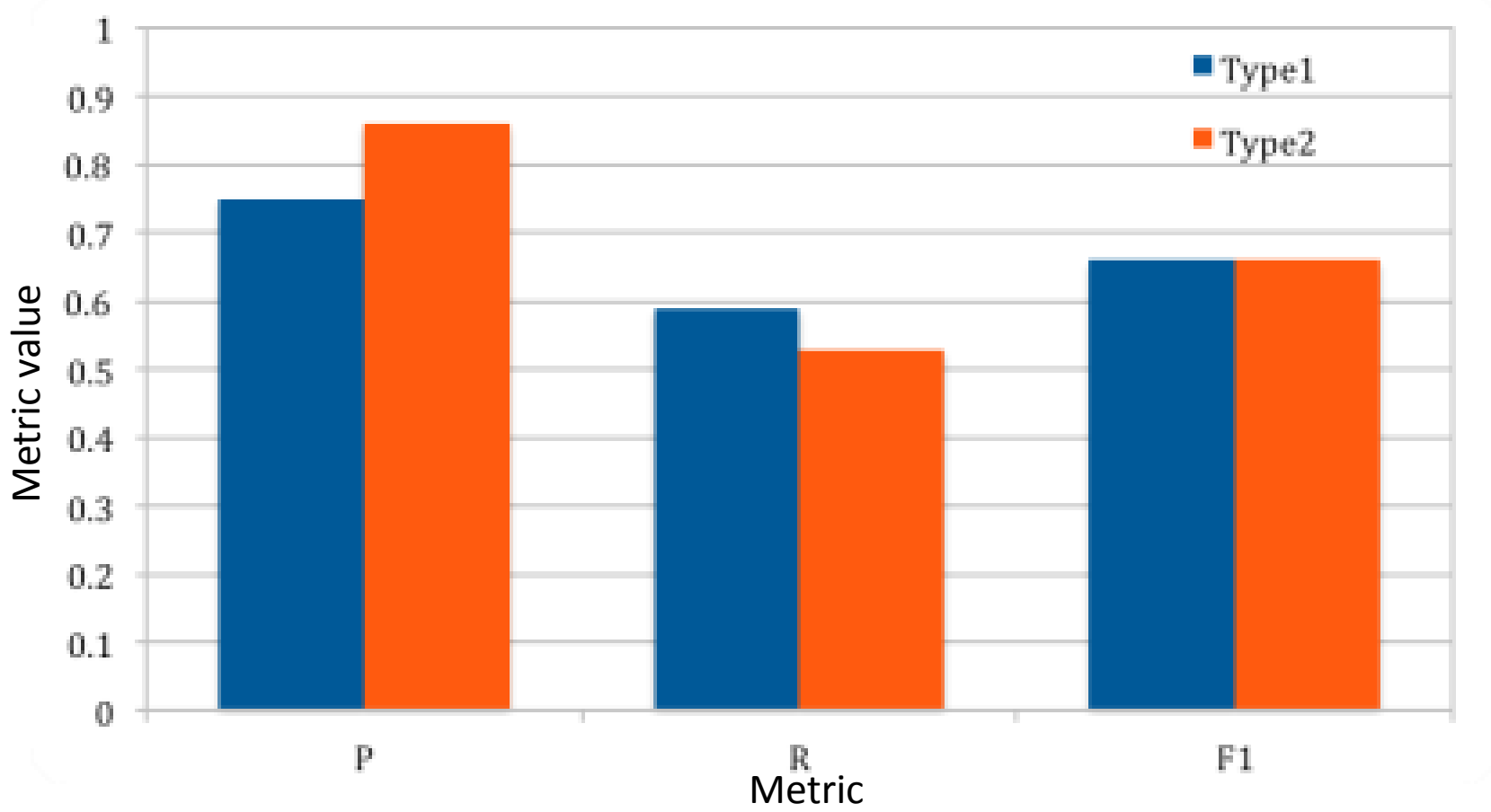

**Figure 4. Polarity assessment of the generated summaries.**

Figure 4 shows the results of the rating of the summaries with respect to their polarity and topic relatedness, respectively. The two summarizers are equivalent in their ability to respond to changes in the value of the polarity slider (60 F1-measure). While raters can detect polarity slanting with relatively high precision (0.73 for Type1 and 0.84 for Type2) there are many cases in which a polarized summary is not recognized by the raters, i.e. recall < 0.6 for both summarizers. We speculate that the low recall figure might be due to the fact that many news collections are polarized per-se (e.g., most passages in a cluster about a military conflict are necessarily heavy with negatively slanted words), offering limited latitude in lexical choices of the summarizers.

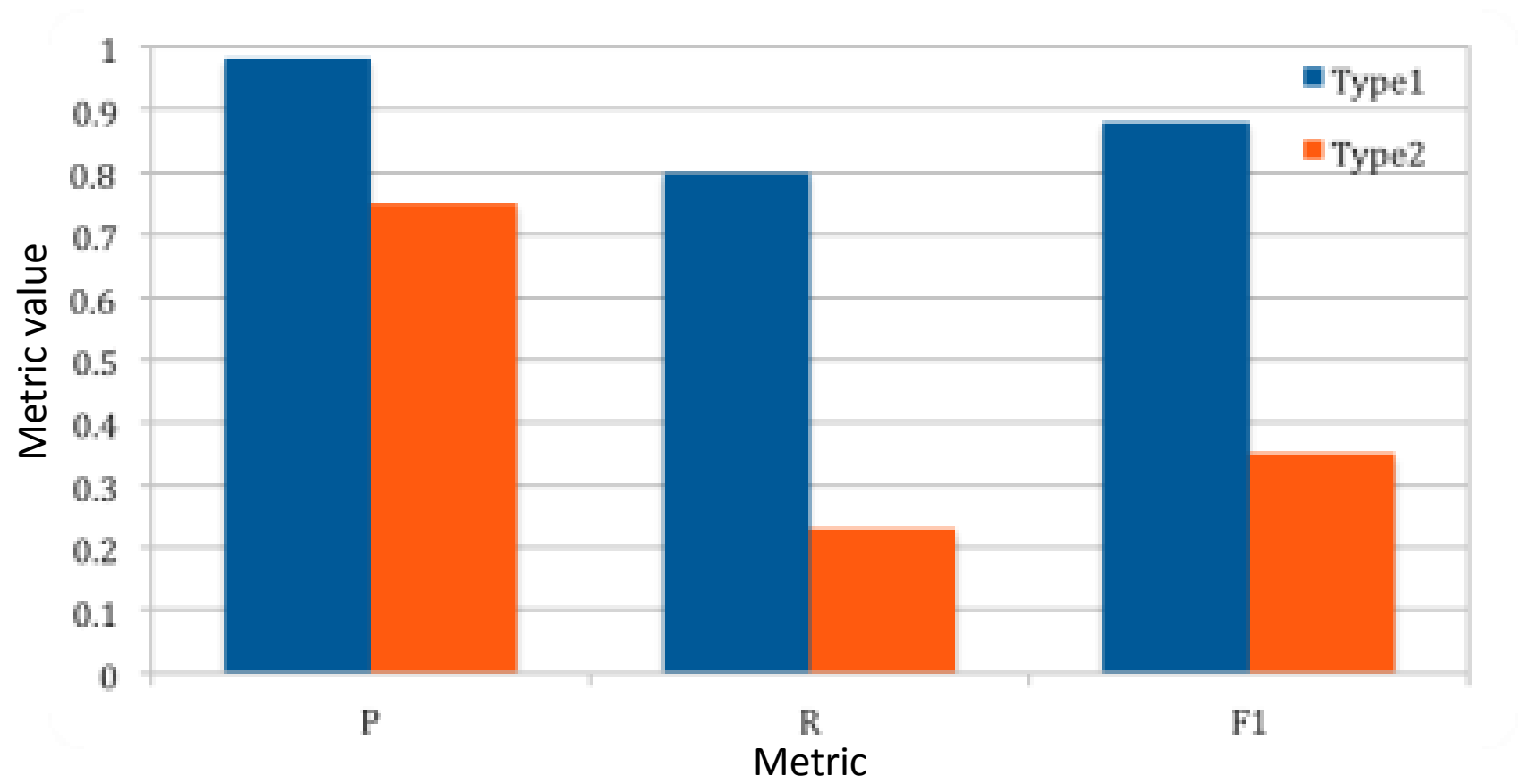

**Figure 5. Topic relatedness assessment of the generated summaries.**

Finally, Figure 5 shows that for topic relatedness Type1 has significantly superior performance than Type2 (0.87 F1 vs. 0.35). The low figure of recall for Type2 (0.22) and relatively high recall (0.75) suggest that in many cases the raters preferred to abstain instead of giving a random answer. This might be due to the fact that being inherently entity-based, the target topics selected for the evaluation were already mentioned in the summaries generated by Type2 even before acting on the control, making it difficult for raters to separate neutral summaries from slanted ones.

## 2.4 Related Entities Panel Evaluation

In D5.2.3 ("Refined diversified news service") [9] we have conducted an evaluation of the quality of the related entities displayed in the related entities panel, labelled (6) in Figure 21. Based on the collected feedback we run another round of evaluation to assess the progress of the integration of Google, JSI and OntoText technology in the latest iteration of DiversiNews.

### 2.4.1 Methodology

The methodology of the evaluation was exactly the same employed for the previous iterations. Two expert raters annotated the relevance of the related entities displayed by the DiversiNews interface for the 31 news collections listed on DiversiNews View1 (show in Figure 13) at the time of the experiment. The annotators were asked to judge the quality of the related entities from the point of view of a user who has an interest in the currently selected collection of news, and who would like to continue browsing related news that might present a different angle on related subjects. Each set of entities has been evaluated on a 5-points Likert scale, where a score of 1 means "completely irrelevant" and a score of 5 means "completely relevant". In the cases in which no entities were displayed, the annotators were instructed to label the example as 0, so as to differentiate this case from the cases in which they were dissatisfied with the results.

### 2.4.2 Main Findings

Figure 6 shows the results of the three rounds of related entities evaluation, which we conducted respectively in February, March and July 2013. The plots show the steady improvement in the quality of the retrieved entities, with the average rating passing from 1.25 in February to 2.87 in July. While the coordination between JSI and OntoText had already produced a significant improvement by the time of the last evaluation, according to the raters there were still some minor issues that would need to be improved to result in a more polished experience. In particular:

- Different spellings of the same entity might appear as different results (e.g., "NYC" and "New Yor City").
- Some of the retrieved entities are too generic to be interesting per-se (e.g., location names).
- Some of the retrieved entities use unusual spelling (e.g., "Meksiko" instead of "Mexico").

All these issues have immediately been addressed after the last round of evaluation.

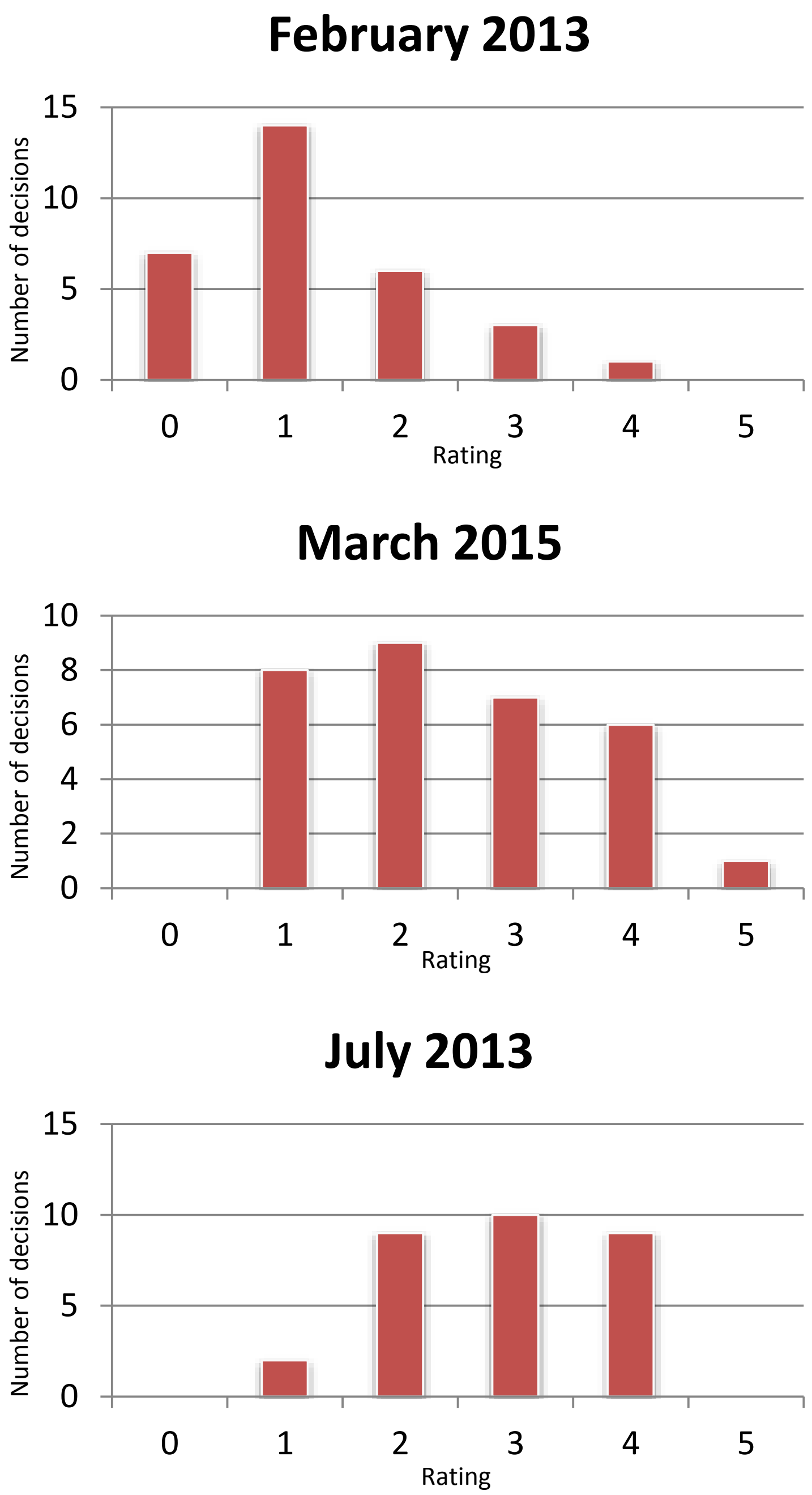

**Figure 6. Progress of the related entities quality evaluation.**

# 3 UX Evaluation

A user experience (UX) evaluation is the process by which the reaction of users to a given product or service are evaluated. As observed in [1], it is non-trivial to evaluate user experience and come up with solid results, since user experience is subjective, context-dependent and dynamic over time. On the other hand, a UX evaluation is a necessary requirement in order to gather qualitative insights and quantitative figures concerning the real potential of any user-facing application.

The goal of the evaluation is both summative and formative, as we have asked the subjects to rate the platform and to provide suggestions on how to improve it. More in detail, these are the main questions that we set off to answer with the UX evaluation:

- General assessment of user experience of DiversiNews:
    - Is the interface self-explanatory?
    - Are the controls ergonomic and accessible?
    - Is the effect of acting on a control predictable and expected?
    - Is the expected workflow clearly exposed?
    - Is the interface responsive?
- Development and Direction:
    - What are the most useful controls from the user standpoint?
    - What are the outstanding issues with the UI that would need to be fixed in order to transform the DiversiNews demo into a real product?
    - Would DiversiNews be deployable as it is (UX-paradigm wise)?
- Perceived utility:
    - To what extent users find the DiversiNews interface useful?
    - Does the interface implement an innovative access pattern to the news?
    - Are news accessed through DiversiNews more understandable?
    - Does DiversiNews make it easier to grasp salient aspects of the news?
    - Does DiversiNews clearly expose diversity?

To this end, we designed an experiment divided in three main parts:

- **Static** evaluation: the subjects have been exposed to snapshots of the DiversiNews interface and asked to understand the role of the UI components without interacting with them. This session is needed to measure the degree to which the interface is **self-explanatory** and will provide useful feedback to improve DiversiNews **accessibility**.
- **Interactive** evaluation: the subjects independently use the interface for a limited time. At the end of this time, they have been asked questions about the quality of the interaction, the responsiveness and the ergonomics of the interface.
- **Perceived utility** evaluation: the subjects answered questions about the utility of the individual components and their potential impact on their news-browsing habits. They answered specific questions about the potential of the different components to highlight and emphasize diversity of opinion in news.

Annex A provides lists all the questions making up the questionnaire, as well as breakdown of the answers provided by the subjects for each of the question. In the remainder of this section, we will provide a bird's eye view of the structure and the results of the evaluation, as well as a commentary on the main lessons learned from the static, dynamic and perceived utility evaluations.

## 3.1 Evaluation subjects

The administered questionnaire contains some questions that we used to profile the potential user base of DiversiNews, as shown by the charts in Annex A.2. Two typologies of users have been involved: 14 casual users (i.e., average users of a news portal) and 2 professional users, i.e., news operators working for a press office. Only one of the subjects is younger than 20, while most of the subjects are in the age ranges 20-29 (62.5%) and 30-45 (31.25%). Concerning their education level, all the participants have a relatively high education level: most of them (75%) either have a bachelor's or a master's degree, while the remaining 25% are evenly split between undergraduates and PhDs. We also profiled the internet usage habits of the subjects. All the subjects declared to use the internet for several hours every day. Two of them (12.5%, the news professionals) declared to use the internet mostly for work, while all the others use it both for work and pleasure. We also asked the subjects what is their preferred way of accessing news. Two thirds of the subjects mostly access the news through online news portals such as Google News, 12.5% declared to read news mostly through social networks (Facebook, Twitter, Google+) and 6.25% from newspapers. 12.5% of the subjects do not really pay attention to news, and surprisingly none of the subjects considers either radio or television as their main source of news.

## 3.2 Static evaluation

### 3.2.1 Objectives

The static evaluation aims at assessing how self-explanatory DiversiNews interface is, and to what extent the different controls embedded in the interface conform to user expectations. These are critical aspect in a UX evaluation, as 1) meeting user expectations, 2) doing what users want to do and 3) providing sufficient information to get the job done are three usability requirements that, if not met, ingenerate frustration in the user **Error! Reference source not found.**.

### 3.2.2 Methodology

The subjects have been asked to study screenshots of the main views of DiversiNews for a limited amount of time, and then two answer questions about their expectation about the functionality implemented by each UI component and its expected behaviour. After forming their own opinion, the users received a detailed explanation of the actual functionality of the components, and asked to confirm the correctness of their initial impression. To avoid biasing the subjects, instead of referring to the interface components by means of a descriptive name (e.g. "summary box" or "related entities panel") we identified each component with a numeric identifier. As already mentioned, the screenshots that we used are shown in Figure 13 (View1: news search / cluster selection) and Figure 21 (View2: diversified news browsing).

### 3.2.3 Main findings

The static evaluation has confirmed that DiversiNews interface is **very clear** and **self-explanatory** from the very **first moments** of usage. The very large majority of the subjects **correctly identified** the **function** and the **behaviour** of all the components even before actually using them. Detailed results are available in A.3 and A.4.

#### 3.2.3.1 View1: News search / cluster selection

More in detail, concerning the news search / cluster selection view of DiversiNews (View1, shown in Figure 13), only two of the users had wrong expectations concerning the function of the search box, while 3 users found the links to the news clusters at the bottom confusing. For the latter, the offsetting part appears to be that when clicking on a specific news title the users would expect to be redirected to the specific news, as opposed to a view of the corresponding news cluster. The wording in the caption of the page ("A tool for the interactive exploration of news") seems to contribute to the misunderstanding.

**Lessons learned**

Even though View1 is considered highly self-explanatory by more than 80% of the subjects, the comments of the not satisfied users suggest that further improvements are possible:

- To make the list of links at the bottom of the view more self-explanatory, visual hints have been added clarifying that the news title is a placeholder for a whole cluster of related news. We added to each title the number of news in the corresponding cluster, with a label reading "Related news in cluster: XX", and moved the link from the news title to this. In this way, we made it clearer that the title is just a label for a collection of related news.
- The wording of the labels on View1 has been corrected to stress even more that the news browsing experience is based on clusters of related news, as opposed to individual news articles.

#### 3.2.3.2 View2: Diversified news browsing

The users immediately understood that View1, the main view of the interface (Figure 21), is a dynamic page, and correctly guessed the relation between the components and their interaction modality.

In particular, 75% of the users understood just by looking at a screenshot that component (1) in Figure 22 is the summary of the collection of related news displayed by component (2), i.e. the contents of the news cluster. The subjects clearly identified (3), (4), (5) and (6) as the interactive elements of the panel (Figure 24). Similarly, the users easily understood which components would have an effect on the ranking of the news (Figure 26). Only 2 of the users did not understand that the list of news in (2) is related to the query submitted through View1. These are the same users that, while evaluating View1, had commented that they would expect the next page to consist of single news as opposed to a collection of related ones. Therefore, it might be reasonable to attribute this misunderstanding to wrong expectations created by View1 rather than deficiencies in the design of View2. On the other hand, four of the other subject said that it took a while to figure out the connection. This finding suggests that UX might be improved by adding visual aids that clarify this point.

Most of the subjects correctly understood that acting on any of the interactive controls on the right would affect the content of the summary and the list of news (Figure 27 to Figure 30). On the other hand, on visual inspection approximately half the subjects were expecting that acting on each interactive control would also have an effect on the other controls, i.e., they imagined that all the controls were insisting on the same space and that changing any of the views would trigger a change of view in all the other controls. When told about the function of each of the panels, a large majority of subjects (from 75% to 93.75%) stated that their function was immediately clear, again confirming the high degree of self-documentation of the interface.

**Lessons learned**

According to the collected feedback, there are two main directions along which View2 can be made even more self-explanatory:

- The connection between View1 and the content of View2 has been made more explicit. This is actually the same issue already raised for View1, which has the responsibility of preparing the users for the effect of the actions that they can perform on it.
- The interface has been made more explicit in stressing the fact that each panel operates on a different dimension of the news collections, and that acting on any of the panels is not supposed to have an effect on the others.

## 3.3 Interactive evaluation

### 3.3.1 Objectives

With the interactive evaluation we assess the reactions of users when they are exposed to the actual interface. An interactive evaluation makes it possible to understand if and how an interface is usable and ergonomic, and to what extent a developer technology has a real potential as an end-user product. By directly using an application, users can also point out technological limitations (e.g., exceedingly long loading time) that limit the potential benefit of using a product.

### 3.3.2 Methodology

We pointed the subjects to the online demo of DiversiNews at http://aidemo.ijs.si/diversinews/, and asked them to use and explore the interface for a limited amount of time (up to 10 minutes). After that, we asked the subjects to comment on the usability of the interface and its components.

### 3.3.3 Main findings

The outcomes of the interactive evaluation have been **extremely positive**, and we collected a lot of very **valuable feedback**. ~81% of the subjects was either very pleased or pleased with the response time of the interface, while the remaining ~19% found it adequate. The vast majority of the users confirmed that the interactive panels behaved as expected. The layout of the elements on the page was considered by and large intuitive. One of the users has commented that two much estate space is not used on a high-resolution display, while another suggested that the controls panel should wrap on the left side of the page, as a consequence of left-to-right reading habits. Another subject suggested that panels (3) [relevant topics] and (6) [relevant entities] should be closer as they are conceptually related.

The subjects confirm that the interface does a good job at enforcing the notion that (1) is a summary of the news in (2). All the users were happy with the ergonomics of (5), and half of the users found all the controls to be intuitive. Some of them suggested several ways in which (3), (4) and (6) could be improved.

### 3.3.4 Lessons learned

- Since the users confirmed that understanding that (1) is a summary of the news (2) is critical for a good UX (Figure 45), we added visual clues to View2 to stress this point.
- The polarity slider (4) has been improved by limiting it to a set of discrete values, as the continuous movement previously implemented 1) made it difficult to replicate a precise configuration and 2) induced in the users the expectation that every micro-adjustment would result in a different summary / ranking of news.
- The visualization of the relevant topics panel (3) has been improved by displaying less specific terms and by reducing the number of overlapping items.
- Maybe due to the wording “Search for related entities”, the related entities panel (6) was expected to behave like a search in the current document as opposed to triggering a new query resulting in the selection of a different set of articles. Hence, we changed the label of the panel to “Browse news involving related entities”. The interface has also been adapted so as to make clear that, while (3) lists terms appearing in the current news cluster, (6) list terms that are not.
- The search box on View1 now triggers a search whenever the user hits the “enter” key.

## 3.4 Perceived utility evaluation

### 3.4.1 Objectives

This evaluation aims at understanding the real potential of DiversiNews as a platform for diversity aware news browsing. In particular, we want to understand if the main pillars of DiversiNews news access paradigm (i.e., allowing users to explicitly control diversity-related dimensions and providing summaries to synthesize diversity and expose it more clearly to the users) are ergonomically sound and appealing to a potential user community.

### 3.4.2 Methodology

After having interacted with DiversiNews during the interactive evaluation, the users have been asked a few questions about their reaction to the platform. The users have been invited to leave extensive comments about all the components of the interface, pointing out the pros and cons of each of them with a special attention towards their potential with respect to diversity awareness.

### 3.4.3 Main findings

The outcome of the perceived utility evaluation confirms that the subject found **summaries** to be an **effective device** to **capture** and represent relevant information and **diversity** of opinion. The users confirmed that the controls implemented by DiversiNews **succeed in modelling different dimensions of diversity** and provide a more rounded, diversity aware paradigm for online news consumption.

~82% of the subjects found that summaries are at least adequate in quality (Figure 58), and confirmed that they are structurally relevant for the implementation of the desired functionalities (Figure 59). Almost 80% of the subjects supports our initial intuition that summarizing collections of related news according to different criteria is an effective way of letting relevant information emerge (Figure 62) and stress diversity in news (Figure 63). Even though most users affirm that they would not accept a news browsing interface that would hide from them the sources and the details of the summarization technology (Figure 61), ~19% of the users was so enthusiastic about DiversiNews concept to say that with an “ideal” summarization technology they would also consider using an interface in which the actual list of news were not displayed (Figure 60). According to the majority of subjects, all the interactive panels implement a functionality that is considered desirable in a news browser and that is found instrumental in easing the discovery of diversity. In this respect, users appear to have especially appreciated the geographic source widget (Figure 66 and Figure 67), while the related entities widget is found to be less effective than the others in letting diversity emerge (Figure 71). We speculate that this latest finding might be due a presentational issue, as the related news panel is the only one that triggers a news collection change as opposed to a reorganization of the already selected one.

The subjects provided useful suggestions on how to improve DiversiNews’ widgets, as detailed in A.6. To further confirm the soundness of the concept and the core design elements, the vast majority of the remarks pertain presentational issues, e.g., using a sparser representation in the relevant topics panel (3) and possibly replacing text with figures, or using per-country bubbles in the geographic source panel (4).

We would like to conclude this section with the open comments collected from the raters that once again confirm how DiversiNews concept was received positively (blue) and how all the negative criticism received (red) is limited to minor presentation issues that we addressed immediately after the evaluation in order to provide a more refined UX:

- **“Excellent tool & project,** but needs some UI adjustments to fly”
- “Make it so that pressing enter also gives results. Now I have to click on the search button.”
- “When I search for something in the search box I don't want to have to press "Search" every time. It should activate when pressing "Enter" or automatically. **I really like the interface and it would be really nice to use it. I would start reading more news, as the summary part seems great.”**
- **“Great page, I like the summaries very much,** the overall design could be improved.”

# 4 Conclusions

We have presented the results of an extensive evaluation of the technological components making up the DiversiNews platform, and of the usability of the interface and utility of the platform by means of a detailed user experience survey.

The outcome of the evaluation has clearly confirmed the validity of all the major design choices leading to the current implementation of DiversiNews. In particular, users are very supportive of the idea of an interface built from the ground up with the objective of stressing diversity of opinions and points of view, and expressed appreciation for a tool that lets them control the criteria based on which the information is presented. The summary-centric approach implemented by DiversiNews is very appealing for both casual and professional users, as it reduces information overhead while making it possible to grasp different opinions by reading just a few sentences after a few small tweaks of the controls of the interface. Users praised the diversity-aware controls that are currently implemented, which were found to be very intuitive from the very first inspection, and suggested us to further extend the controlled dimensions by adding even more controls.

Concerning the quality of the generated summaries, we have observed that the quality of the generated summaries is considered to be adequate for the objectives of DiversiNews in ~70% of the cases, an extremely good result considering the difficulty of open-domain multi-document summarization on real-world data. A comparison of the two summarizers that DiversiNews can employ has shown that one of them (Type1, based on well-established, state-of-the-art technology) provides slightly better results in terms of Informativeness and sensitivity to polarity adjustments, while significantly outperforming the other summarizer (Type2, based on an experimental entity-driven approach to summarization) concerning its ability to focus on specific topics.

The extremely useful feedback that we collected from the raters of DiversiNews components and the subjects of the UX study has already been to use, resulting in a further improved diversified news browsing experience.

# References

[1] Effie Law, Virpi Roto, Marc Hassenzahl, Arnold Vermeeren and Joke Kort. *Understanding, Scoping and Defining User Experience: A Survey Approach*. In Proceedings of Human Factors in Computing Systems conference, (CHI 2009).

[2] Ben Shneiderman, Cathérine Plaisant, Maxine Cohen and Steven Jacobs. *Designing the User Interface* (5th Edition). Pearson Addison-Wesley, 2009.

[3] Ronald A. Fisher. *Statistical Methods for Research Workers* (Twelfth ed.). Oliver and Boyd, 1954.

[4] Katja Filippova and Yasemin Altun. *Overcoming the Lack of Parallel Data in Sentence Compression*. To appear in Proceedings of Empirical Methods for Natural Language Processing (EMNLP 2013).

[5] Katja Filippova. *Multi-sentence compression: finding shortest paths in word graphs.* In Proceedings of the 23rd International Conference on Computational Linguistics (COLING 2010).

[6] Enrique Alfonseca, Daniele Pighin and Guillermo Garrido. *HEADY: News headline abstraction through event pattern clustering.* In Proceedings of the 51st Annual Meeting of the Association for Computational Linguistics (ACL 2013).

[7] Delia Rusu, Mitja Trampus, and Andreas Thalhammer. *Diversity-aware Summarization.* Deliverable D3.2.1 of the RENDER project, 2013.

[8] Enrique Alfonseca and Mitja Trampus. *Diversified News Service.* Deliverable D5.2.2 of the RENDER project, 2012.

[9] Enrique Alfonseca, Daniele Pighin, and Mariana Damova. *Refined Diversified News Service.* Deliverable D5.2.3 of the RENDER project, 2013.

[10] Mitja Trampus, Flavio Fuart, Jan Bercic, Delia Rusu, Luka Stopar, and Tadej Stajner. *DiversiNews, a stream-based, on-line service for diversified news*. In proceedings of *SiKDD 2013*.

The deliverables of the RENDER project can be found at project's web page:
http://render-project.eu/resources/deliverables/

# Annex A UX Evaluation Questionnaire and Per-question Results

In this annex we show the complete text of the UX evaluation questionnaire that was administered to the subjects of the UX evaluation experiment. For each question, we include a chart visualizing the distribution of the answers provided by the 16 subjects. The corresponding question is included in the caption of the chart. Aggregated answers to single-selection questions are visualized as pie charts; in the case of multiple-selection questions, the results are shown as a histogram; finally, for open questions we explicitly list all the answers provided by the subjects. The order of the figures matches that of the questions in the questionnaire.

## A.1 Disclaimer

In this experiment, you will be asked to interact with and evaluate the prototype of a web-based interface for exploring news. The duration of the experiment is of approximately 45 minutes.

Please, answer to all the questions **in the same order** in which they are presented.

**Refrain** from reading the next question until you have not answered the current one.

**Never** go back to fix a previous answer.

Please, answer each question honestly and, in the case of open questions, please provide as much information as possible in the available time.

The most important aspect for us is your first impression. Do not try to give the “right” answer. For example, it might happen that reading the possible answers to a question would change your initial impression. In such cases, please do an effort to tell us what you though **before** reading the options.

## A.2 Subject profiling questions

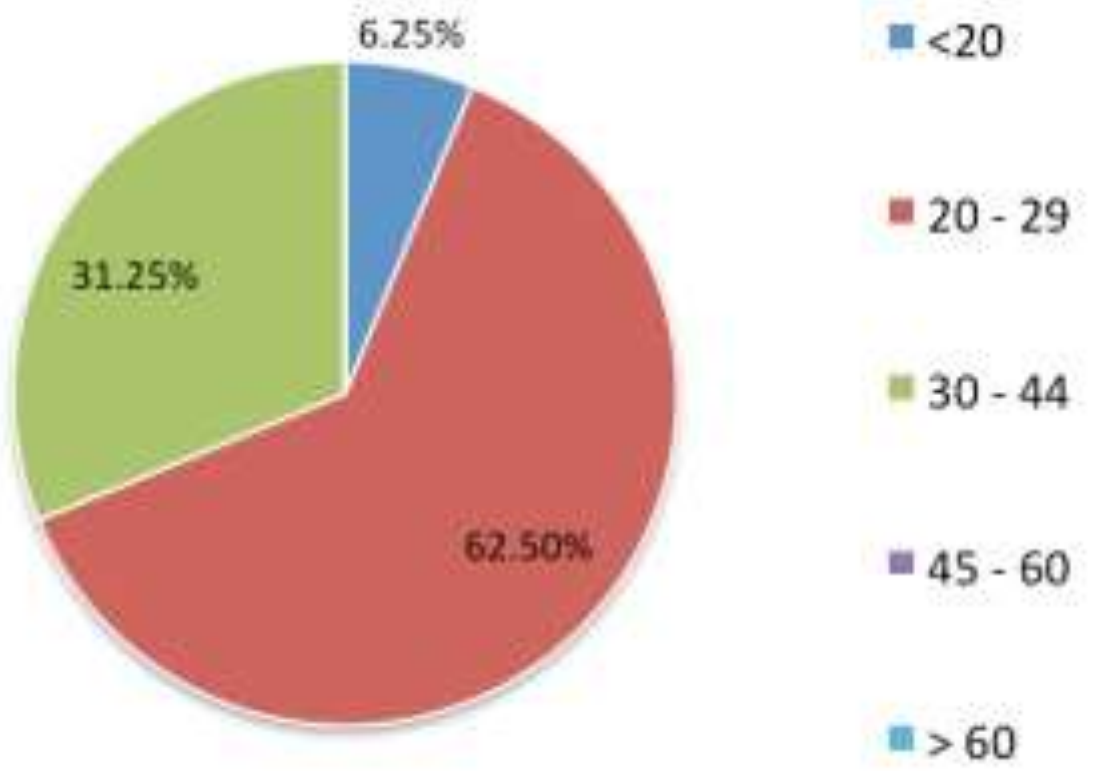

**Figure 7. What is your age?**

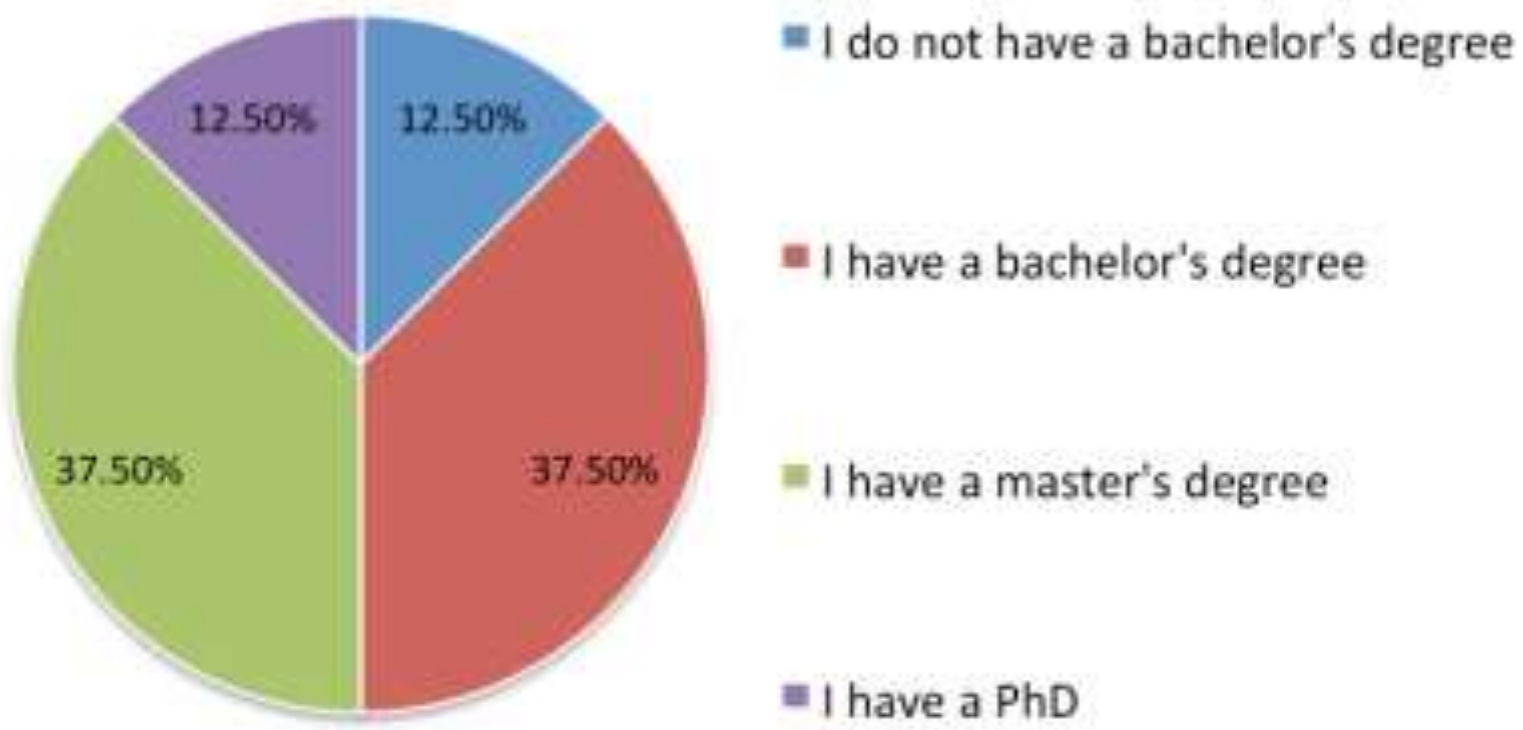

**Figure 8. What is your education level?**

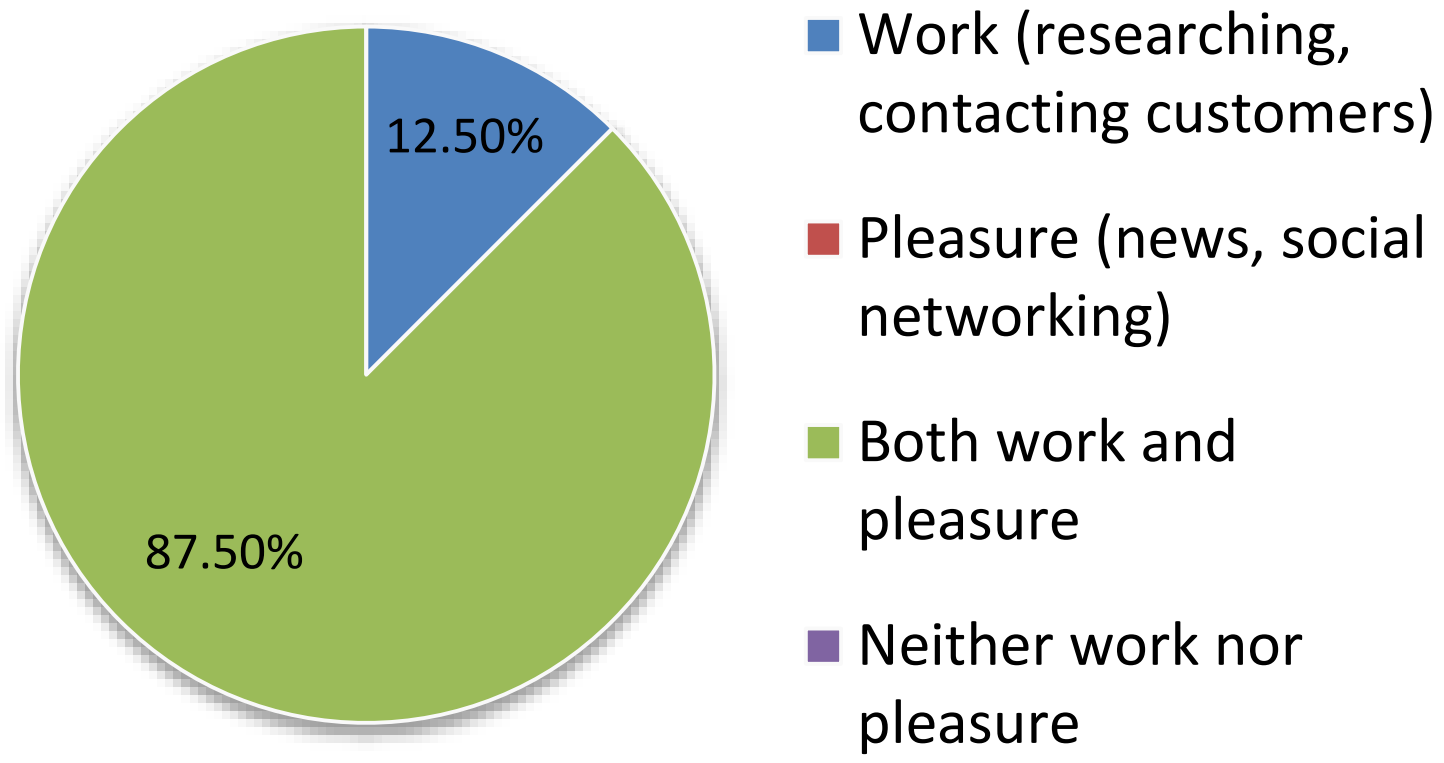

**Figure 9. What is your main usage of the Internet?**

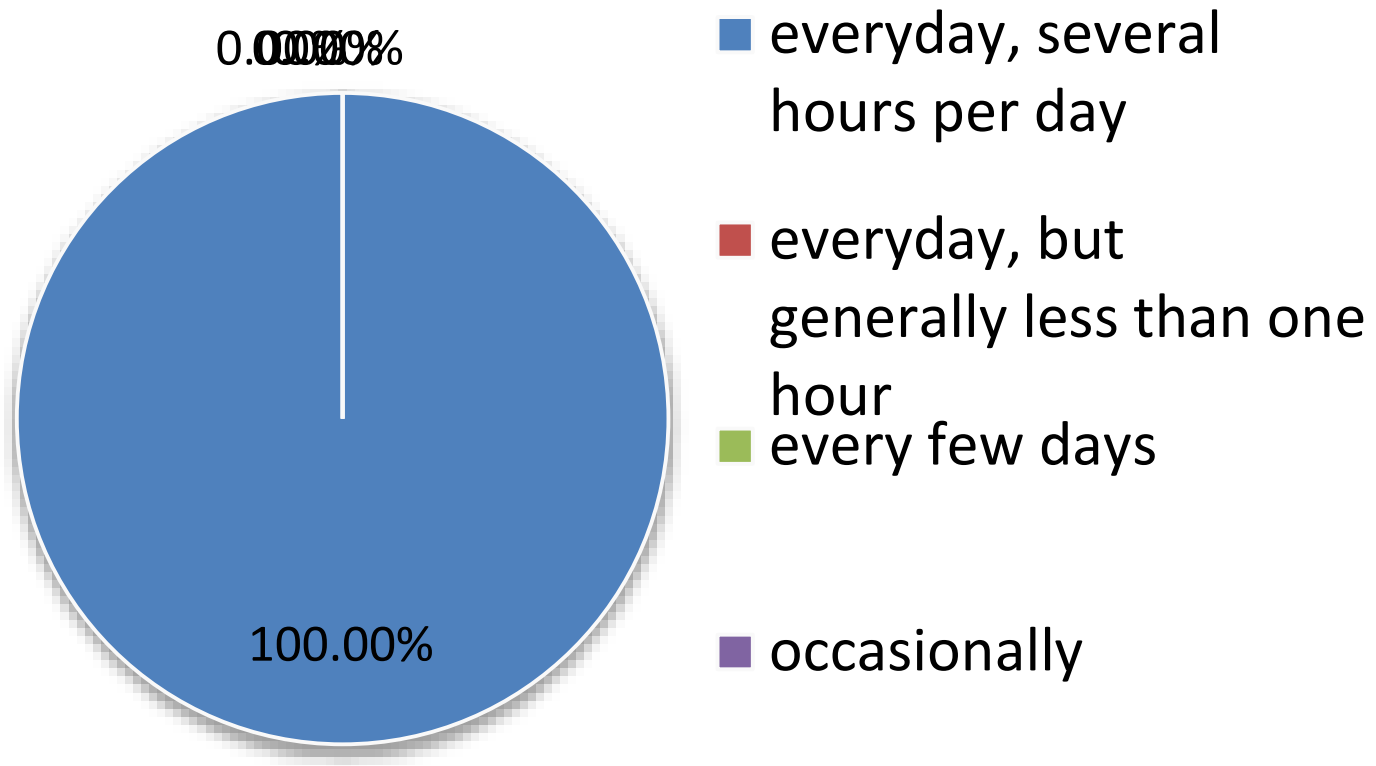

**Figure 10. How much time do you spend on the Internet?**

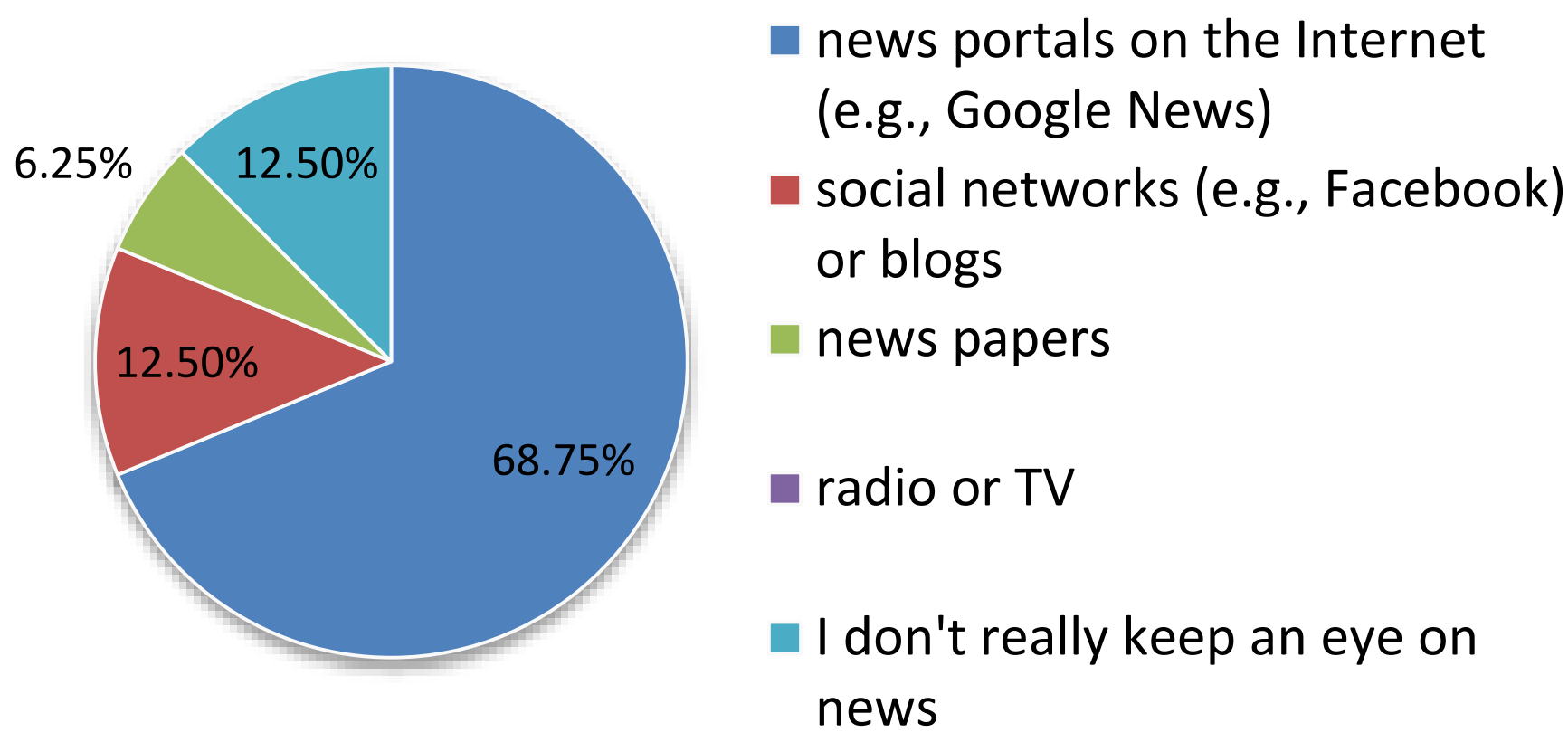

**Figure 11. What is your preferential news channel?**

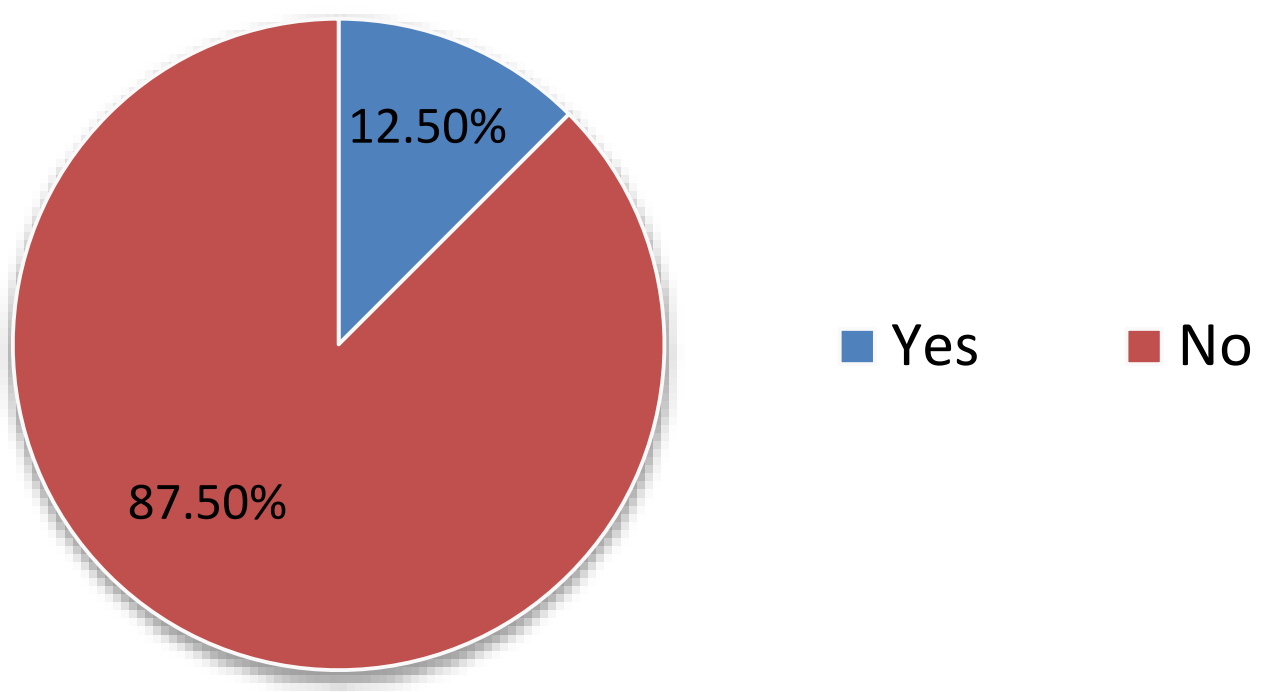

**Figure 12. Are you a news professional? For example: a journalist, a journalism student, a news or public relations agent**

## A.3 Static evaluation of View1

**Estimated time: 5 minutes.**

Please, look for 1 minute at Figure 13 which is a static snapshot of the interface of the news-browsing service. In the remainder of the exercise, we will refer to this view as to View1.

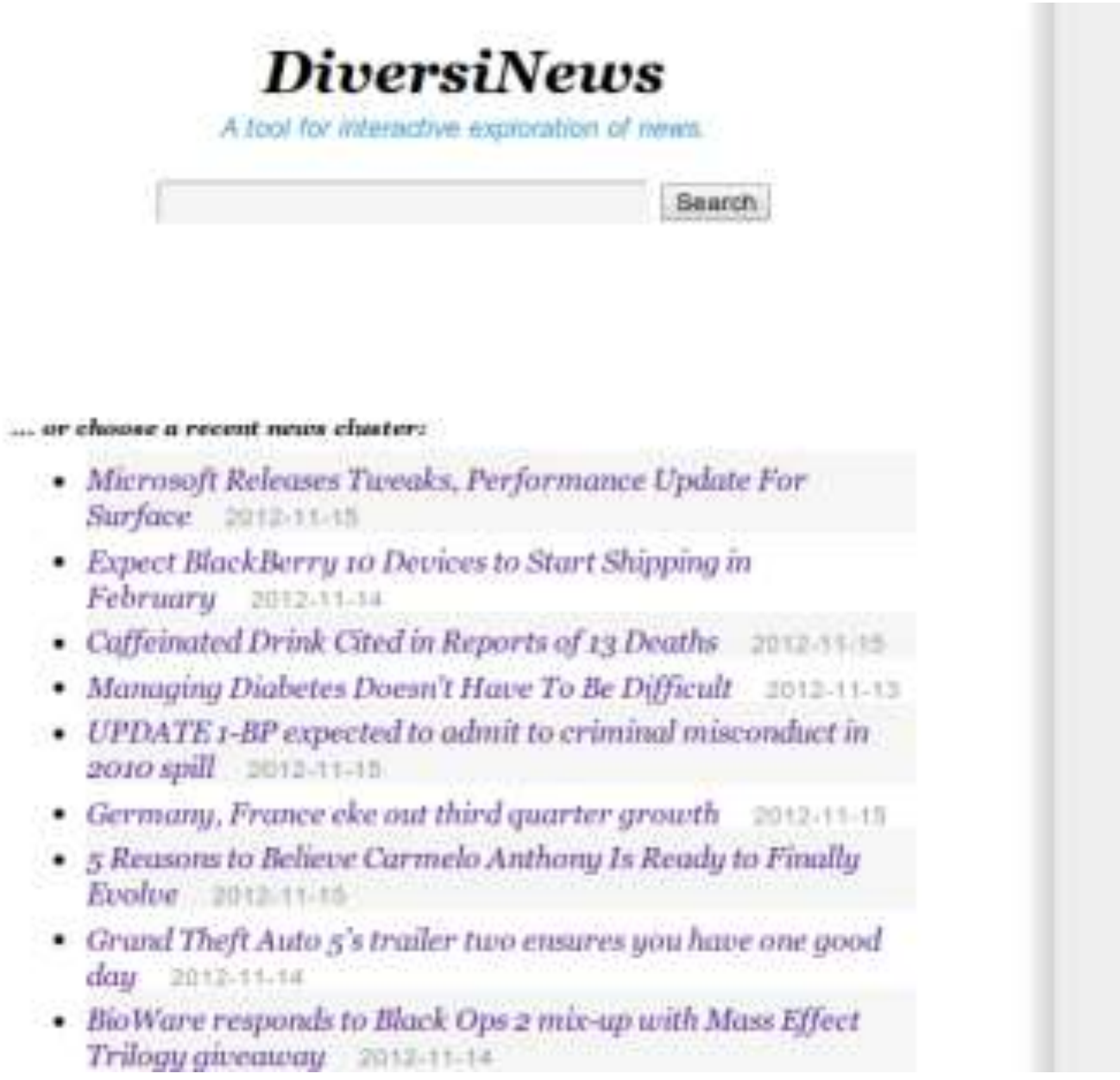

**Figure 13. Static view of DiversiNews View1.**

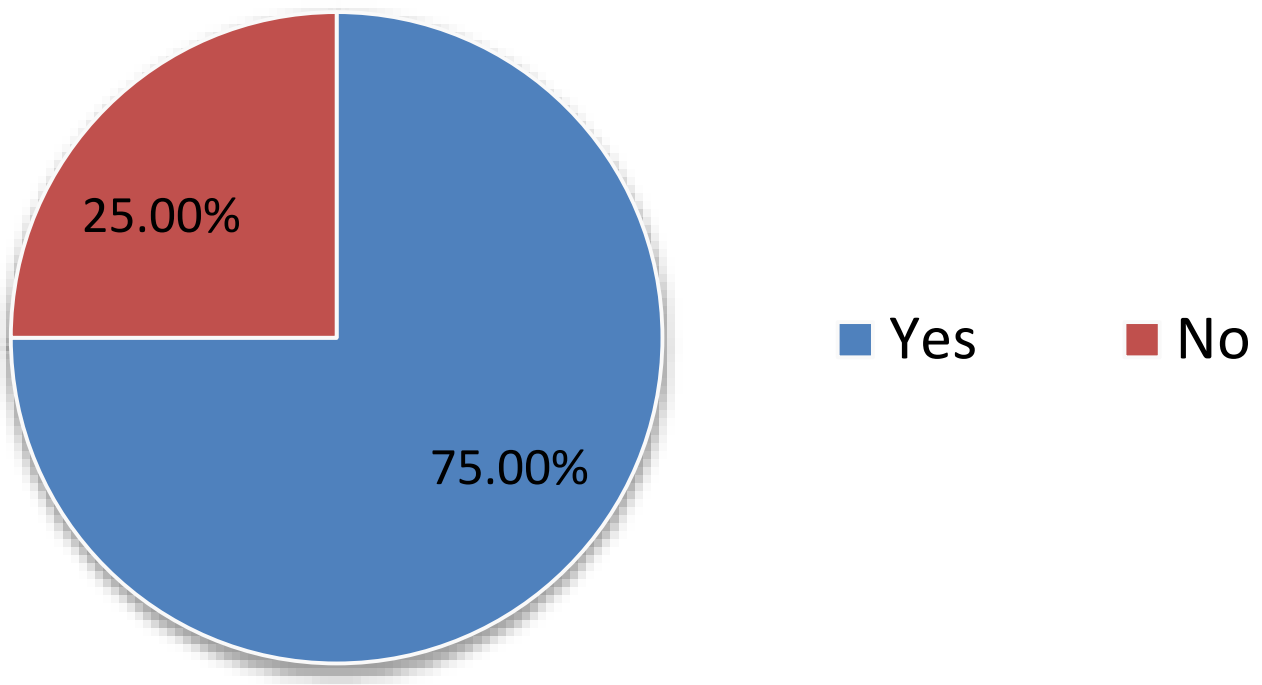

**Figure 14. Did you understand what the search box at the top is for? Say "Yes" only if you have a reasonable expectation of what would happen if you typed something in the text-box and clicked on "search".**

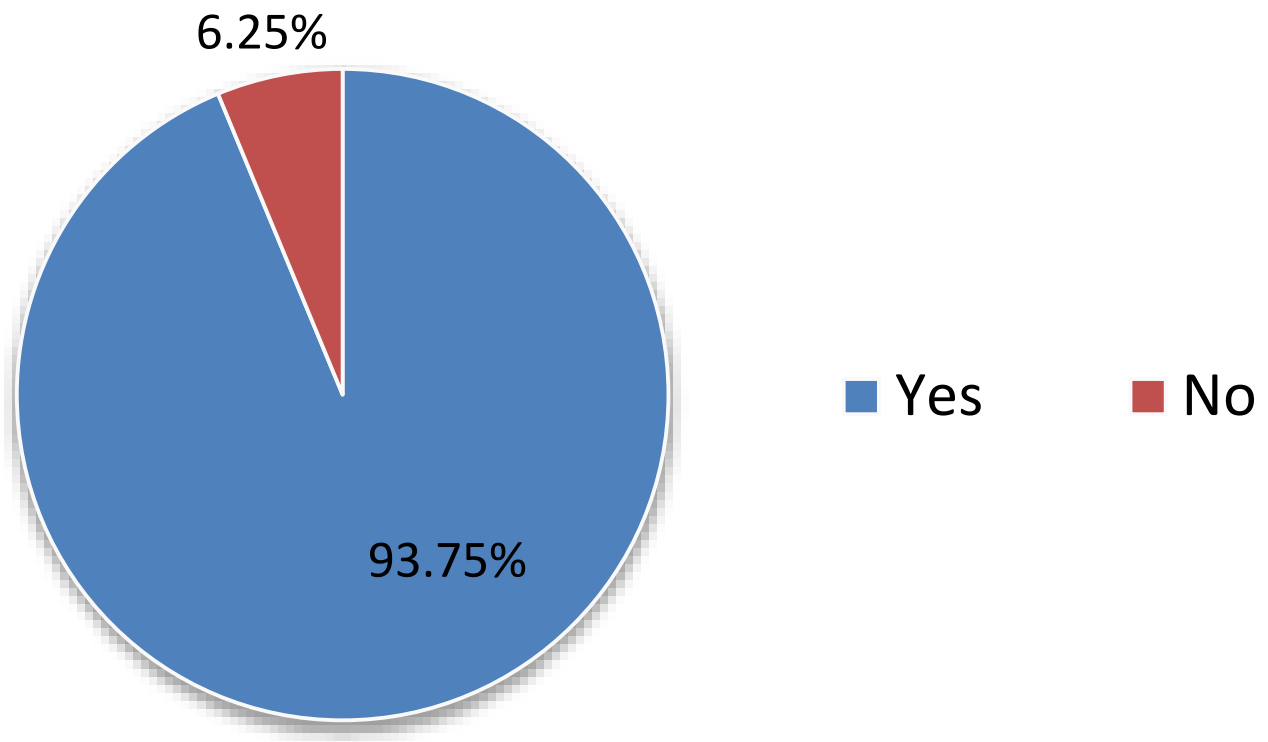

**Figure 15. Did you understand what the list of links at the bottom are for? Say "Yes" only if you have a reasonable expectation of would happen by clicking on any of the links, "No" otherwise.**

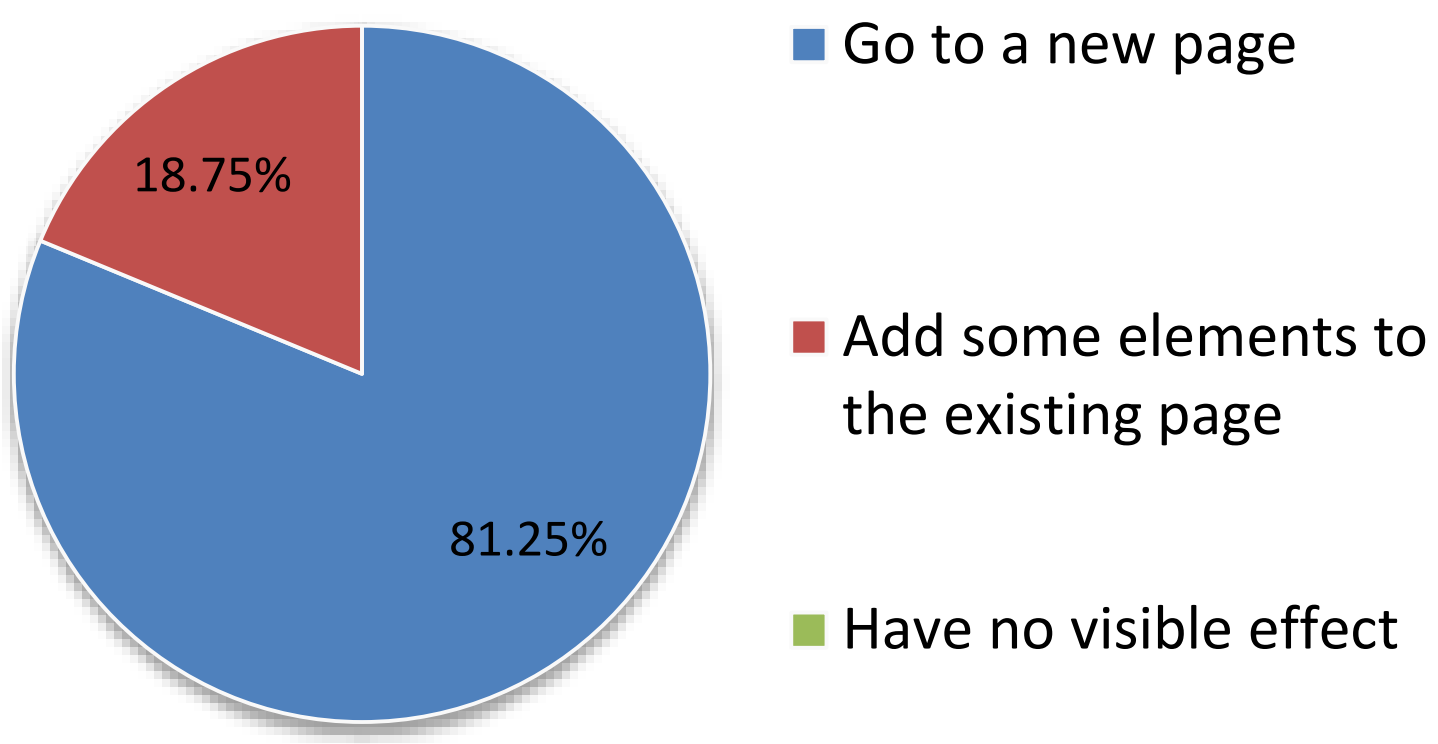

**Figure 16. What is your expectation: typing some words in the text box and clicking on "search" would:**

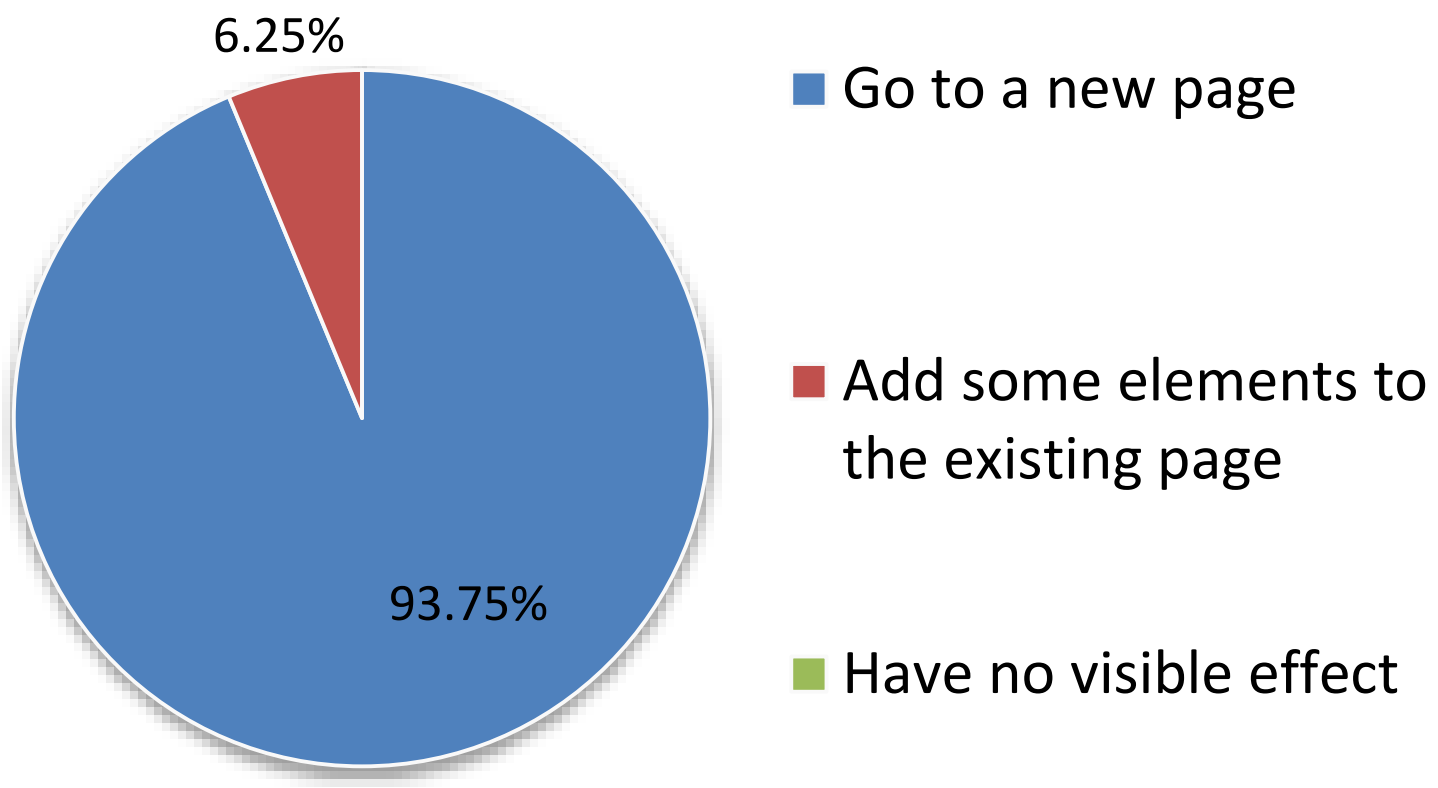

**Figure 17. What is your expectation: clicking on one of the links at the bottom would:**

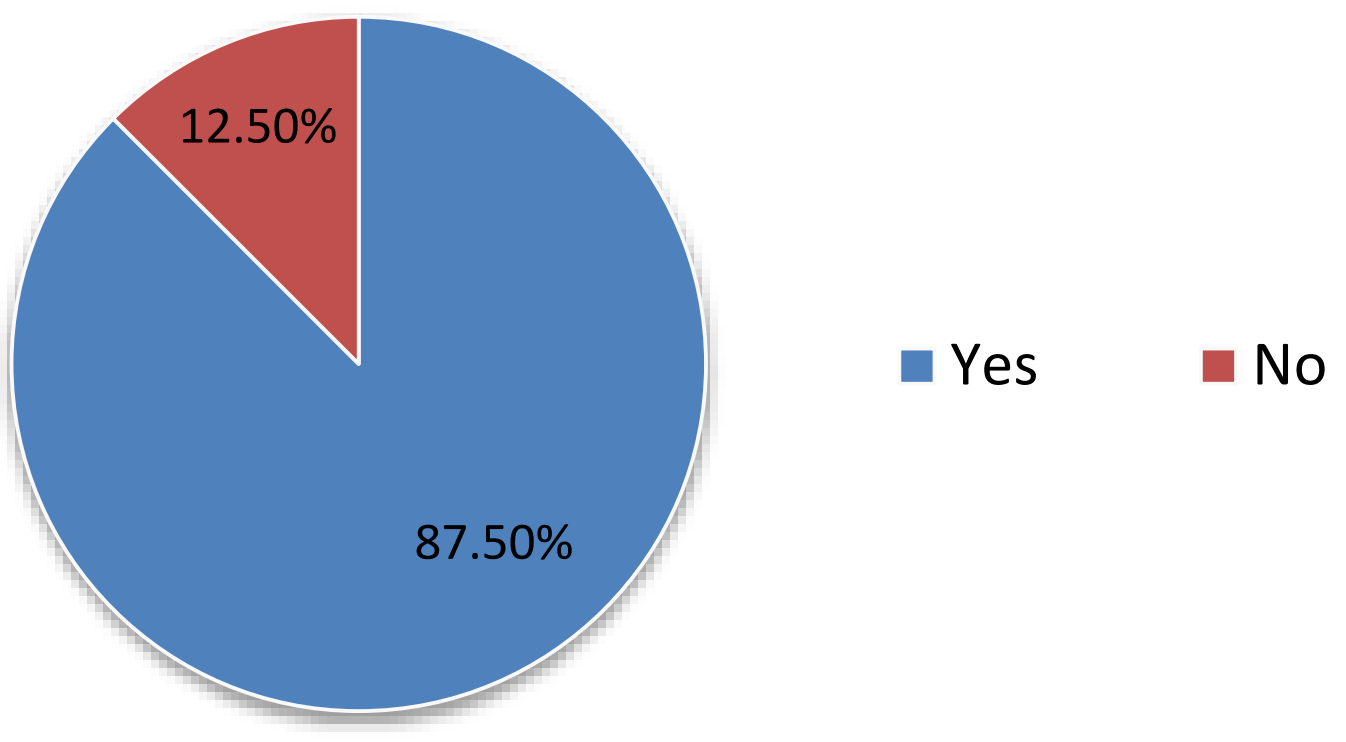

**Figure 18. Typing for one or more keywords in the text box and clicking on "Search" will display a new page from which you can browse a selection of news. These are chosen based on their relevance with respect to the keywords that you searched for. Is this in line with your expectations?**

**If you answered "No" to the previous question, please tell us shortly what you imagined would happen instead. What gave you the wrong impression? Was some component of the interface misleading?**

- First I thought the above mentioned would happened. But then I read "or choose a recent news cluster". I then got confused if I am searching specific news or some kind of clusters.
- I thought it will react just like an `I'm Feeling Lucky` button and go directly to the most recent related news. This is the "A tool for interactive exploration of news" sentence which make me think that.

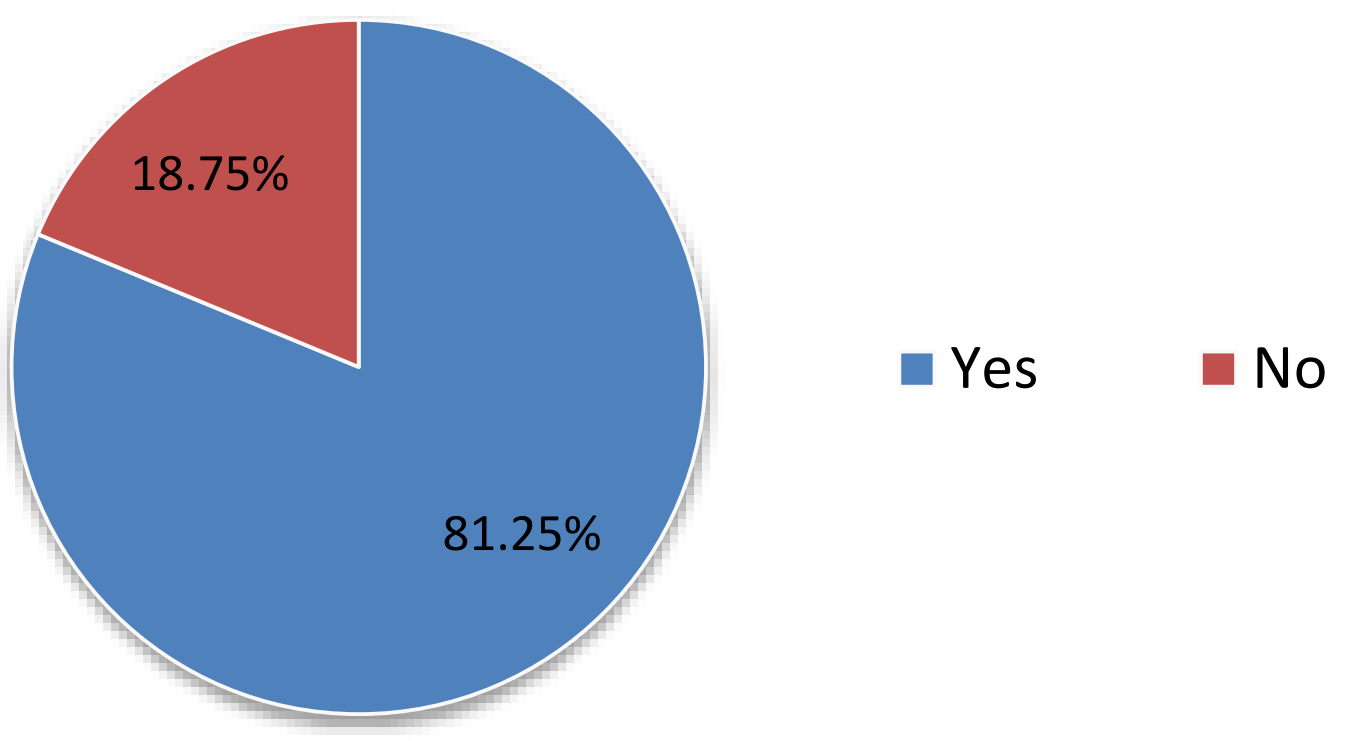

**Figure 19. Clicking on any of the links at the bottom will display a new page from which you can browse a selection of news. These news are all very similar to the one on whose title you clicked on. Is this in line with your expectations?**

**If you answered "No" to the previous question, please tell us shortly what you imagined would happen instead. What gave you the wrong impression? Was some component of the interface misleading?**

- I thought it will just go into the concerned web page.
- I would like if I click on a link at the bottom to show me exactly the news with that title not some that are similar.
- I expected the original news articles as well, not only the related news.

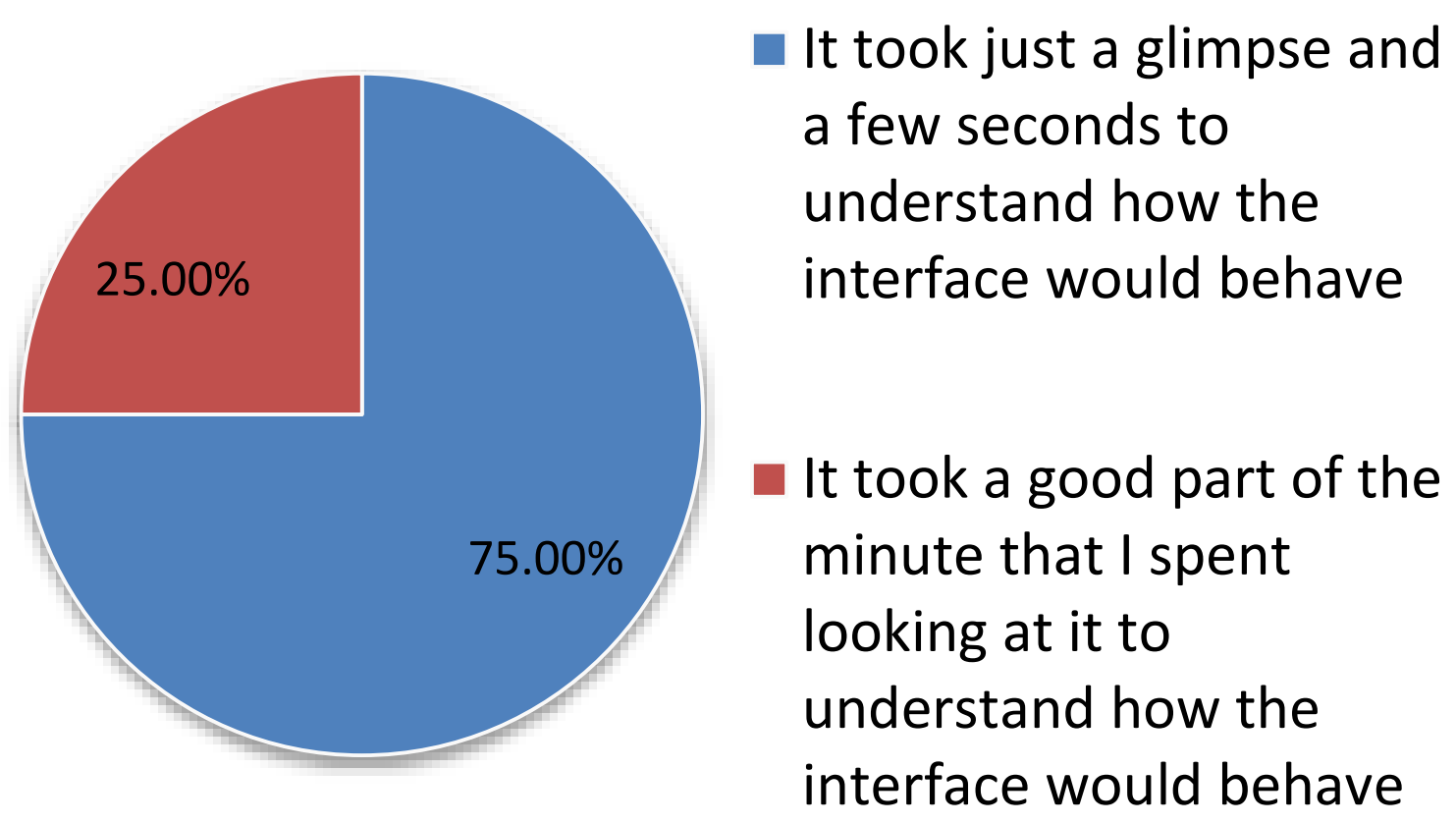

**Figure 20. If you answered "Yes" to the two questions above, which one of the following two statements is true?**

## A.4 Static evaluation of View2

**Estimated time: 10 minutes.**

What is shown below is an example of what you would see after searching for some keyword in the news repository, or after selecting one of the pre-defined collections of news. We will refer to this view as View2. (The six red numbers are not actually part of the interface, we will just use them to indicate different parts of the interface). Please, look at the interface for **2 minutes** and then answer the questions below.

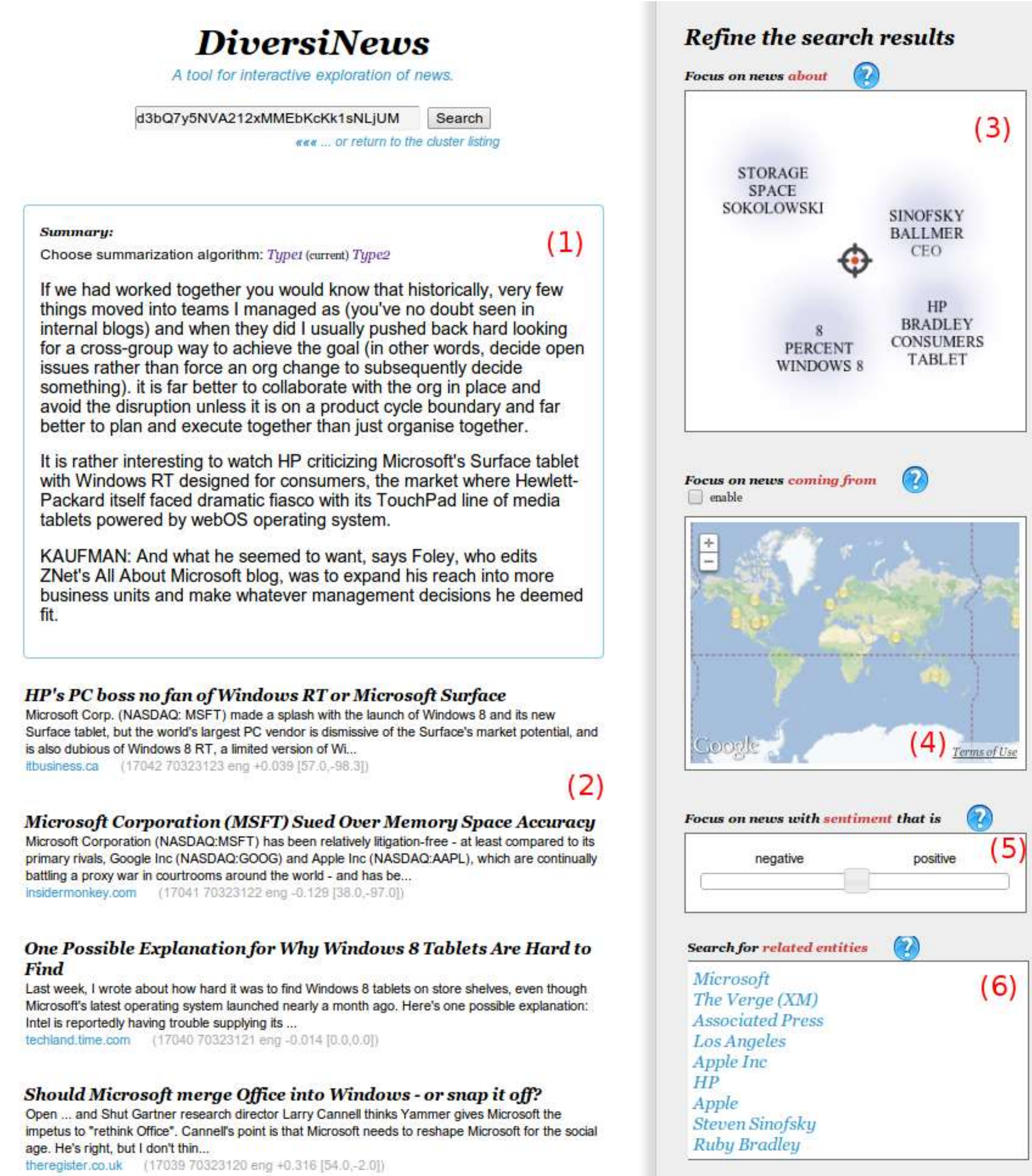

**Figure 21. Static view of DiversiNews View2.**

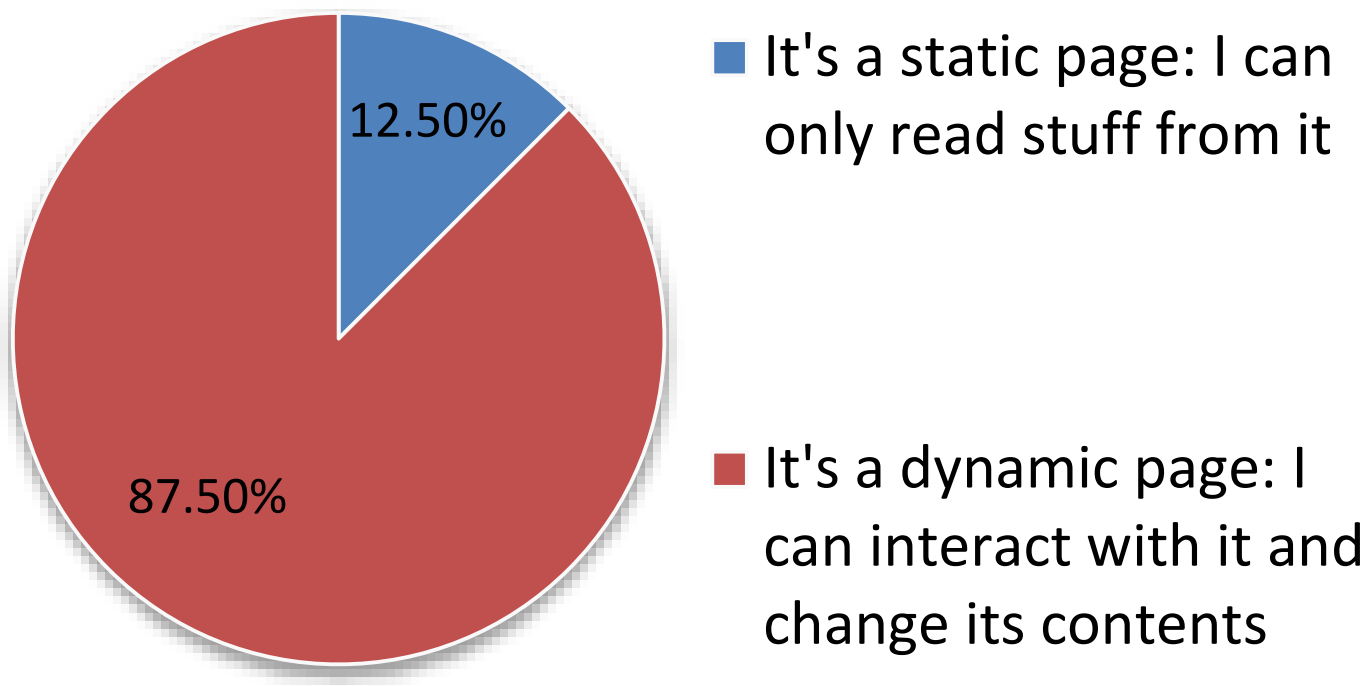

**Figure 22. Based on your first impression, is the one shown a static or a dynamic page? In other words, do you think you can just "read" what's on the page, or can you "act" on the page to change its content?**

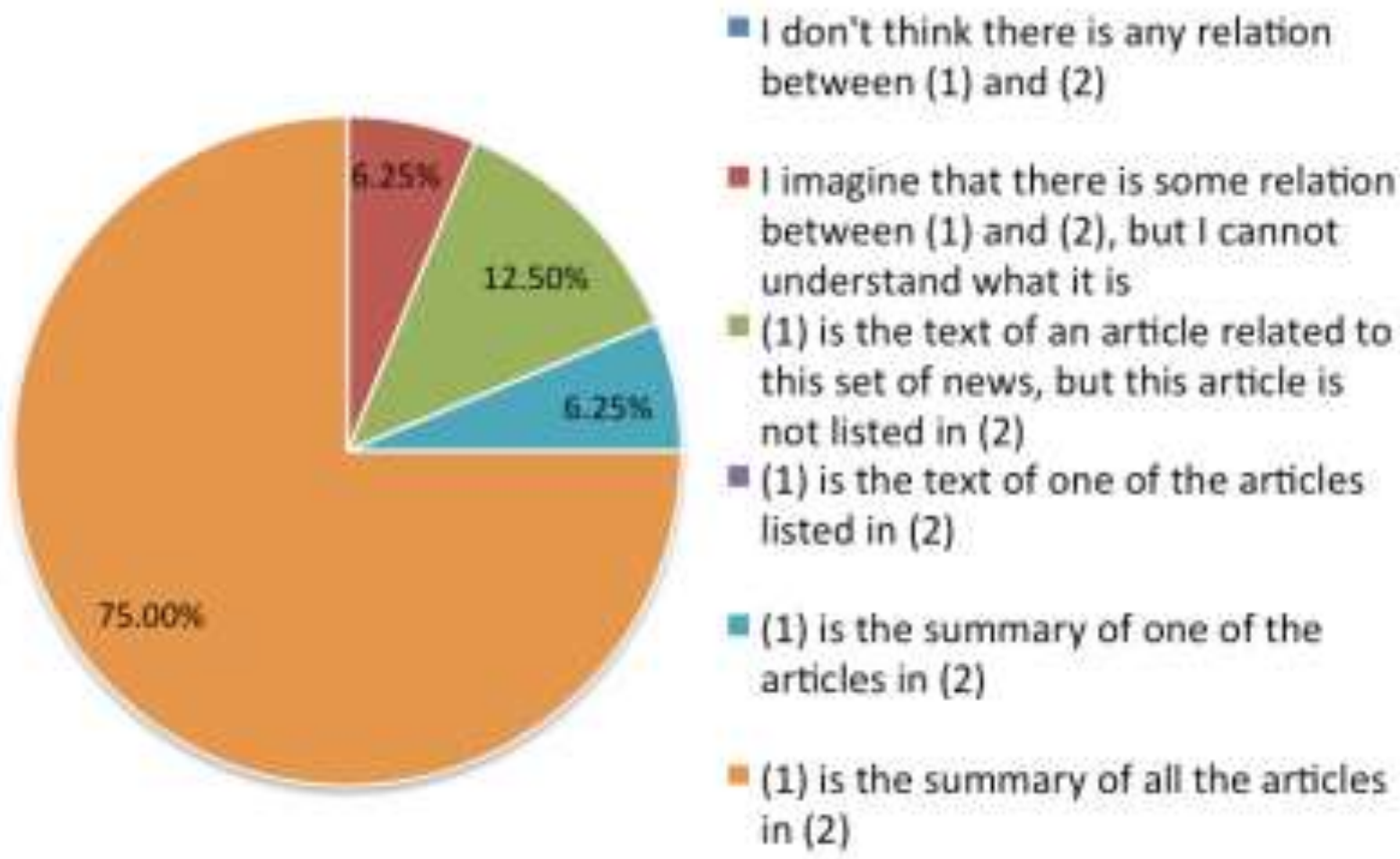

**Figure 23. What was your first impression concerning the relation between elements (1) and (2)?**

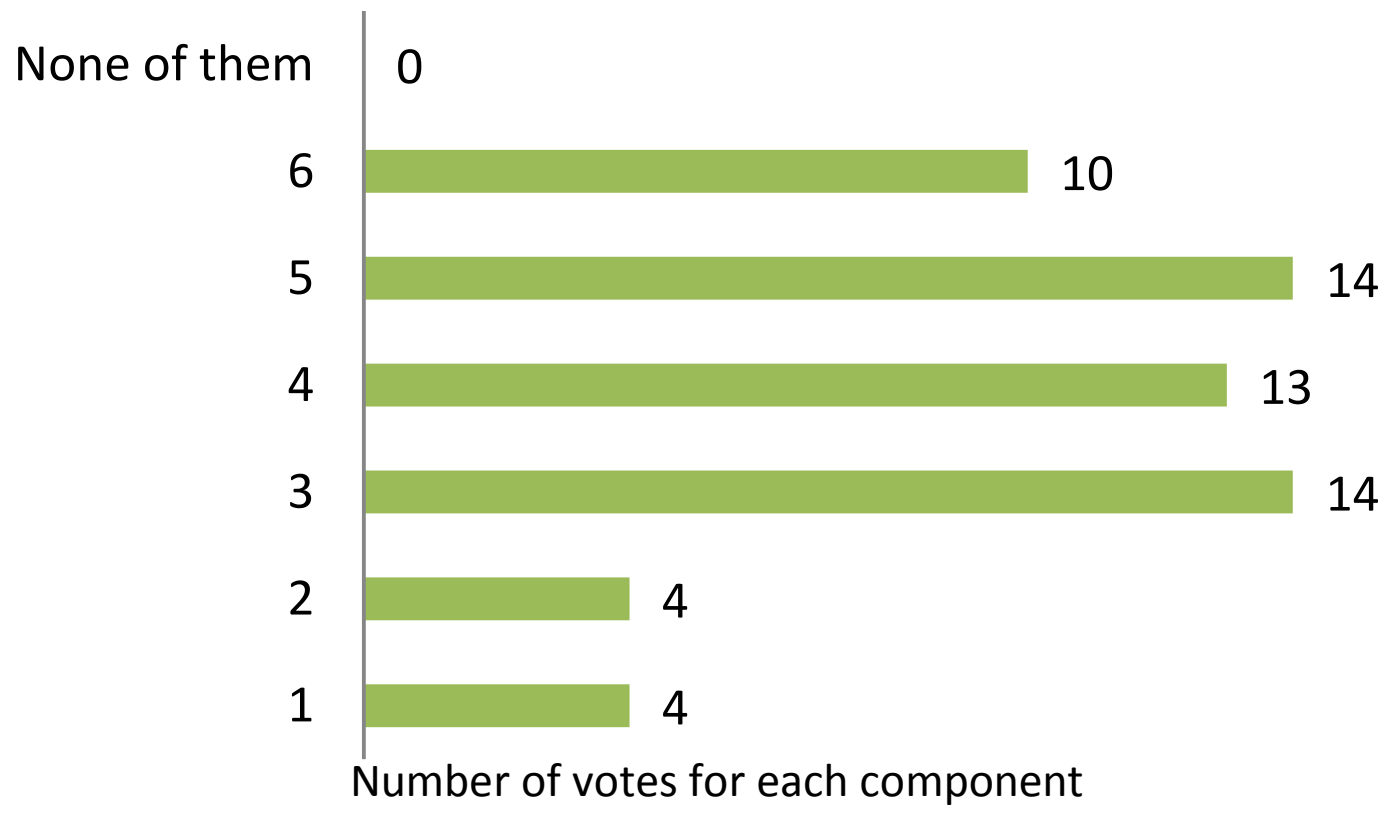

**Figure 24. Based on your first impression, which of the numbered elements (1-6) - if any - do you expect to be interactive? In other words, which elements are going to accept your input? Please, mark all the numbered components that you expect to be interactive.**

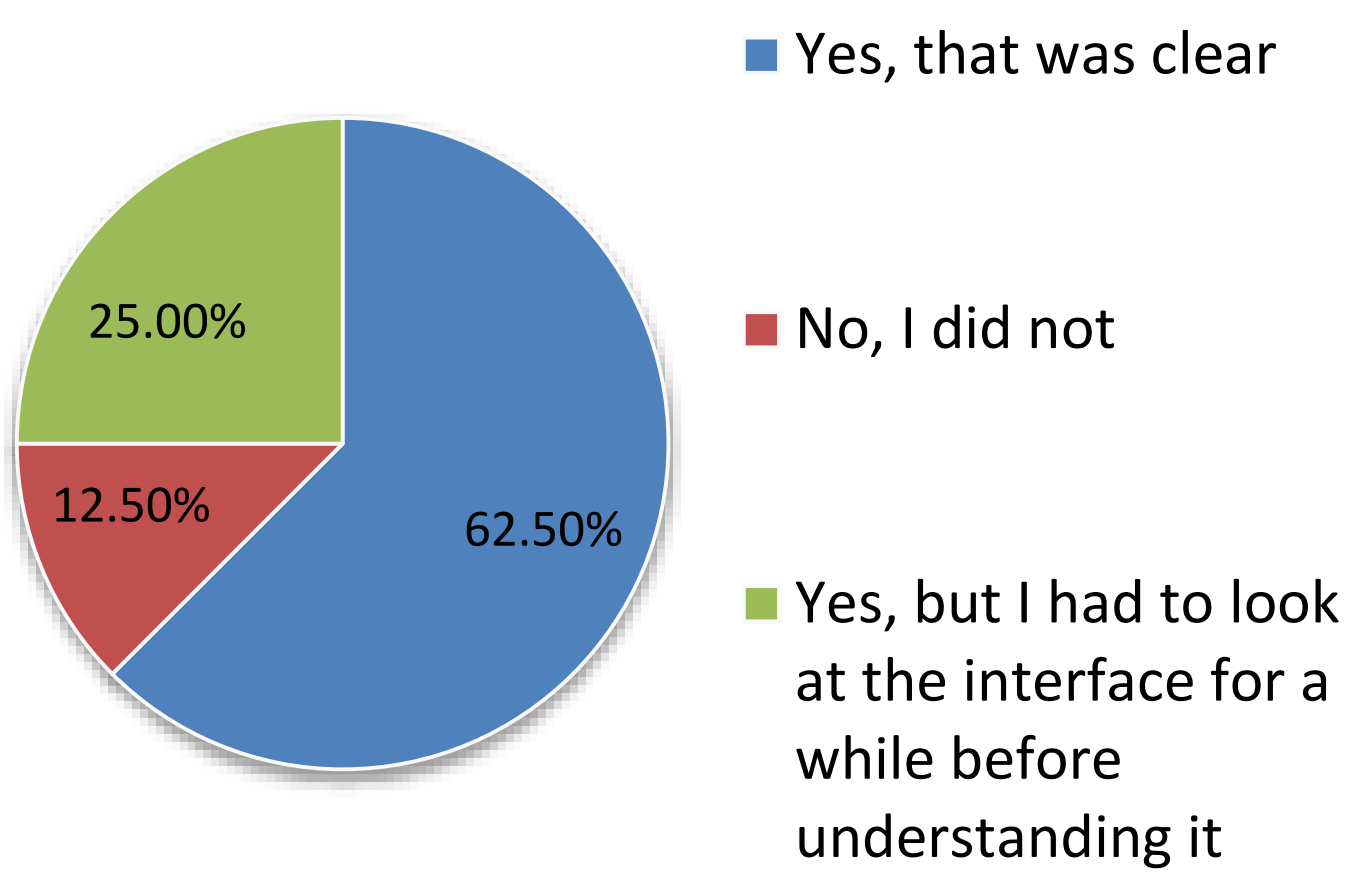

**Figure 25. At a first glance, did you understand that in (2) you were seeing a list of the news fetched based on your query in View1?**

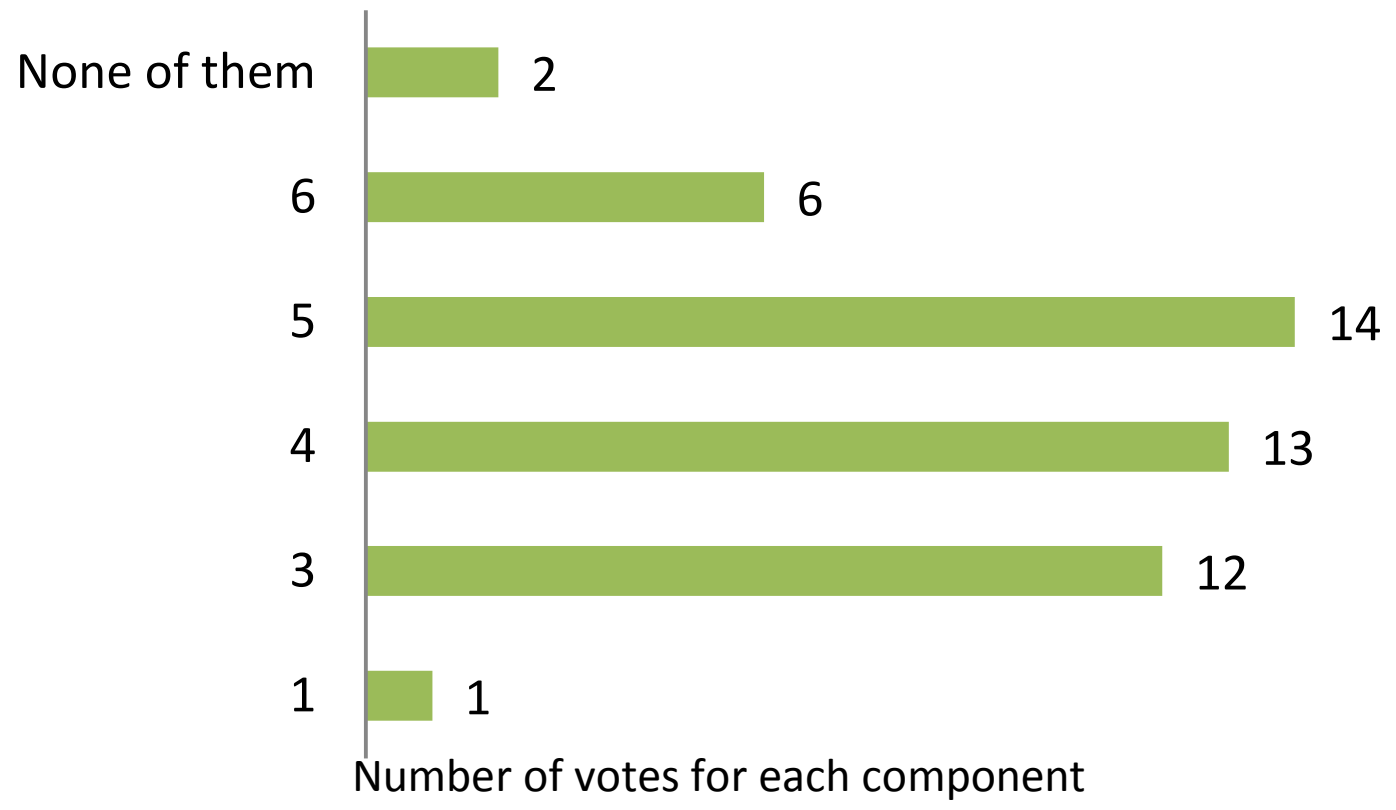

**Figure 26. Based on your first impression, which elements (if any) do you expect to have an effect on the ranking of the news in (2)? Please, select all those that apply.**

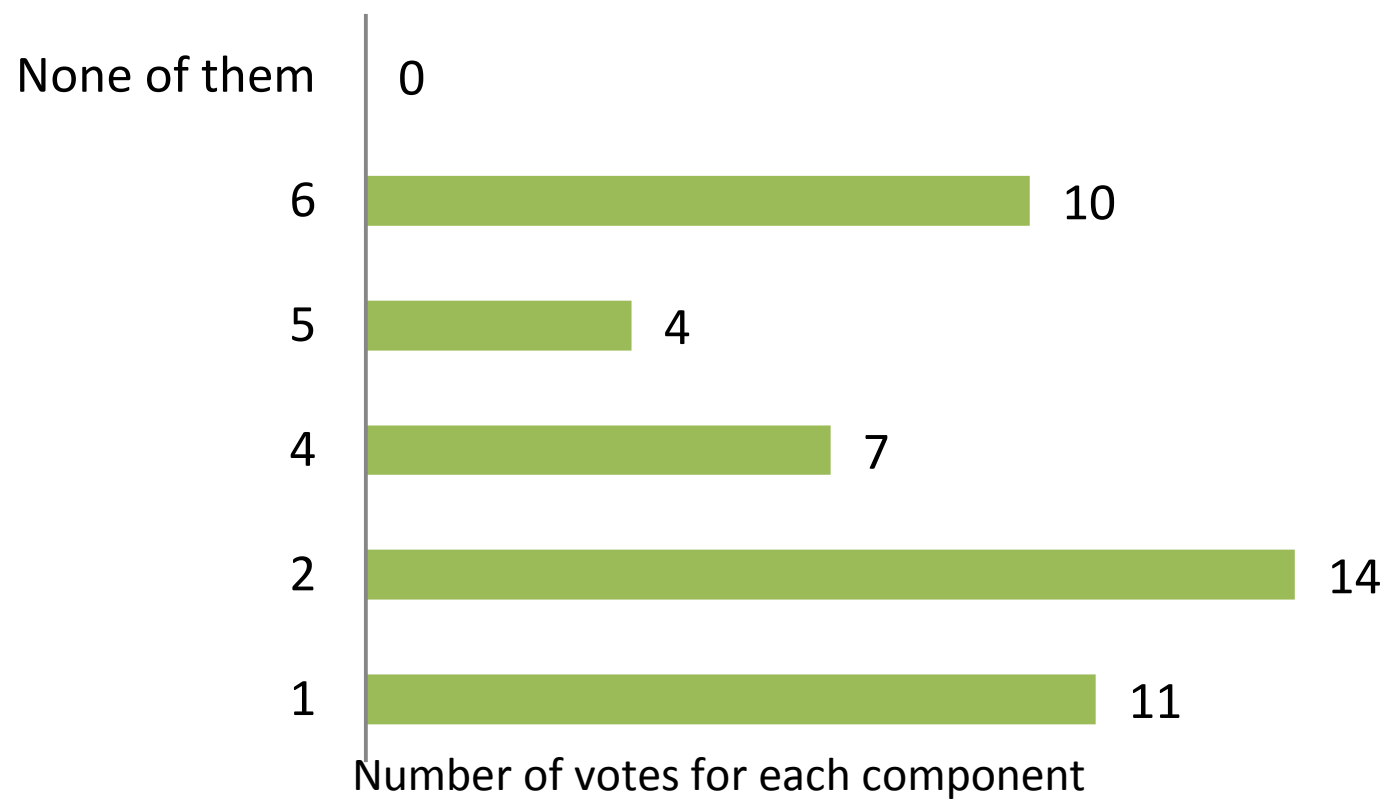

**Figure 27. Now focus on (3). This is actually an interactive element. Please select from the list the elements whose contents you would expect to change when you click somewhere in (3). Please, select all those that apply.**

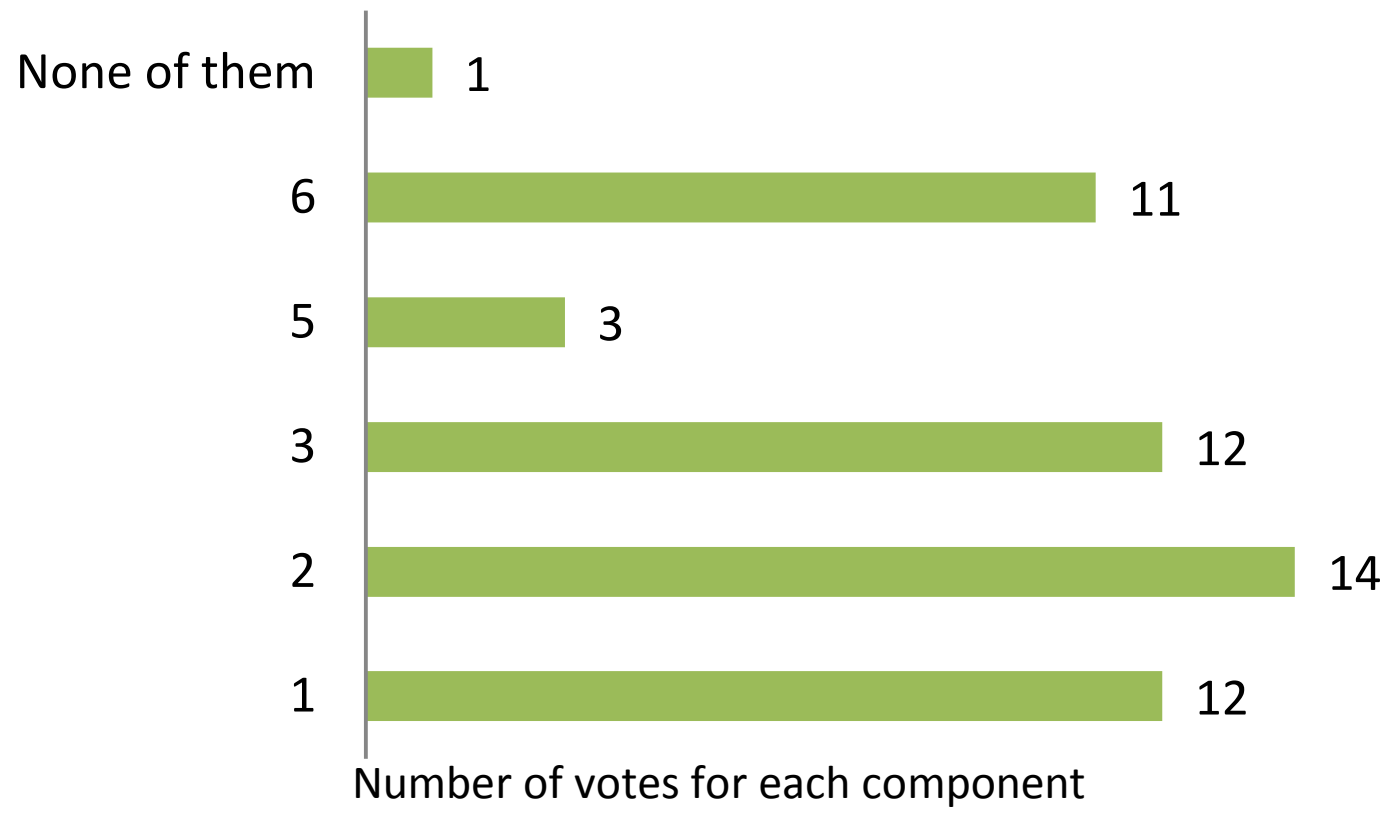

**Figure 28. Now focus on (4). This is also an interactive element. Please select from the list the elements whose contents you would expect to change when you click somewhere in (4). Please, select all those that apply.**

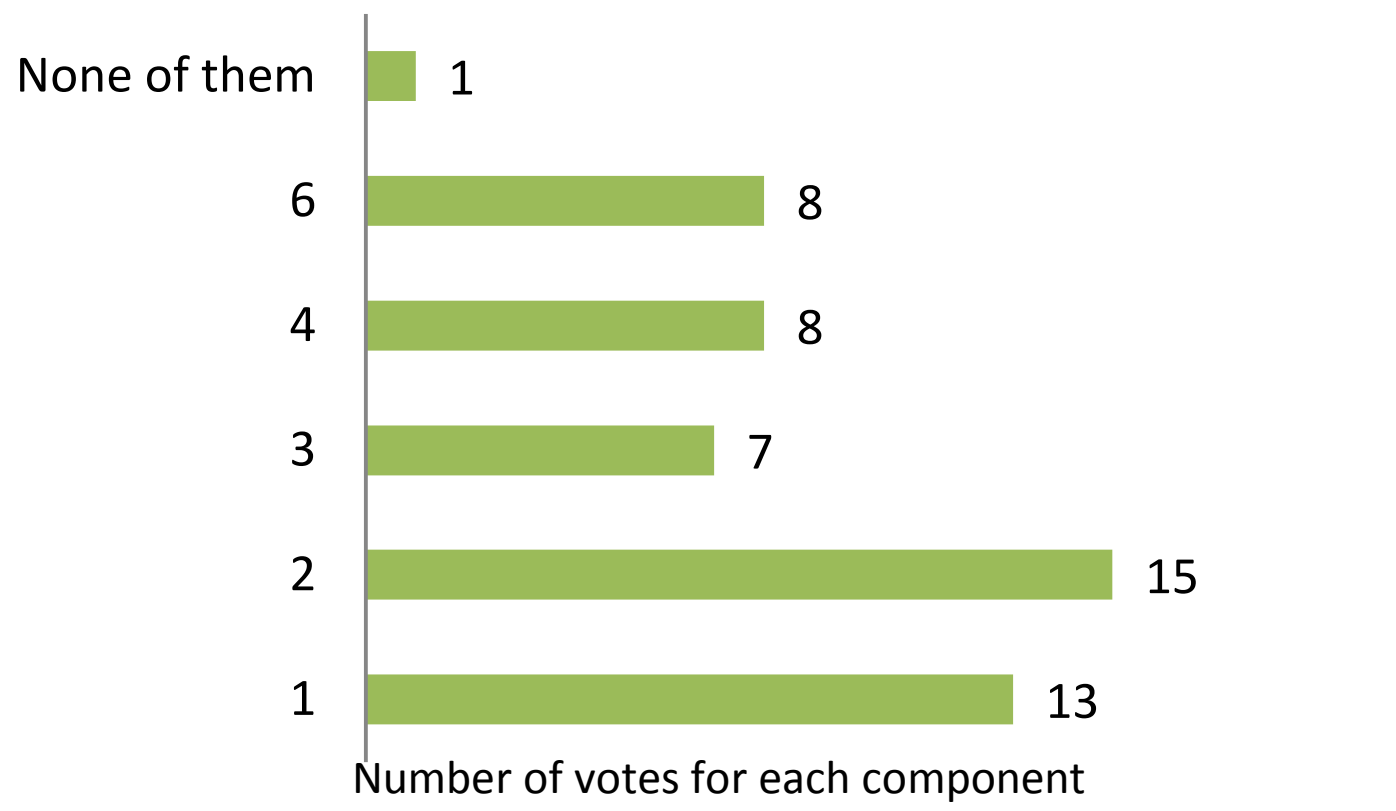

**Figure 29. Now focus on (5). This is also an interactive element. Please select from the list the elements whose contents you would expect to change when you change the position of the slider in (5). Please, select all those that apply.**

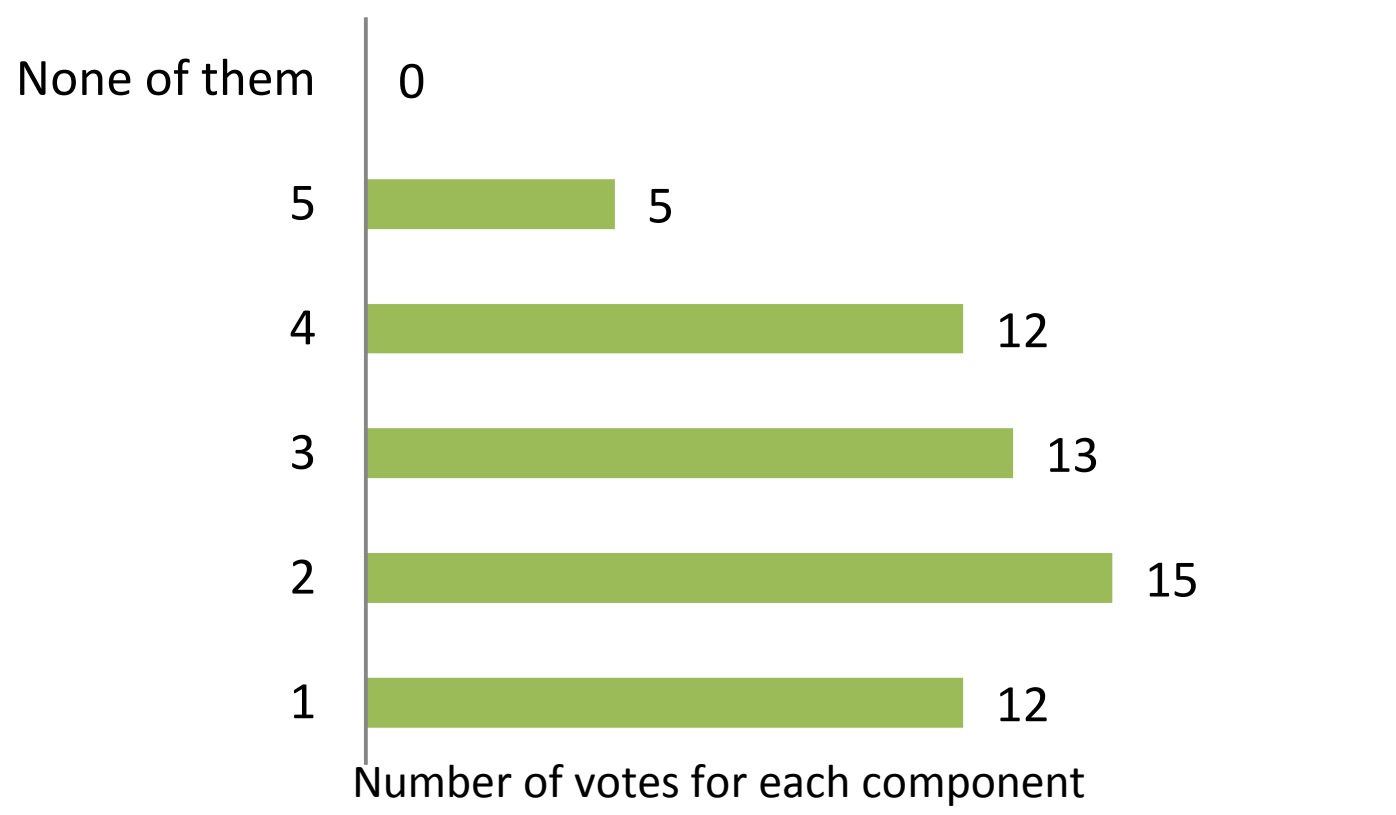

**Figure 30. Now focus on (6). This is also an interactive element. Please select from the list the elements whose contents you would expect to change when you click on one of the names listed in (6). Please, select all those that apply.**

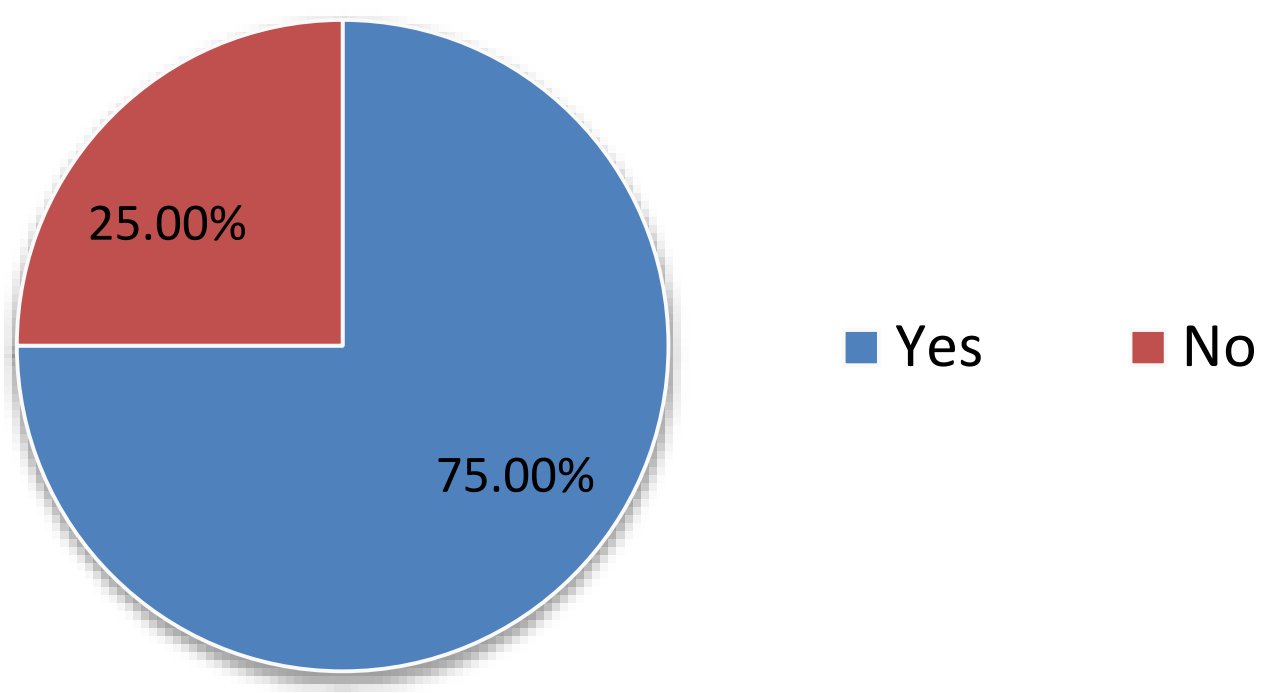

**Figure 31. The element (1) is an automatically generated summary of the news listed in (2). Did you understand it when looking at the page for the first time?**

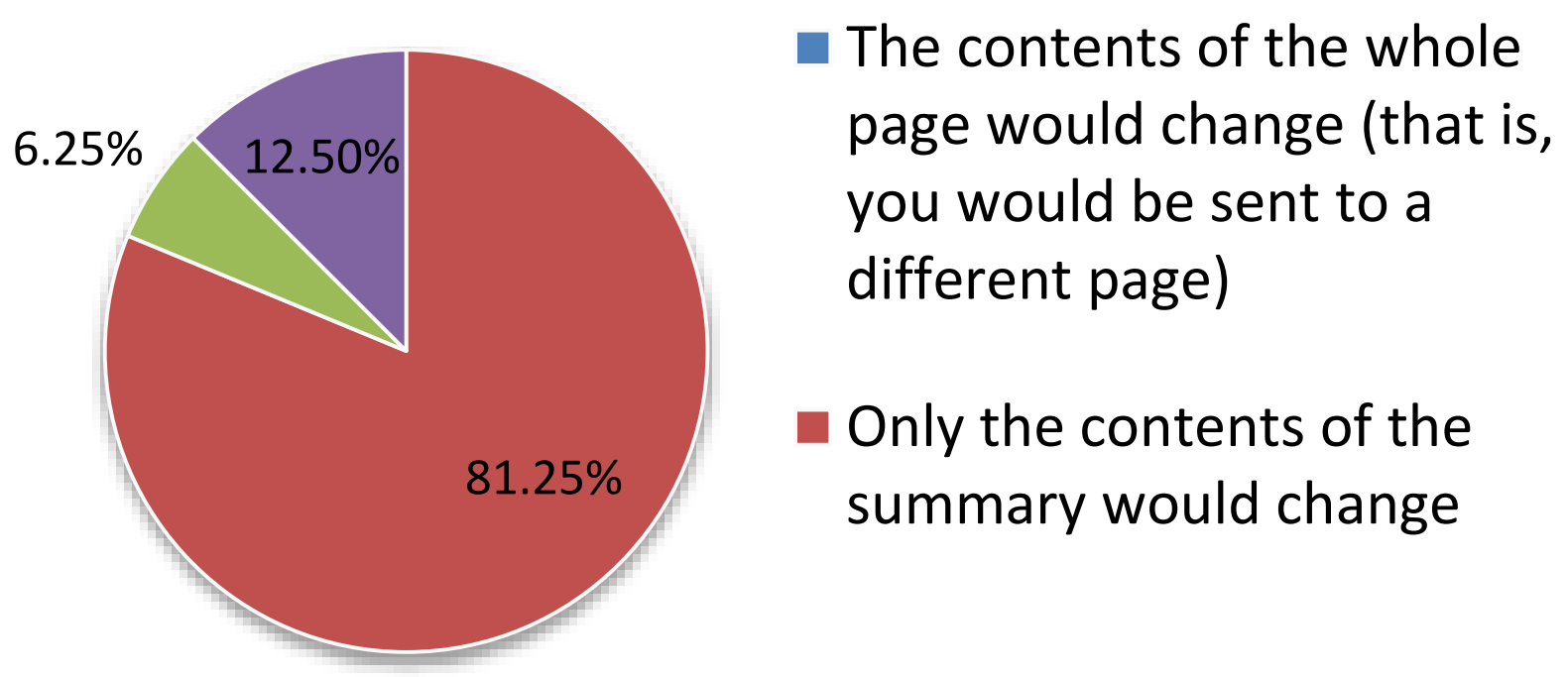

**Figure 32. Near the top of (1) there is line of text saying "Choose summarization algorithm". What do you think would happen if you clicked on any of the two links "Type1" and "Type2"?**

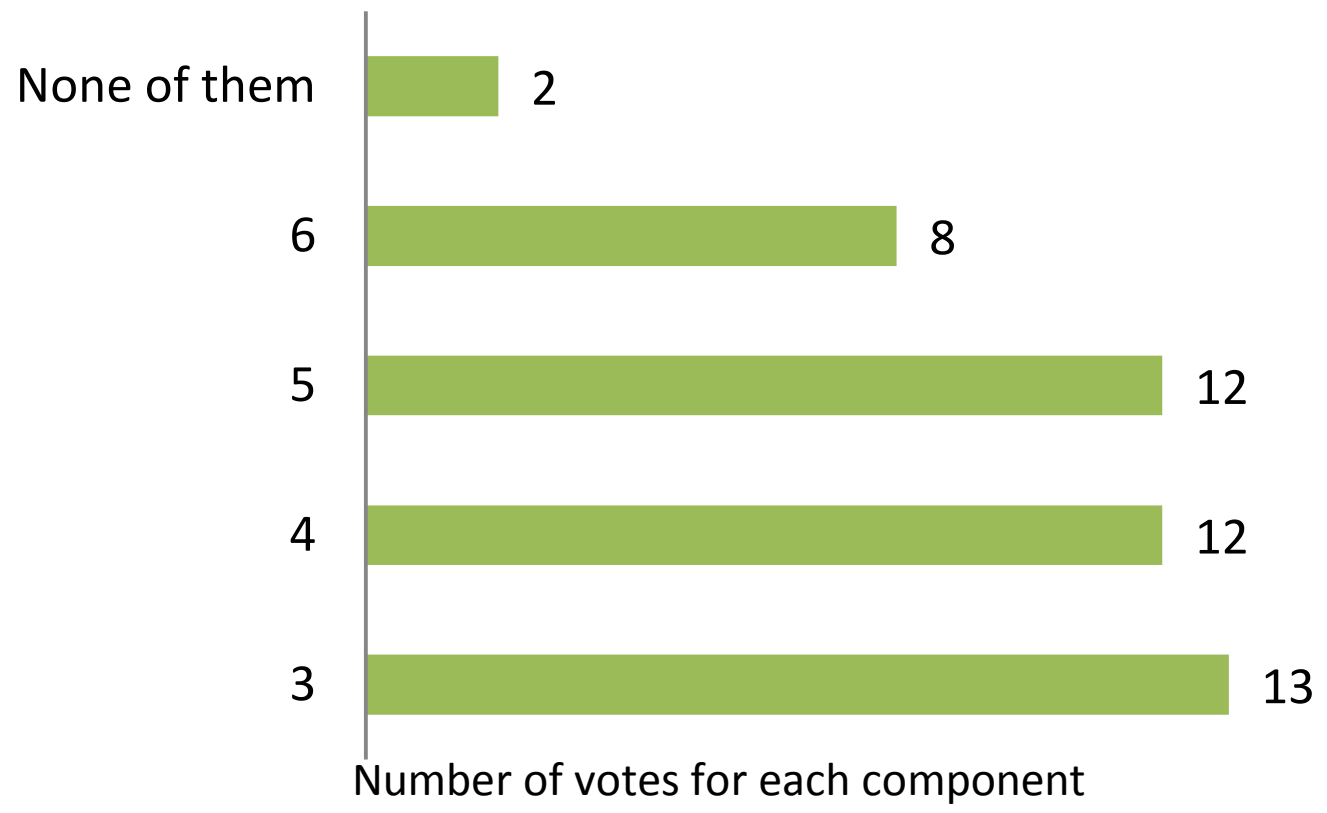

**Figure 33. Which of the other elements in the page - if any - do you expect to affect the content of the summary in (1)? Please, select all those that apply.**

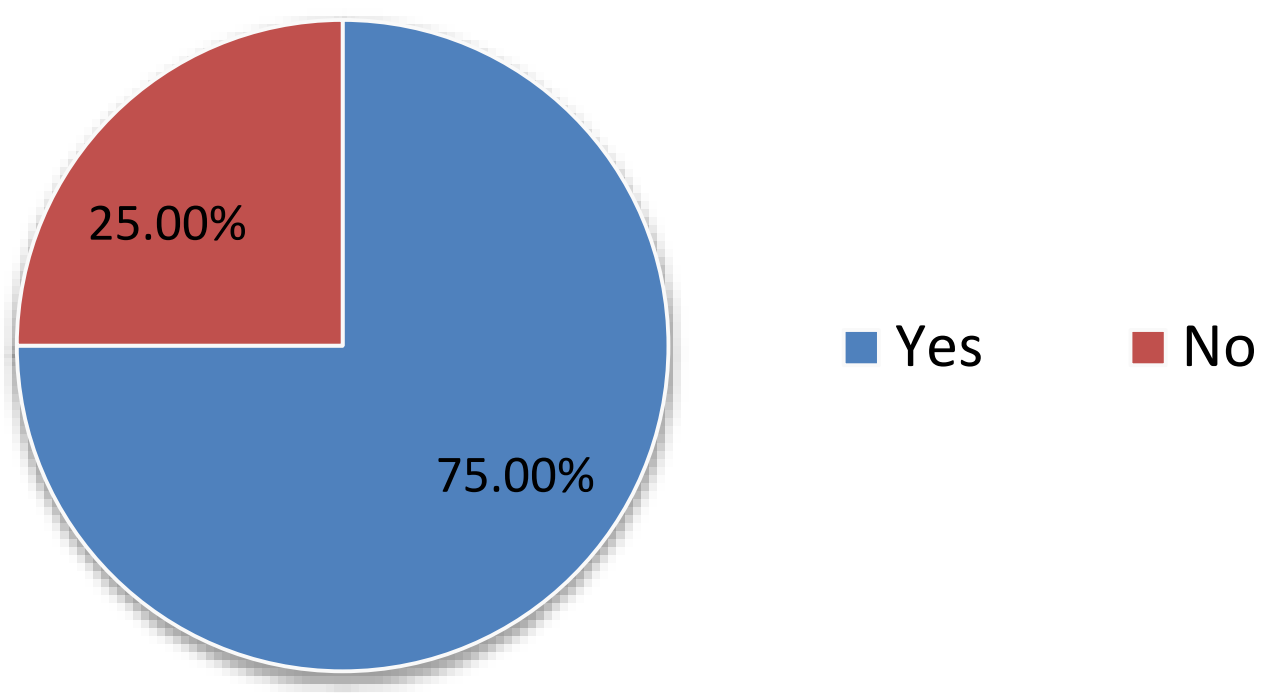

**Figure 34. Acting on (3) makes it possible for you to give more relevance to the news for which some specific topics are more relevant. Was it clear from the beginning?**

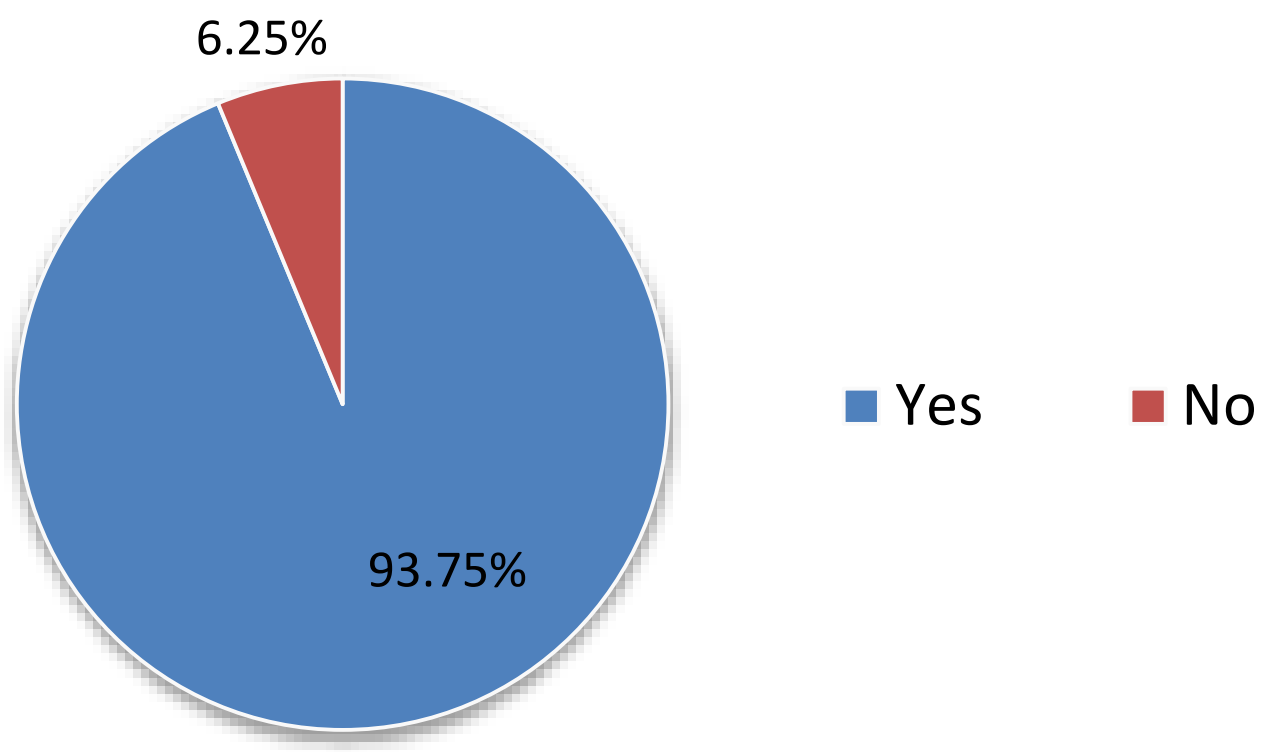

**Figure 35. Acting on (4) makes it possible for you to give more relevance to the news published closer to the selected point on the map. Was it clear from the beginning?**

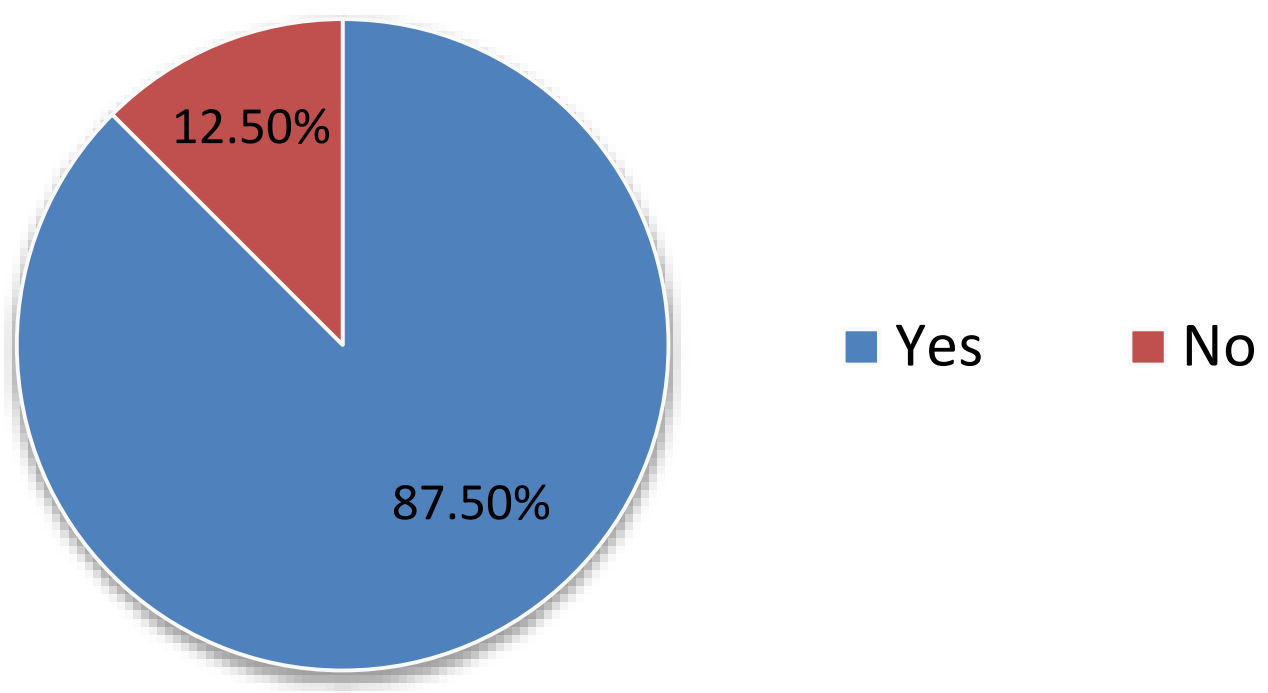

**Figure 36. Acting on (5) makes it possible for you to give more relevance to the news which have a more positive/negative take on the events. Was it clear from the beginning?**

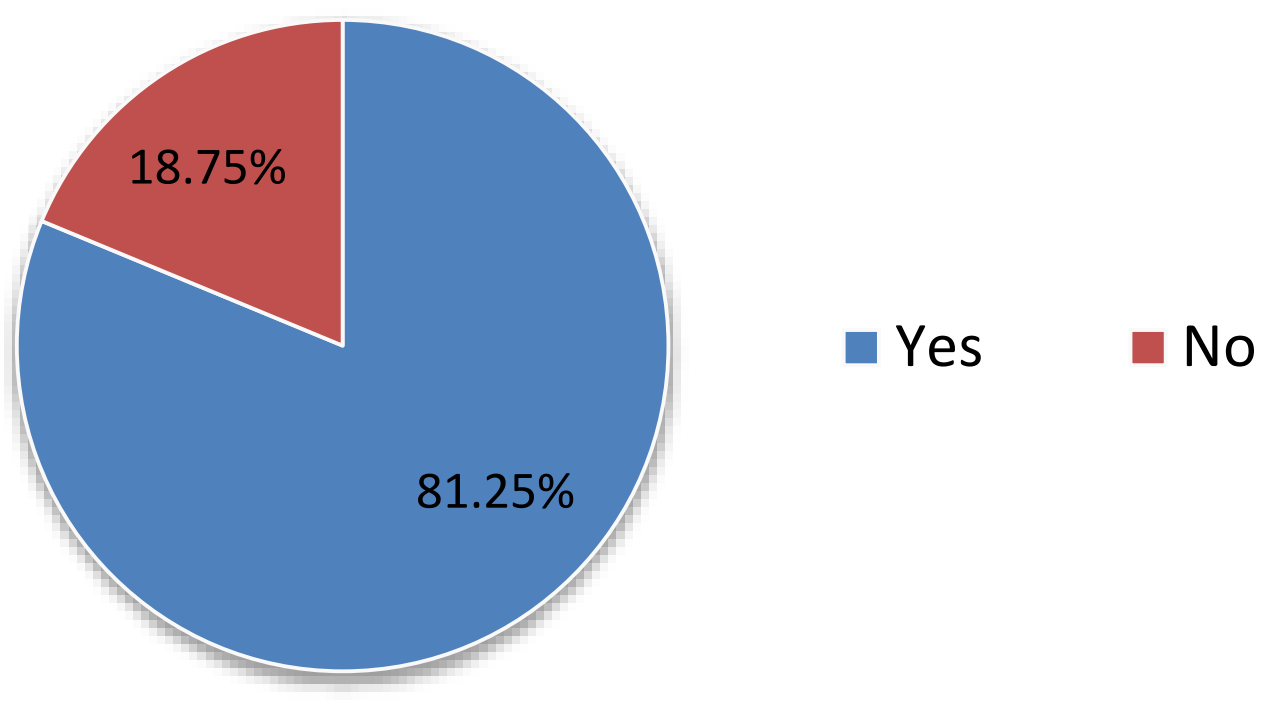

**Figure 37. Acting on (6) makes it possible for you to browse news involving entities (organizations, notable people, locations) which are related to the collection of news that you are currently visualizing. Was it clear from the beginning?**

## A.5 Interactive evaluation

**Estimated time: 20 minutes.**

After looking at the interface "on paper", now you are going to play with it for a while and let us know what you think about it. Spend **10 minutes** playing with the interface, acting on all the controls and observing the effect that they have. Read the summaries that are generated, try to understand how the different controls affect the summaries and the ranking of the news. When you are done, please answer the following questions.

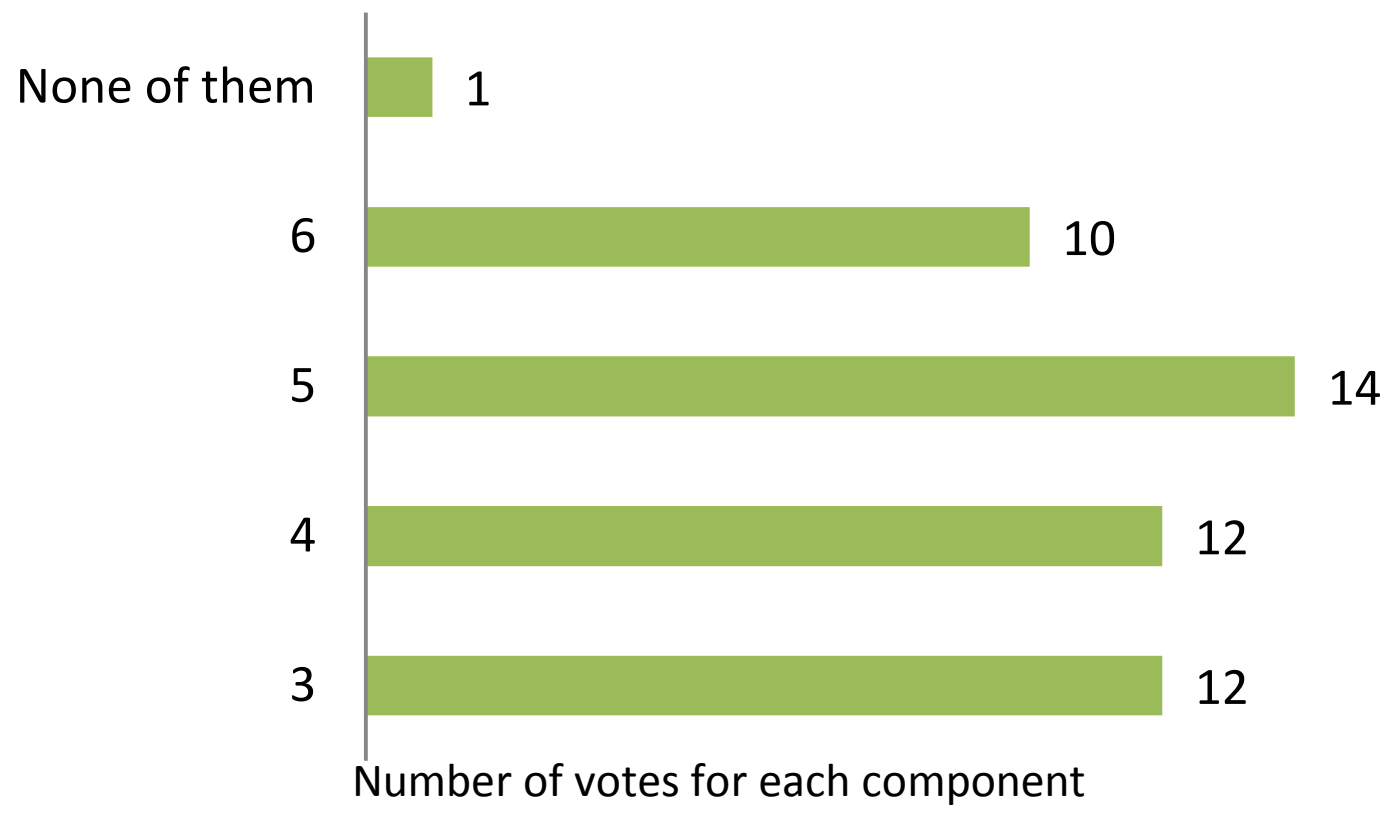

**Figure 38. Which of the controls behaves according to what you were expecting before using the interface? Please, select all those that apply.**

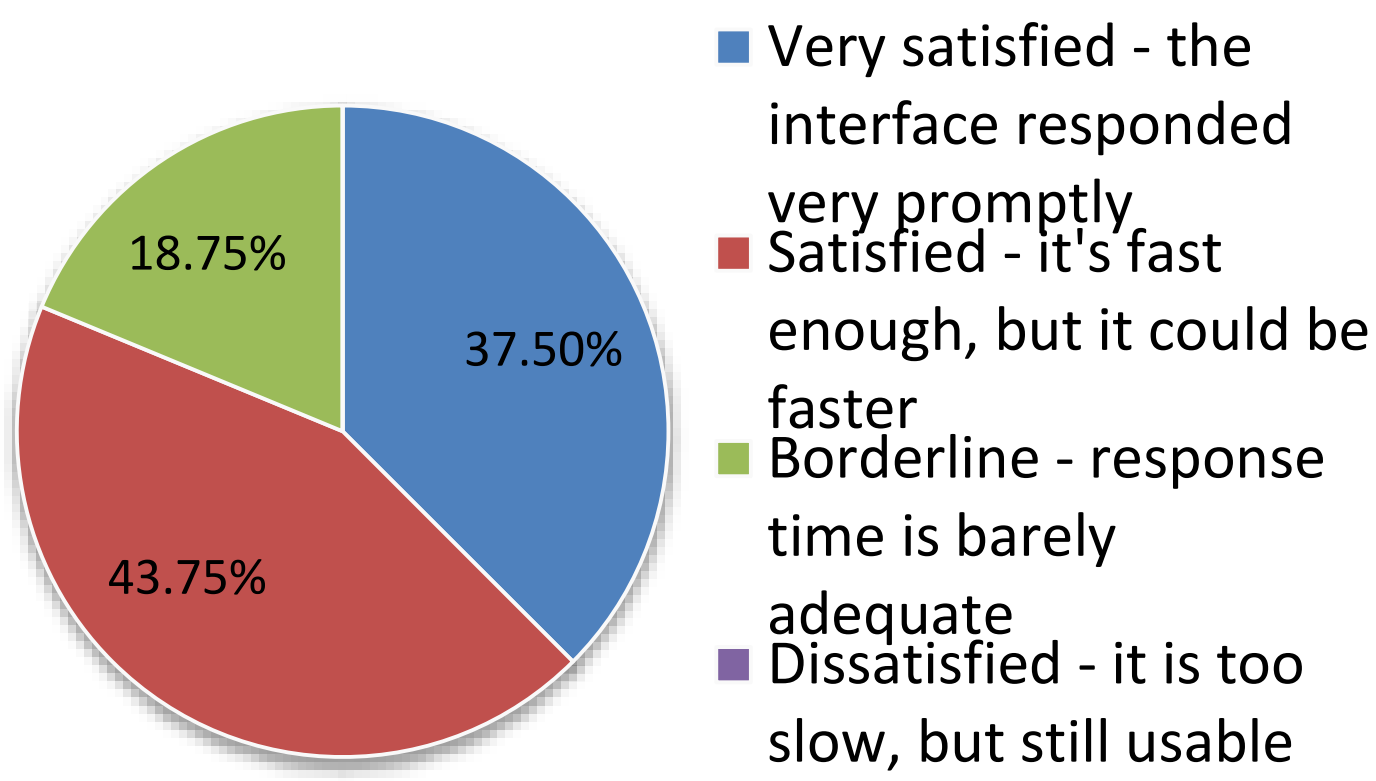

**Figure 39. As you should have seen, every time you act on any of the controls on the right the content of the summary and the ranking of the news (both on the left) change. How happy are you with the response time of the system?**

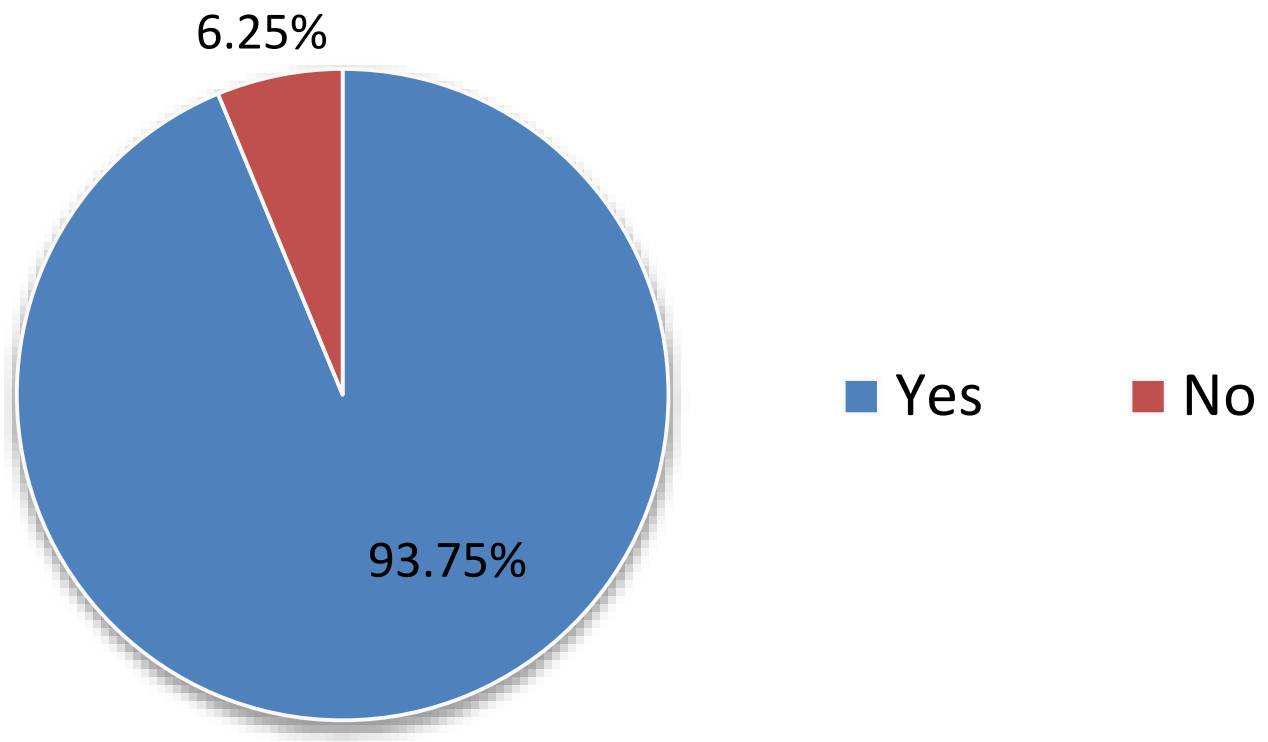

**Figure 40. Did you understand what the question mark icons are for?**

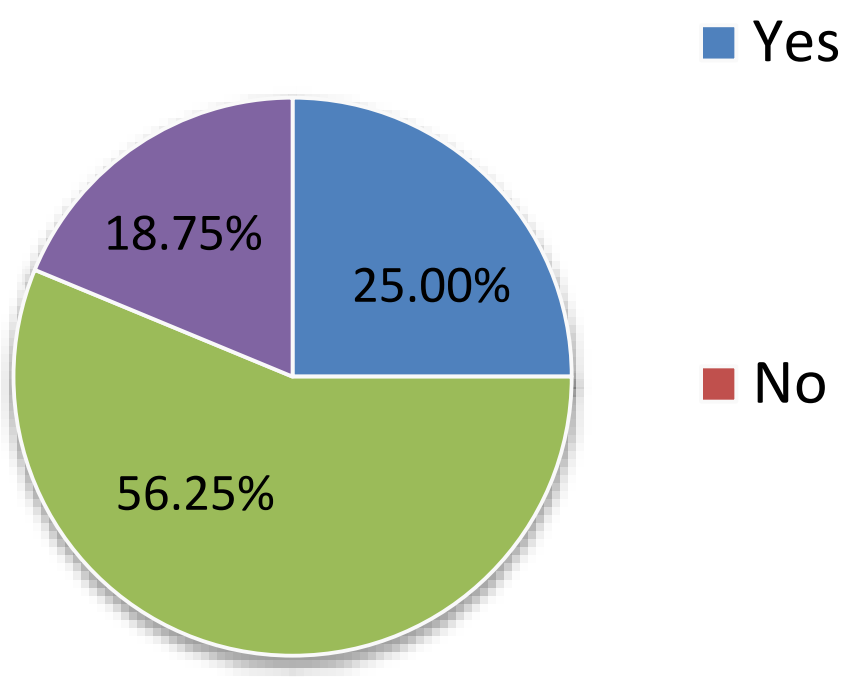

**Figure 41. Did you find the information provided by the question mark icons useful to understand how to use the interface?**

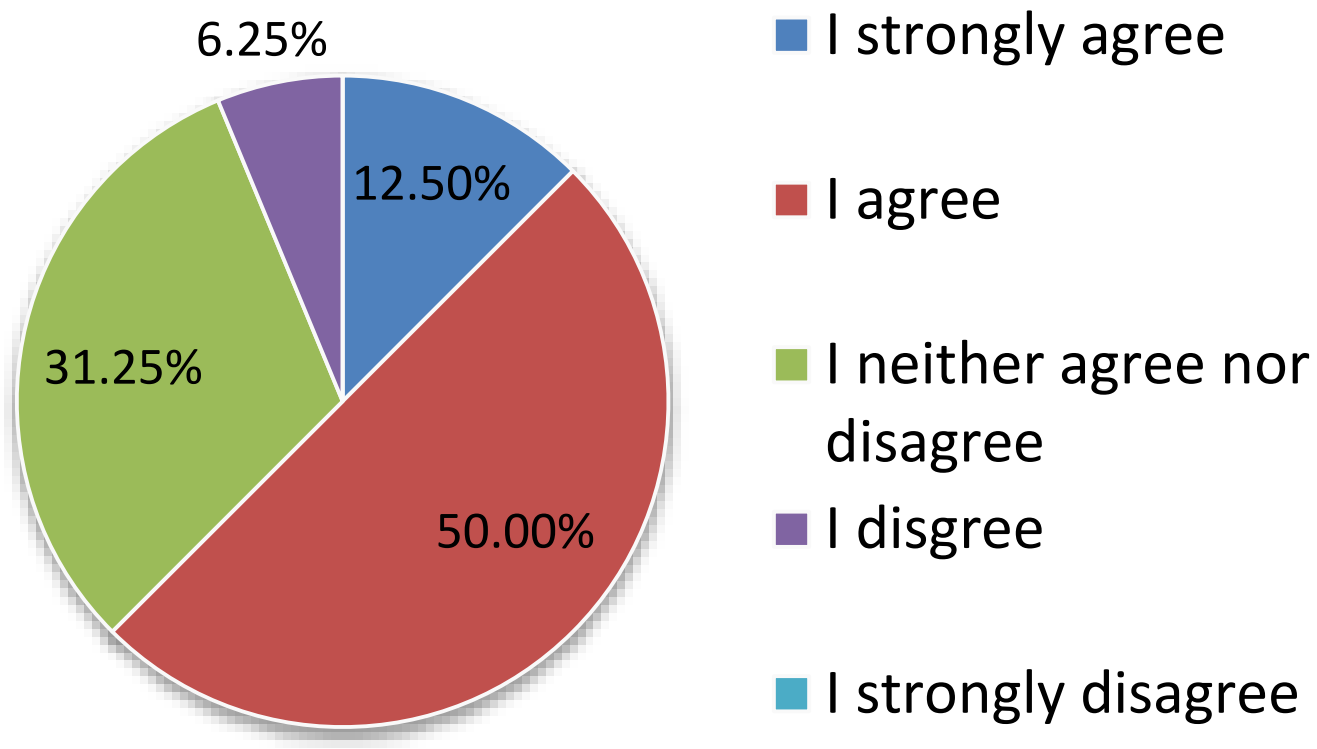

**Figure 42. The positioning of the elements in the interface is logical and intuitive.**

**Do you have any other suggestion on how to improve the layout?**

- Make it clearer that (1) is a summary. The canvas of the page is huge in my chromebook (lots of grey space on the right)
- (6) should have similar interface as (3) to make its function more intuitive
- (4) map is overkill considering number of sources
- I would change the ordering of 3-6 as 4, 3, 6 and 5 from top to bottom. it can be more intuitive to read from left to right. I.e., having controls on the left that determine the content. For me and on a first sight, the elements right side are often just read-only properties of the main content.

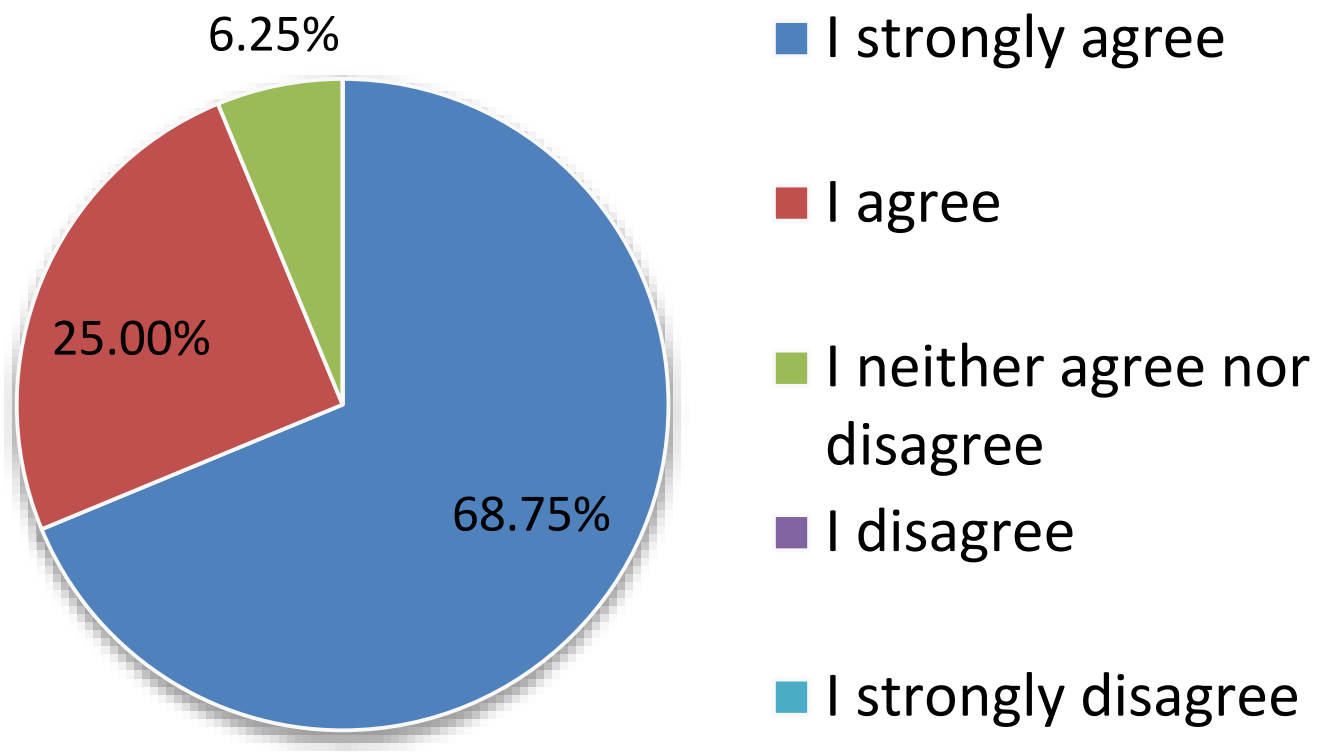

**Figure 43. After using the interface for a while, it becomes very clear that (1) is a summary of the news in (2).**

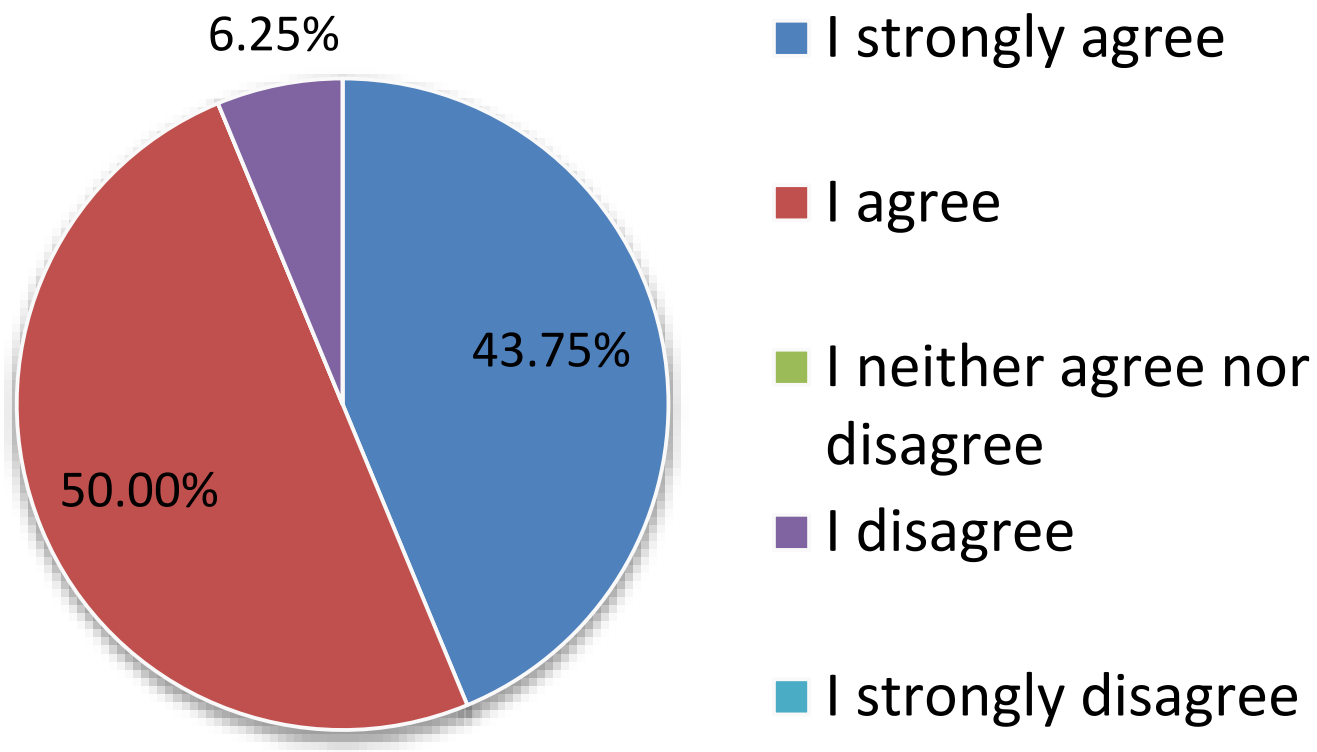

**Figure 44. For a user of the interface, it is important to know that (1) is a summary of the news in (2).**

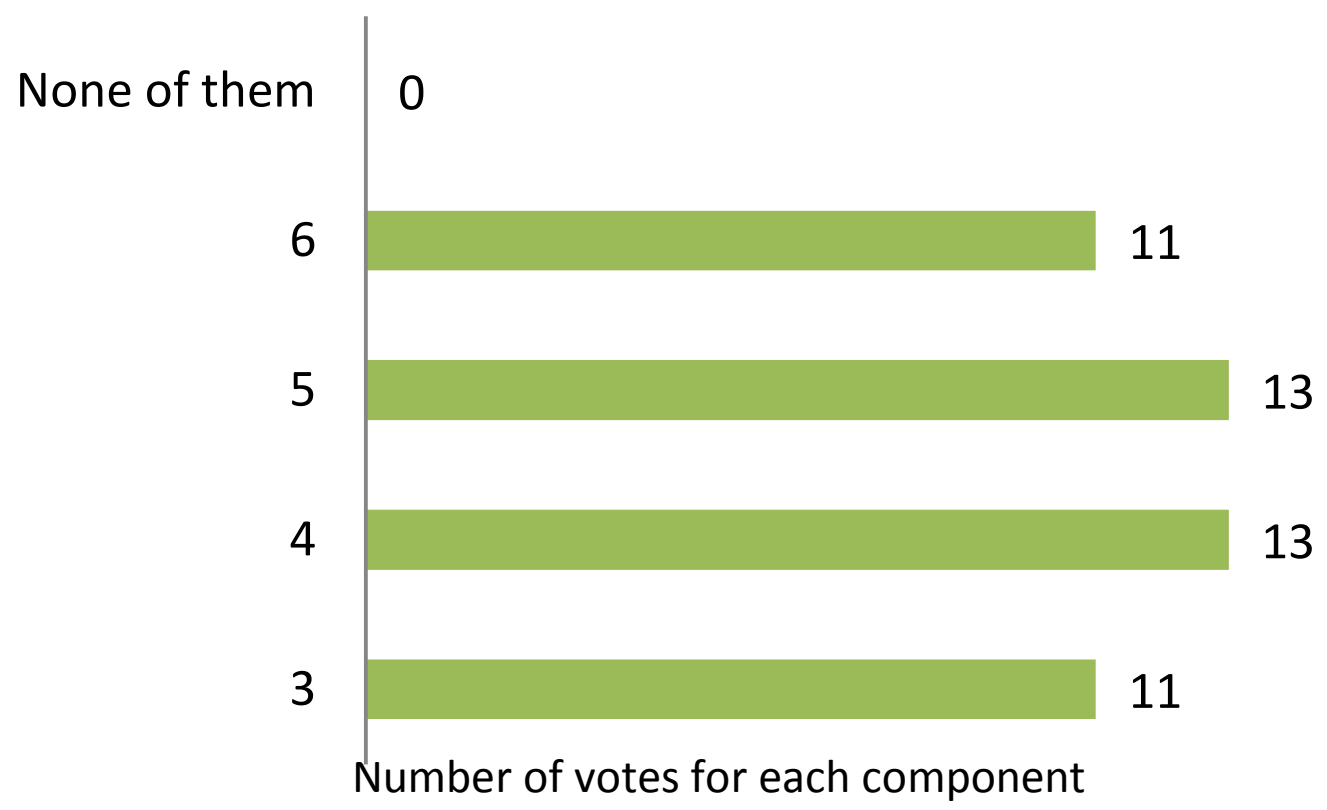

**Figure 45. Which interactive elements of the interface are adequately intuitive, according to you? Please, mark all that apply.**

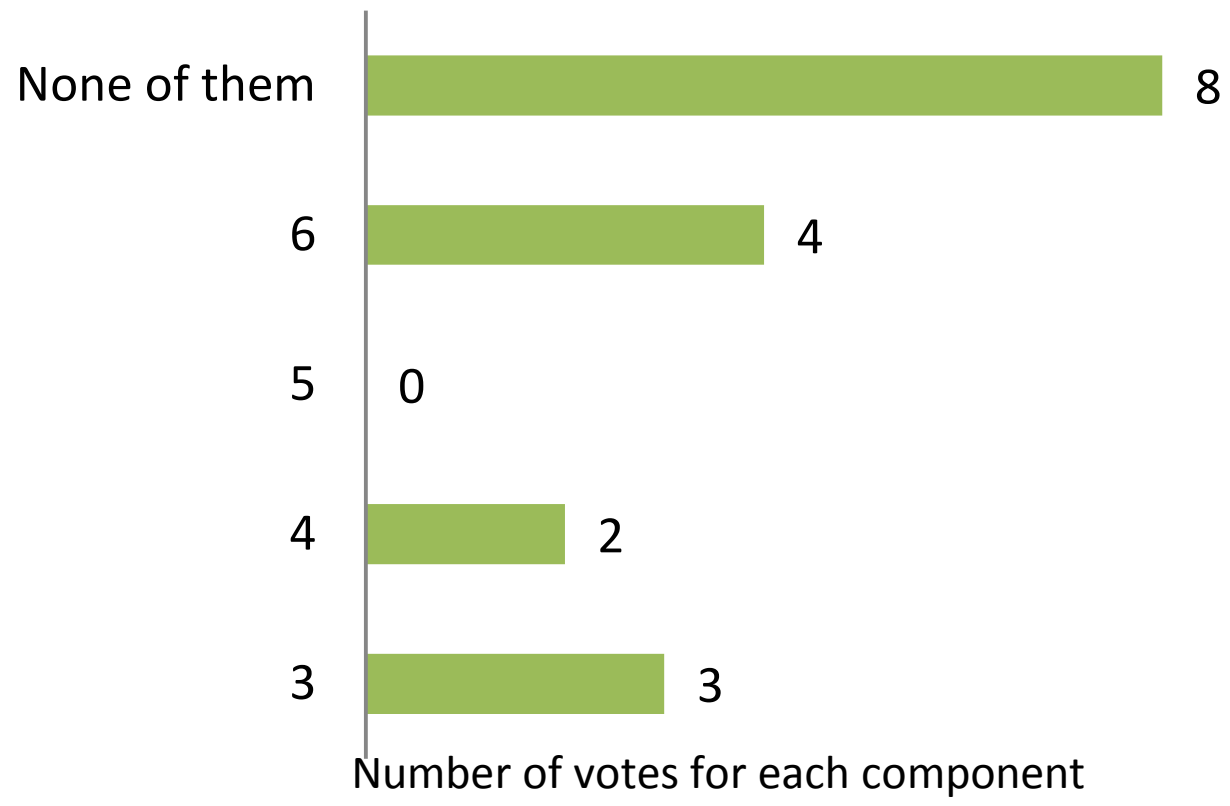

**Figure 46. Which interactive elements of the interface could be more intuitive, according to you? Please, mark all that apply.**

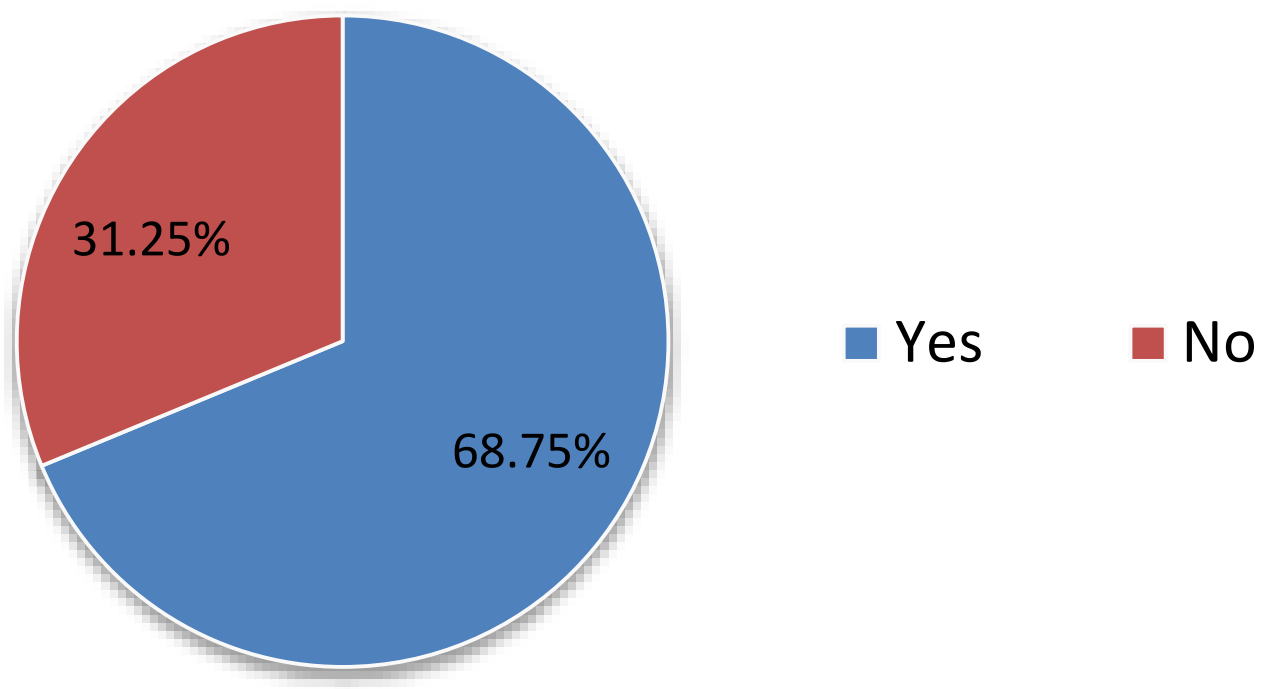

**Figure 47. Concerning (3), did you understand why some of the terms are listed together and some are not?**

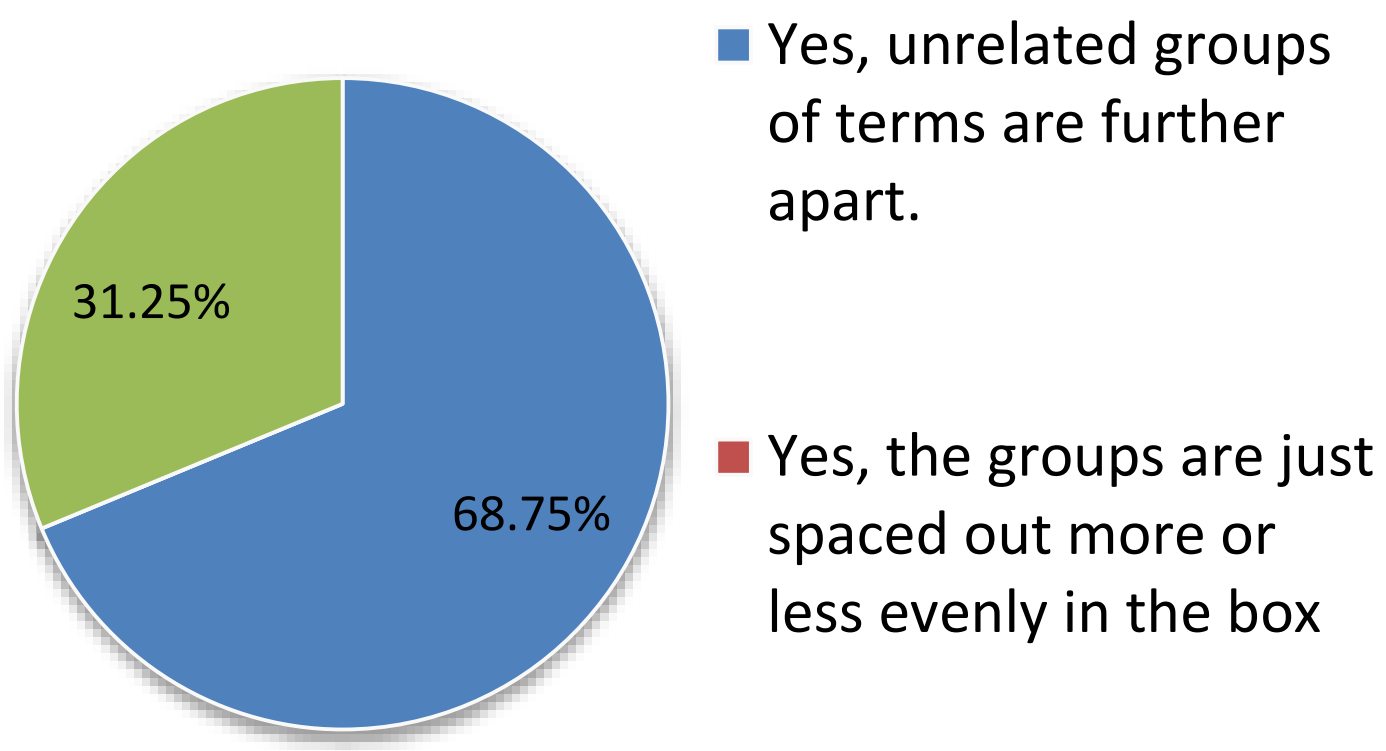

**Figure 48. Concerning (3), do you think that you understood why some groups of words are further than others?**

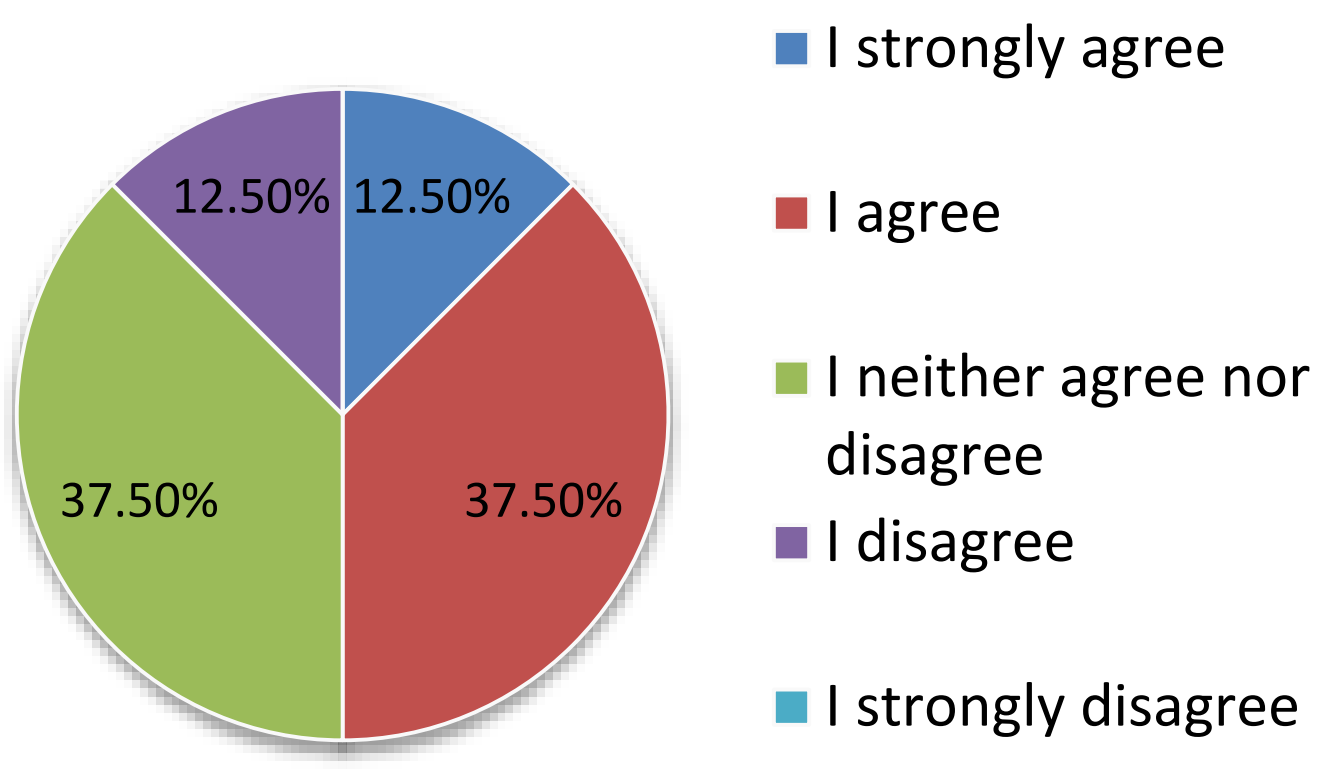

**Figure 49. Concerning (3), the panel lists relevant terms with respect to the current news collection.**

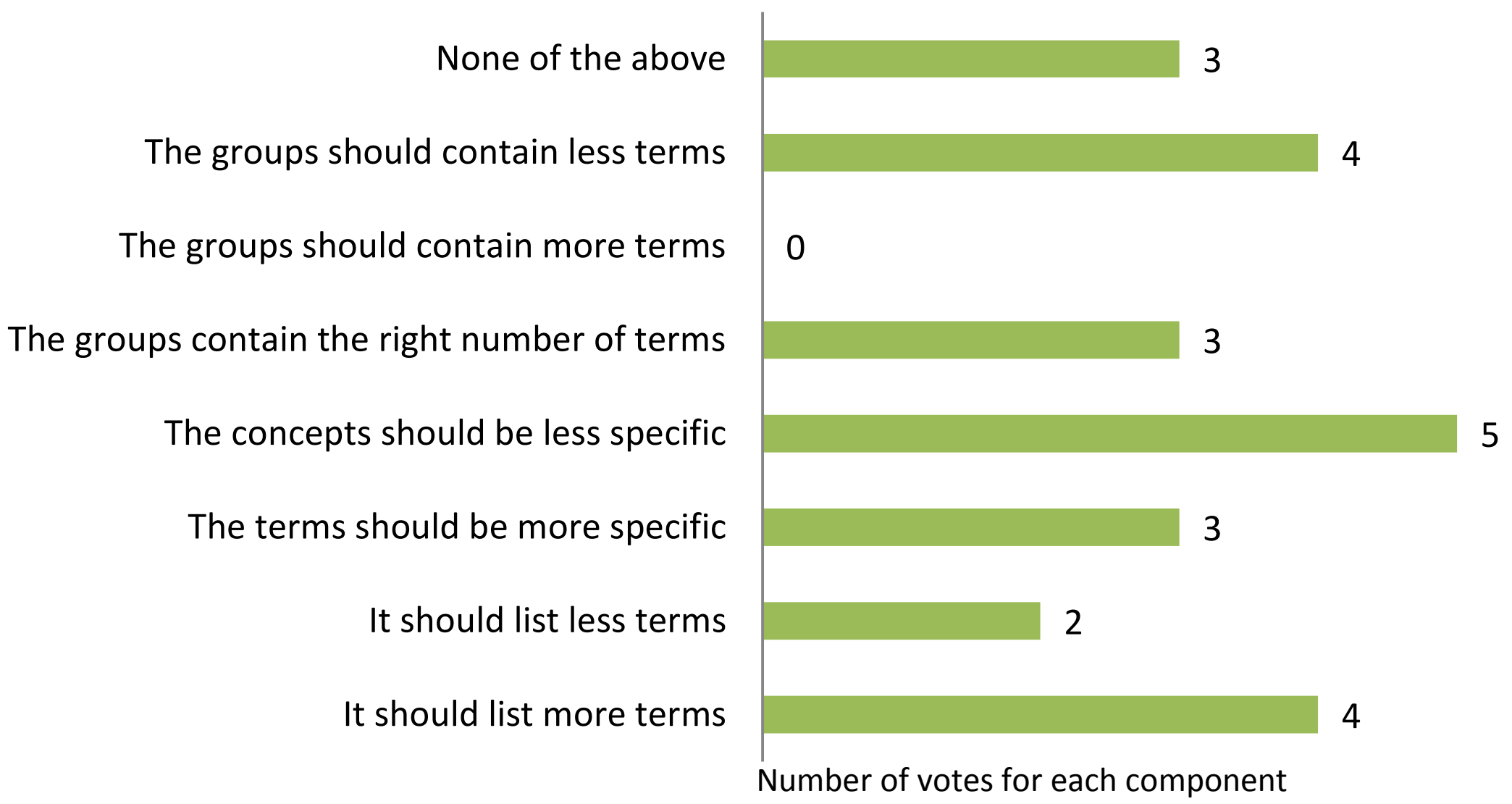

**Figure 50. Concerning (3), how do you think that the panel could be improved? Please, mark all that apply.**

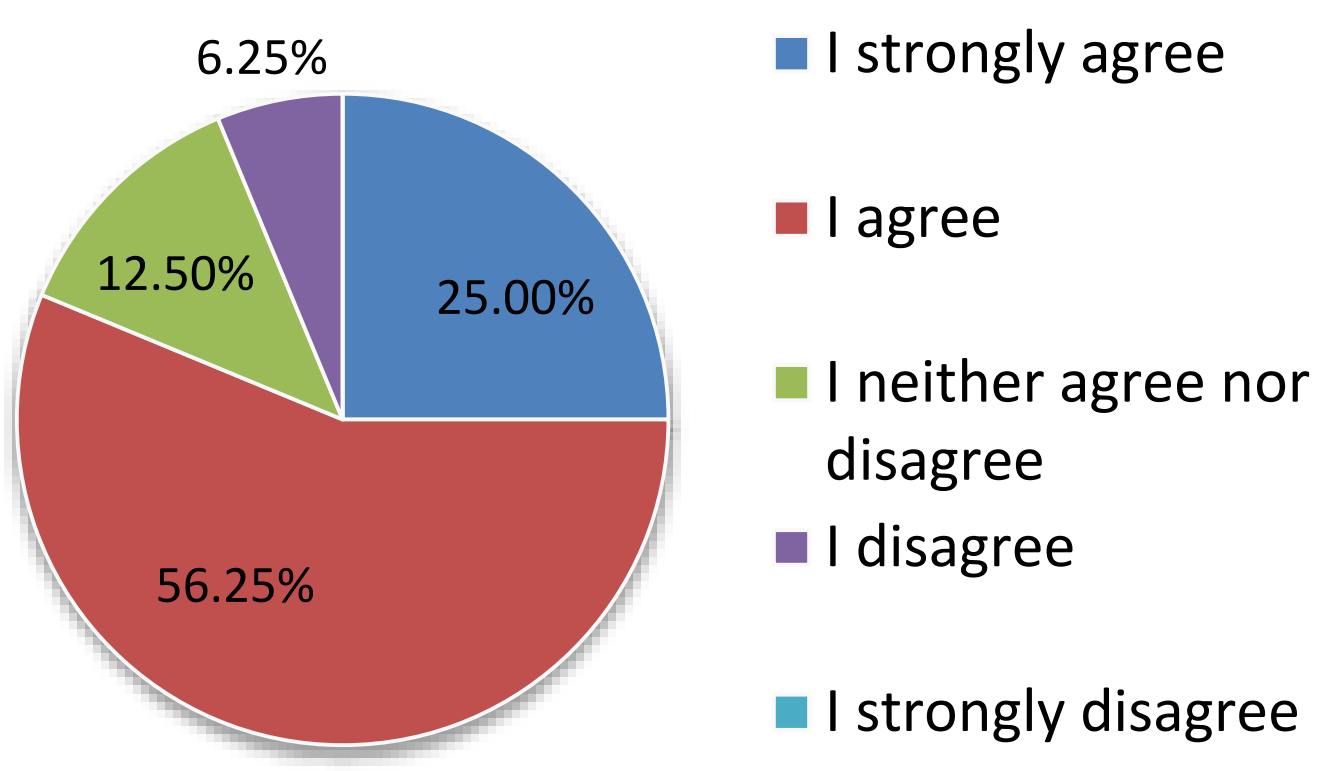

**Figure 51. Concerning (3), the interaction with the panel is intuitive.**

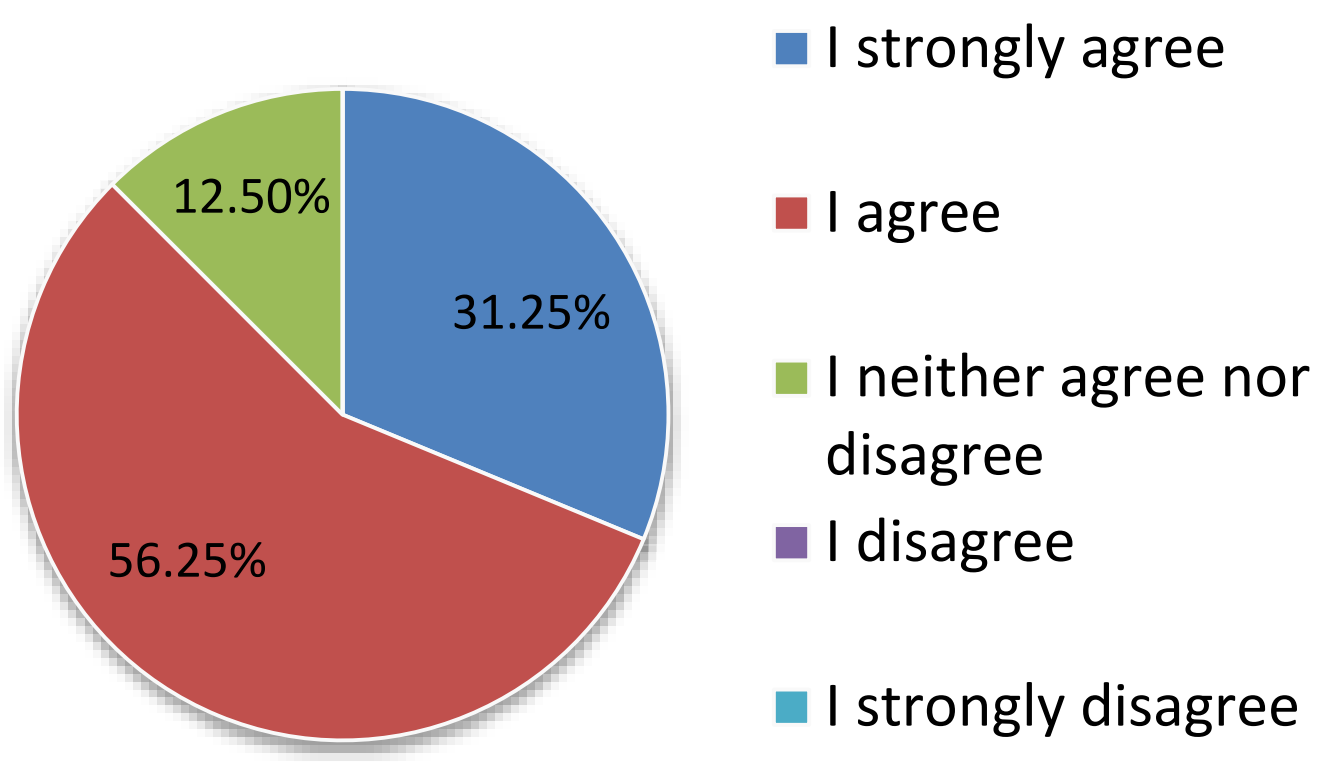

**Figure 52. Concerning (3), moving the target icon in the panel had a noticeable effect on the content of (1) and the ranking of news in (2).**

**If you have some comments about the interaction with (3), please add them here.**

- Clustering is not intuitive.
- It should also change the related entities (number 6).
- Regarding above question: What are the groups in the answers? In general, more items should be listed to be useful. The few shown elements were not relevant for me at all. E.g. searching for 'austerity' gives me only unrelated topics, perhaps too specific but also not obviously connected to economics, politics e.g. on a national level. The chosen visualization technique is hardly usable when having further topics displayed. Overlaps cause un-readability.
- So if moved the target on none of the suggestion, the contents of 1 and 2 changed and I wasn't sure to which one the information displayed was more related.
- I have found the terms listed for the clusters not very meaningful. Some kind of summarization mechanism to generalize each cluster could help. In addition, I would like to have an option to go to the original settings.

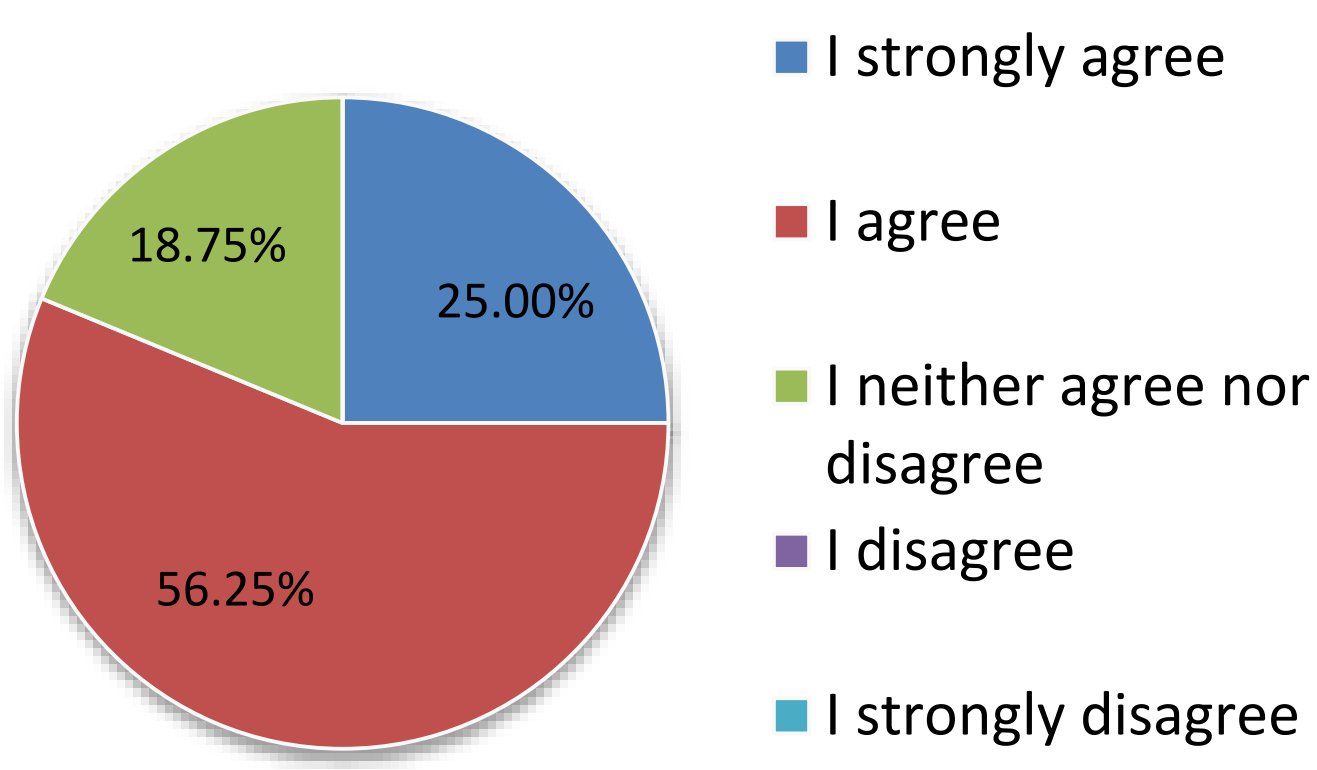

**Figure 53. Concerning (4), the interaction with the panel is intuitive.**

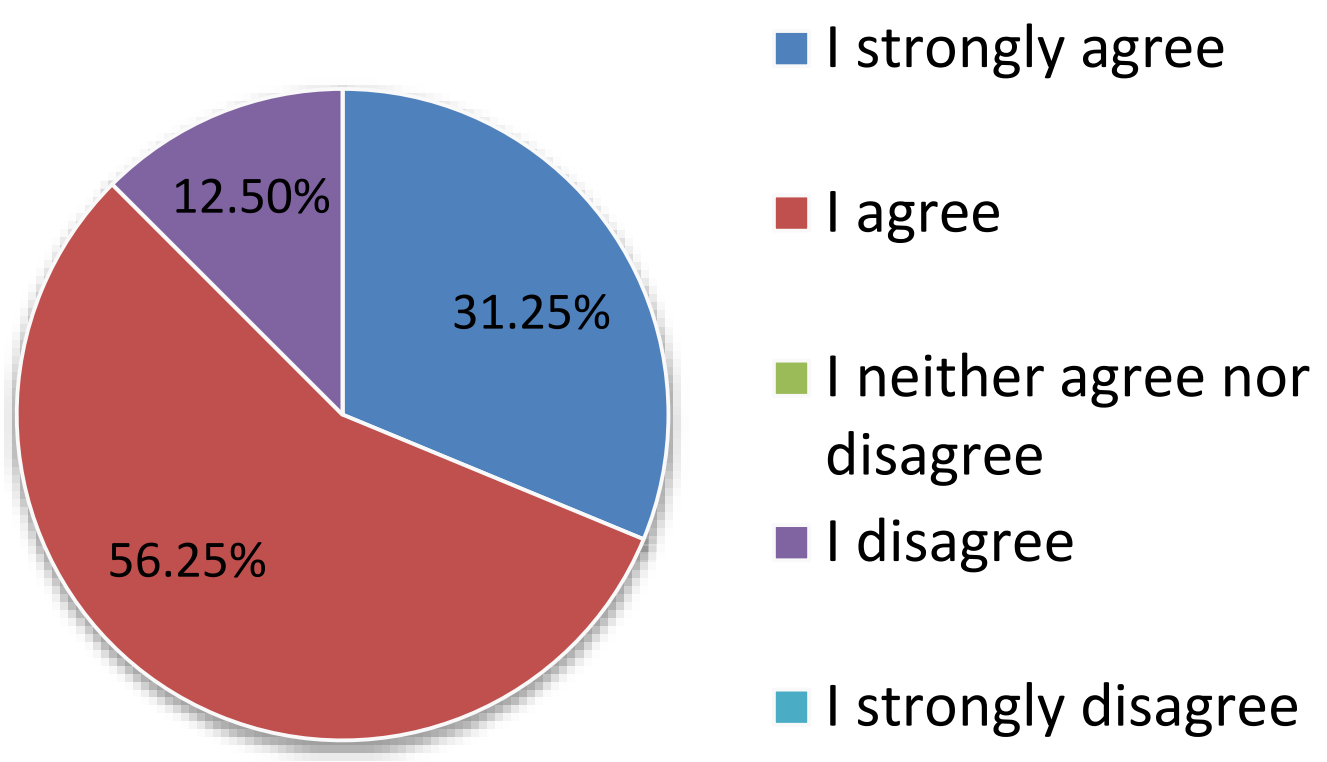

**Figure 54. Concerning (4), moving the red balloon icon in the panel had a noticeable effect on the content of (1) and the ranking of news in (2).**

**If you have some comments about the interaction with (4), please add them here.**

- I quiet liked that the yellow dots display articles, but if there will be a large set of news with articles all over the world, what will it look like?
- You are using a yellow dot for each news article - it was not clear without looking at the help what they mean. Also, it would be nice if you "zoom in" where the articles are. For example, most of the articles are from Europe and the US, but I'm still seeing Asia and even Antarctica. Because the panel is small, I think it would be very nice to focus it where it's important.
- Moving the red balloon icon in the panel had a noticeable effect However, I didn't experience any valuable change. I wanted to switch to a Chinese perspective on European topics, but all I got was articles from English newspapers (also the opinion within the articles was rather UK burdened)
- I though zooming on an area would filter out news that are not from that area.
- Why is there an enable/disable option only for 4? I have not noticed this option at the beginning. It could be explained in the info box appearing with the question mark.

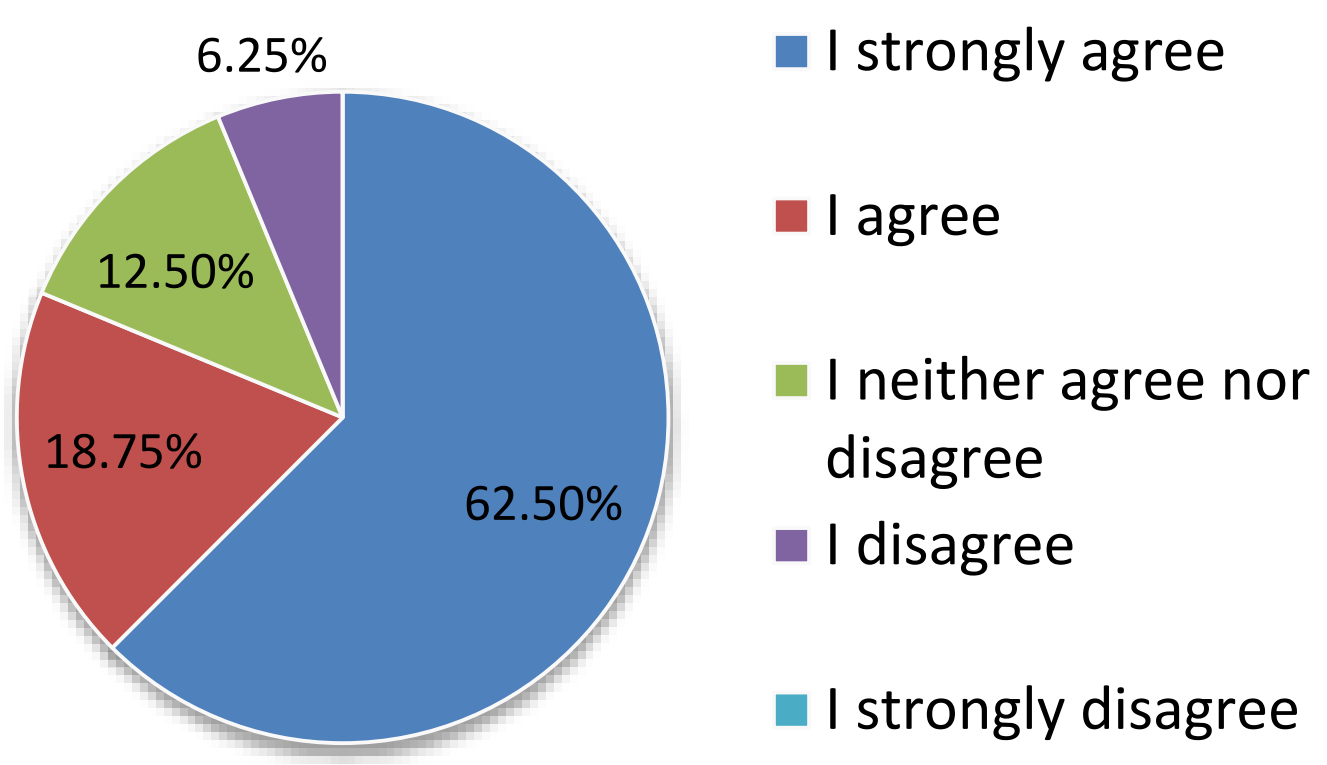

**Figure 55. Concerning (5), the interaction with the panel is intuitive.**

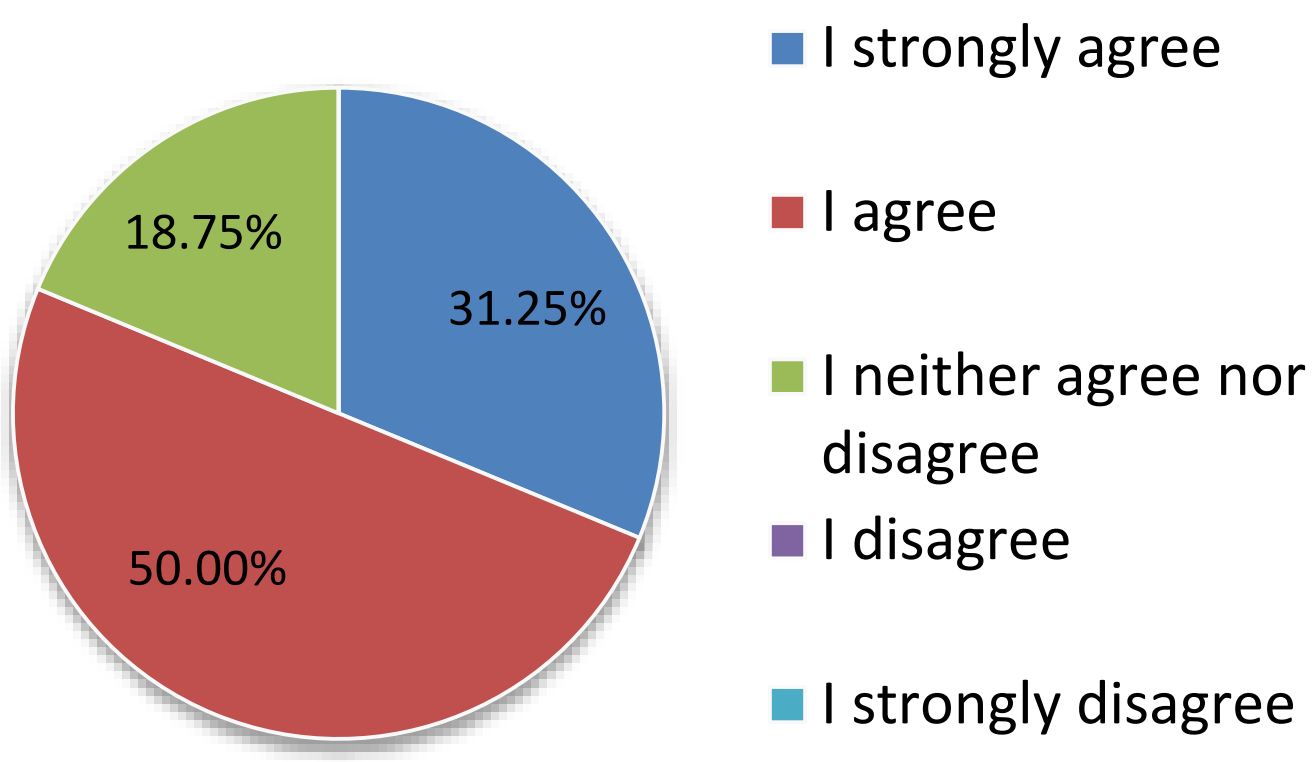

**Figure 56. Concerning (5), moving the slider in the panel had a noticeable effect on the content of (1) and the ranking of news in (2).**

**If you have some comments about the interaction with (5), please add them here.**

- Positive and Negative are subject to the user judgement. Not sure how it interferes with pos/neg meanings of keywords within the articles (I guess)
- I think the mechanism for the sentiment detection should be improved. I am not impressed with the ranking. There should also be an indicator of origin so that I can easily go back to obtain neutral news. Without such indicator, it is very difficult to find mid-point of the bar.
- It might be better to have a few discrete places where you can put the bar, as it would be easier to reproduce the results. Adding a color (red -> green) might also help.
- Sentiment analysis does not look that precise. Also, I don't see why I would like to use it as a parameter? I want to read the news objectively no matter whether an imprecise algorithm says they are good or bad.

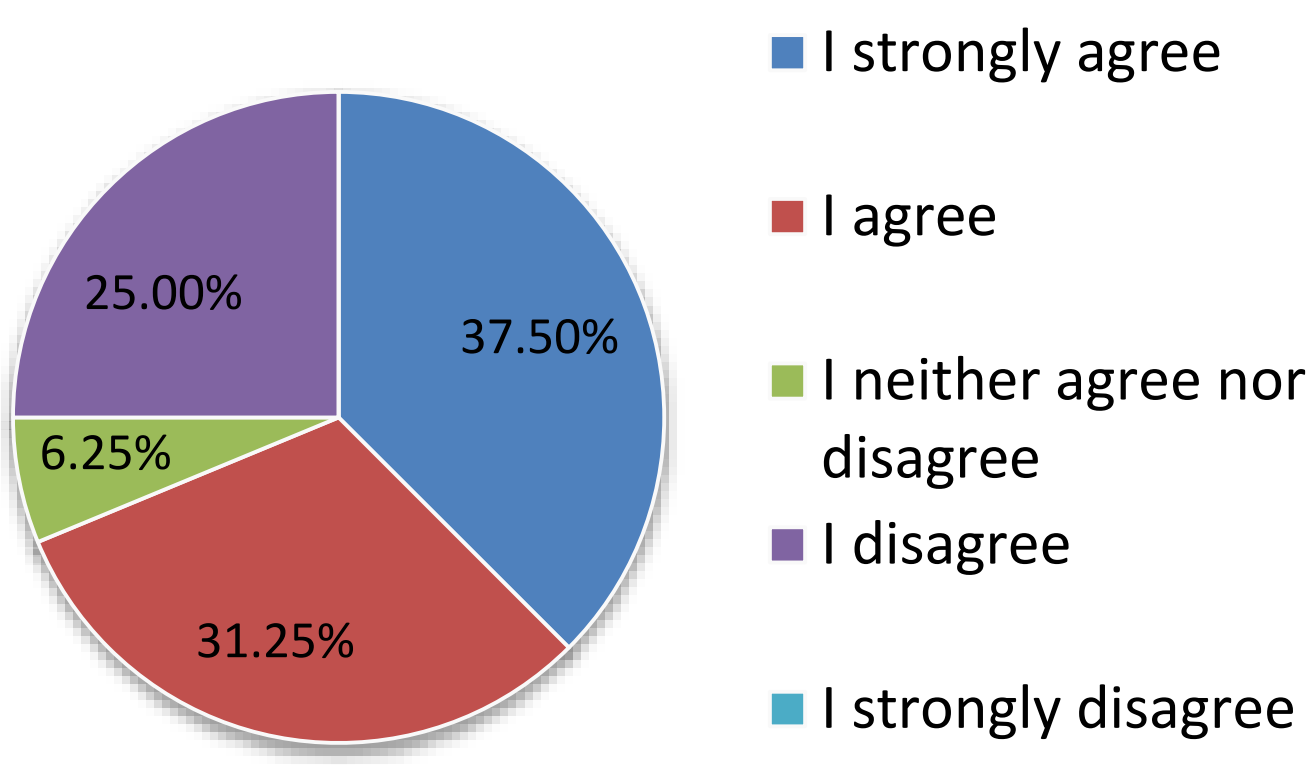

**Figure 57. Concerning (6), the interaction with the panel is intuitive.**

**If you have some comments about the interaction with (6), please add them here.**

- It is not clear whether you search with 6 for new topics or focus specific topics in the current view. I don't think that suggesting queries can be useful. I though that clicking on them would focus the result on those concepts, somehow like (3) does. Maybe a google-like "did you mean...?" would be more helpful.
- Again, how do I go back to the original mode?
- When I change the topic or the area I would expect this to change as well.

## A.6 Perceived utility evaluation

**Estimated time: 10 minutes.**

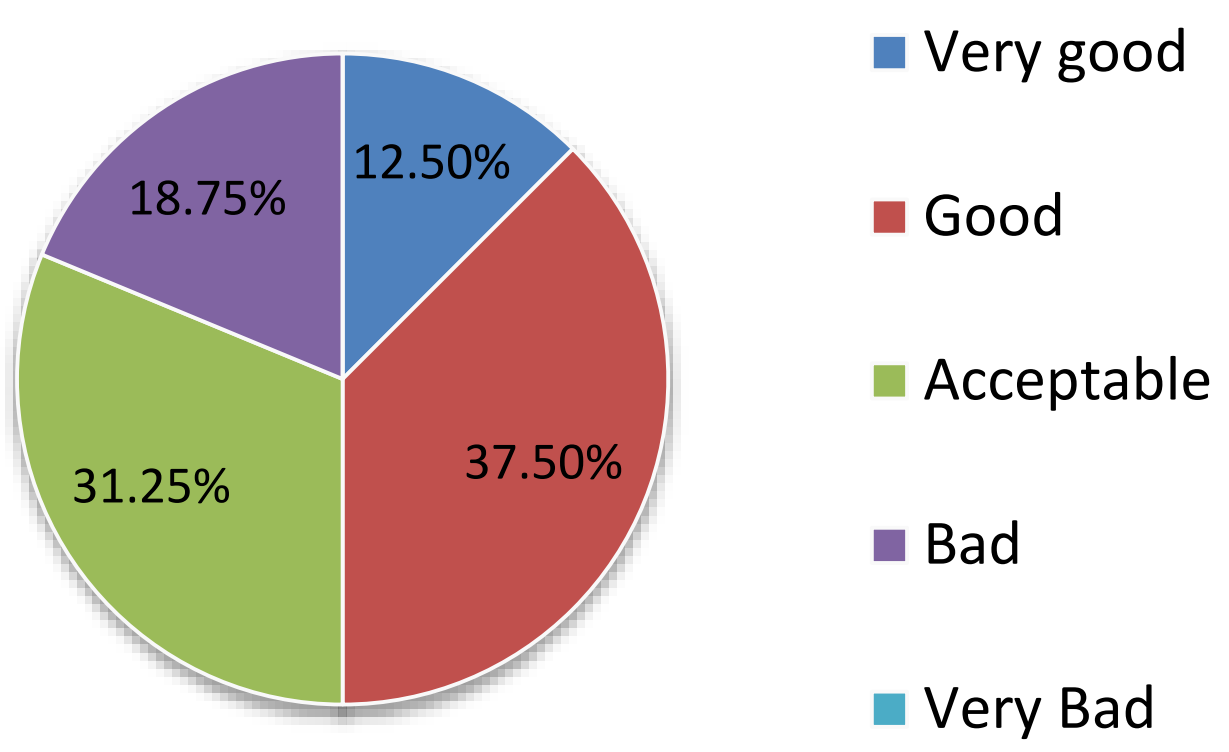

**Figure 58. What is your general impression about the quality of the generated summaries?**

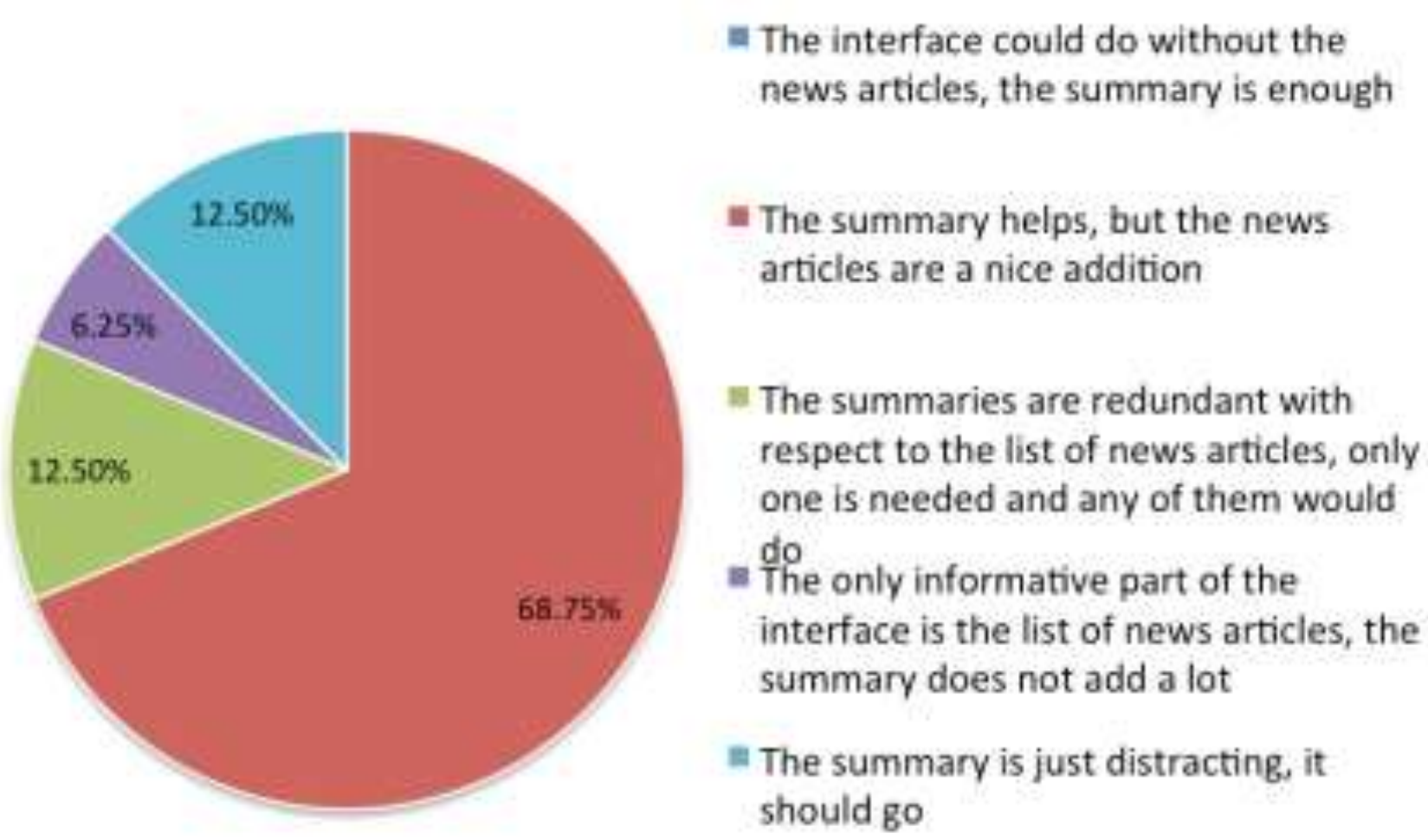

**Figure 59. Based on the summaries that you have read, you would say that:**

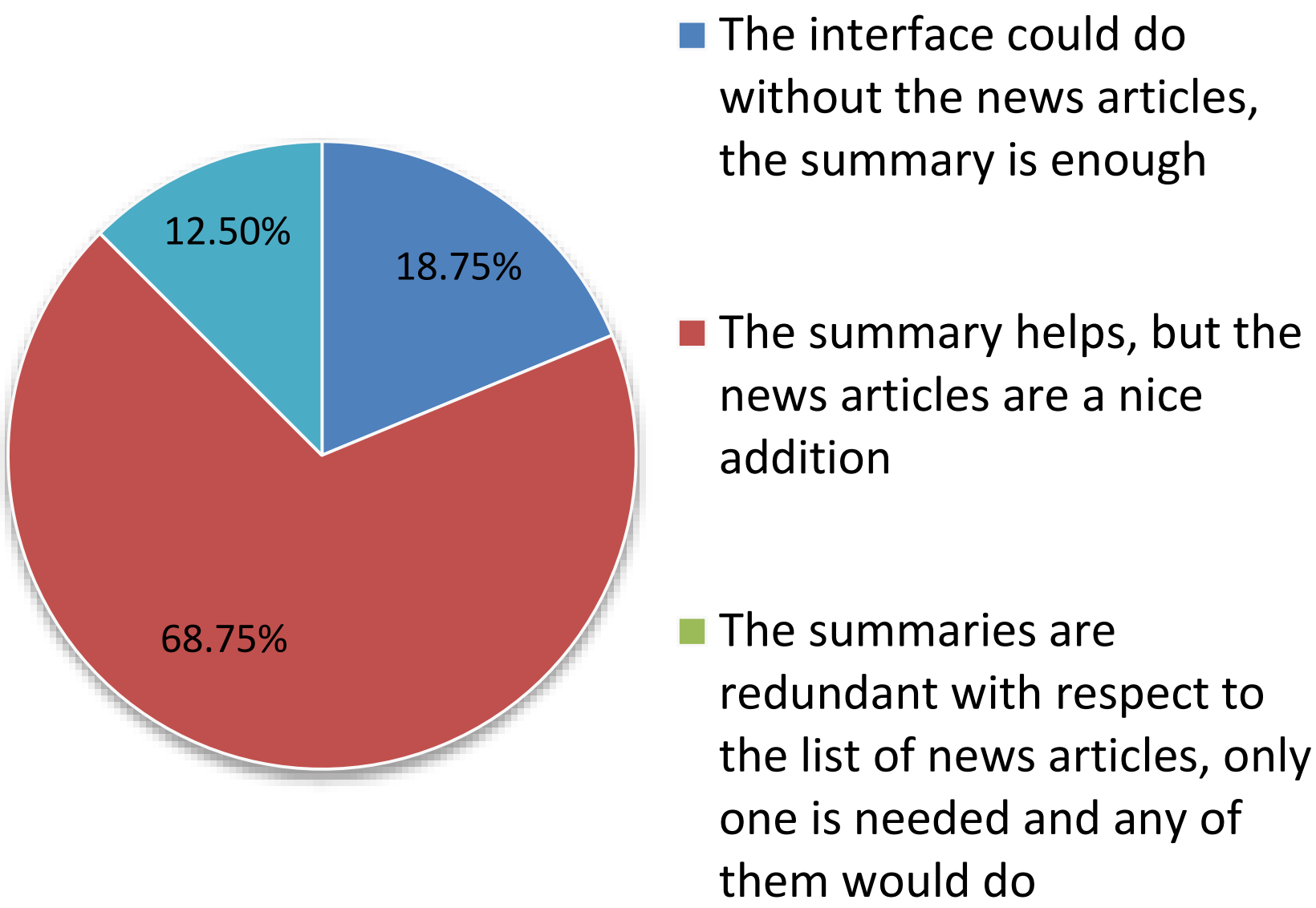

**Figure 60. Now, imagine you had a perfect summarizer that would do a great job at compressing the text in the news collection. In this case, you would say that:**

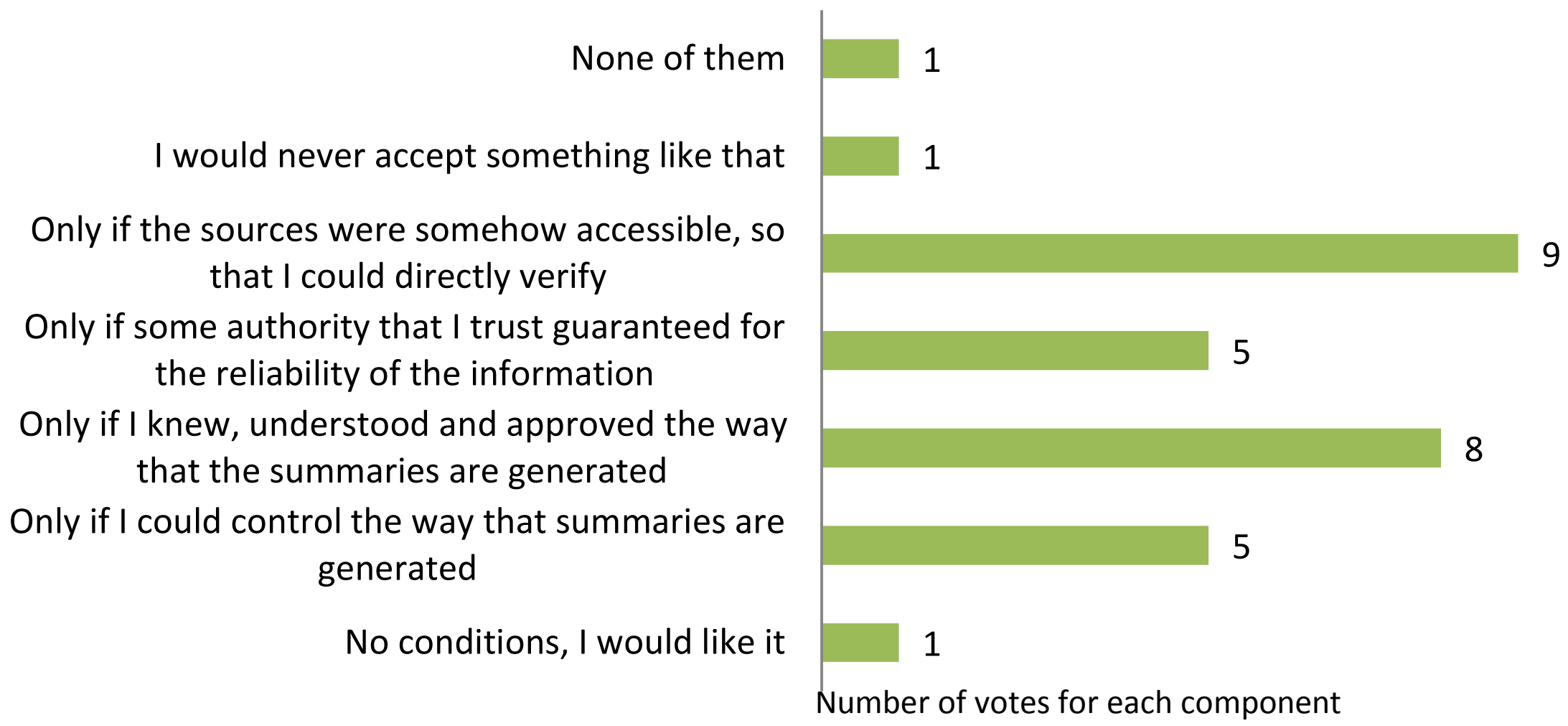

**Figure 61. Under what conditions would you trust a news browser that shows summaries of sets of related news? Please, mark all that apply.**

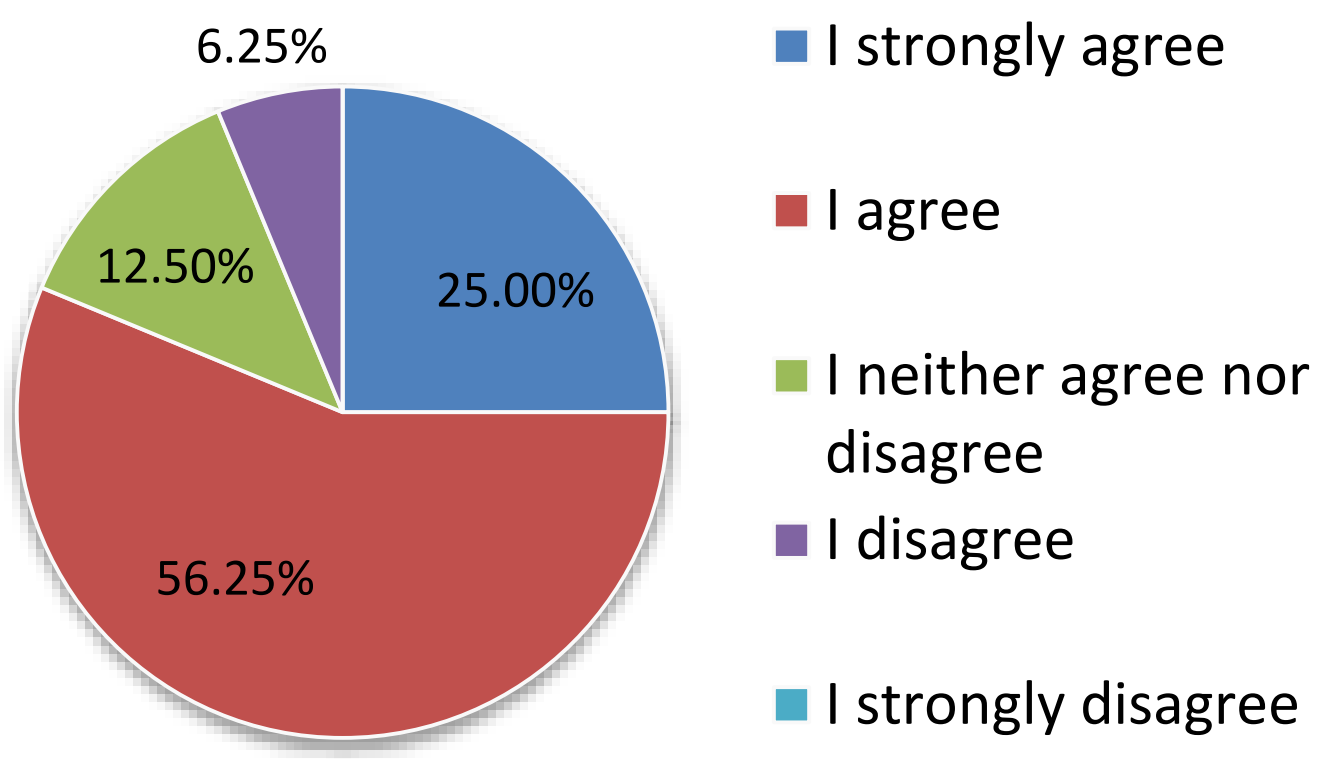

**Figure 62. The summaries are a good way of letting relevant information emerge.**

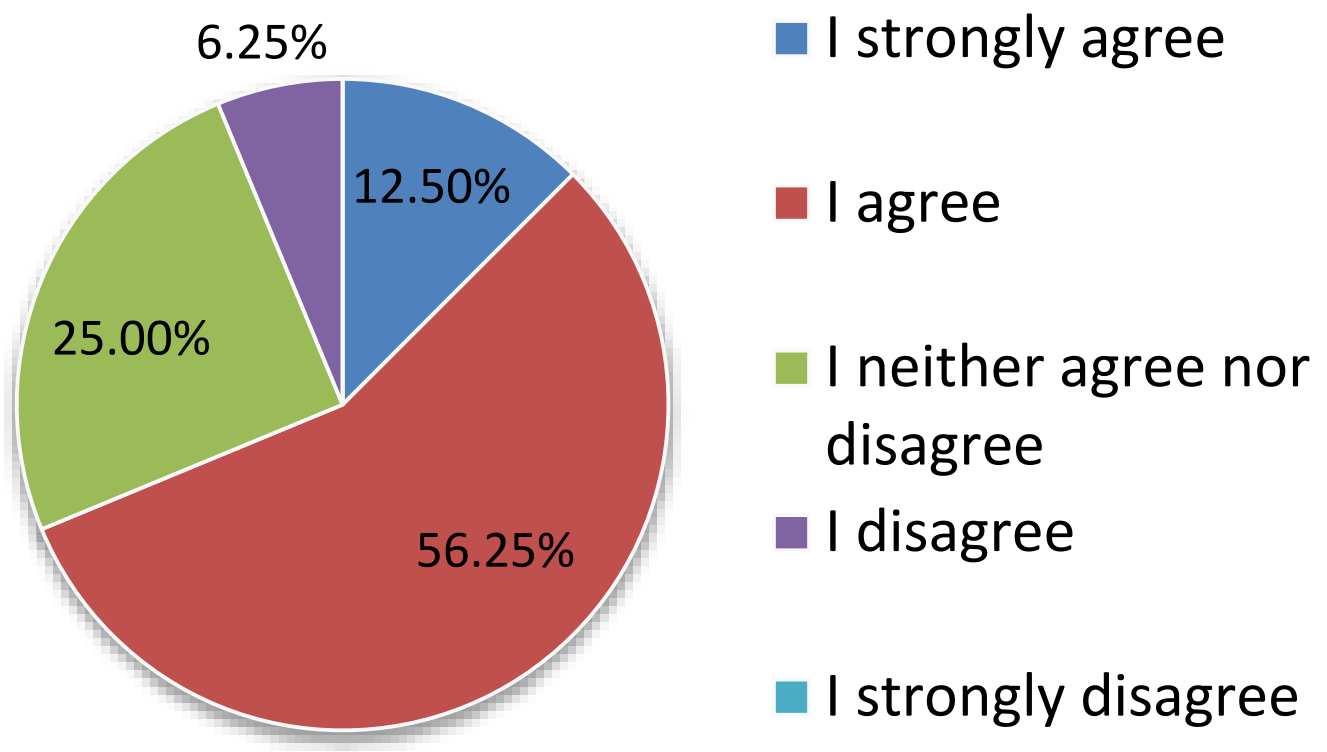

**Figure 63. Generating summaries according to different criteria is a good way of letting diversity of opinion and points of view emerge.**

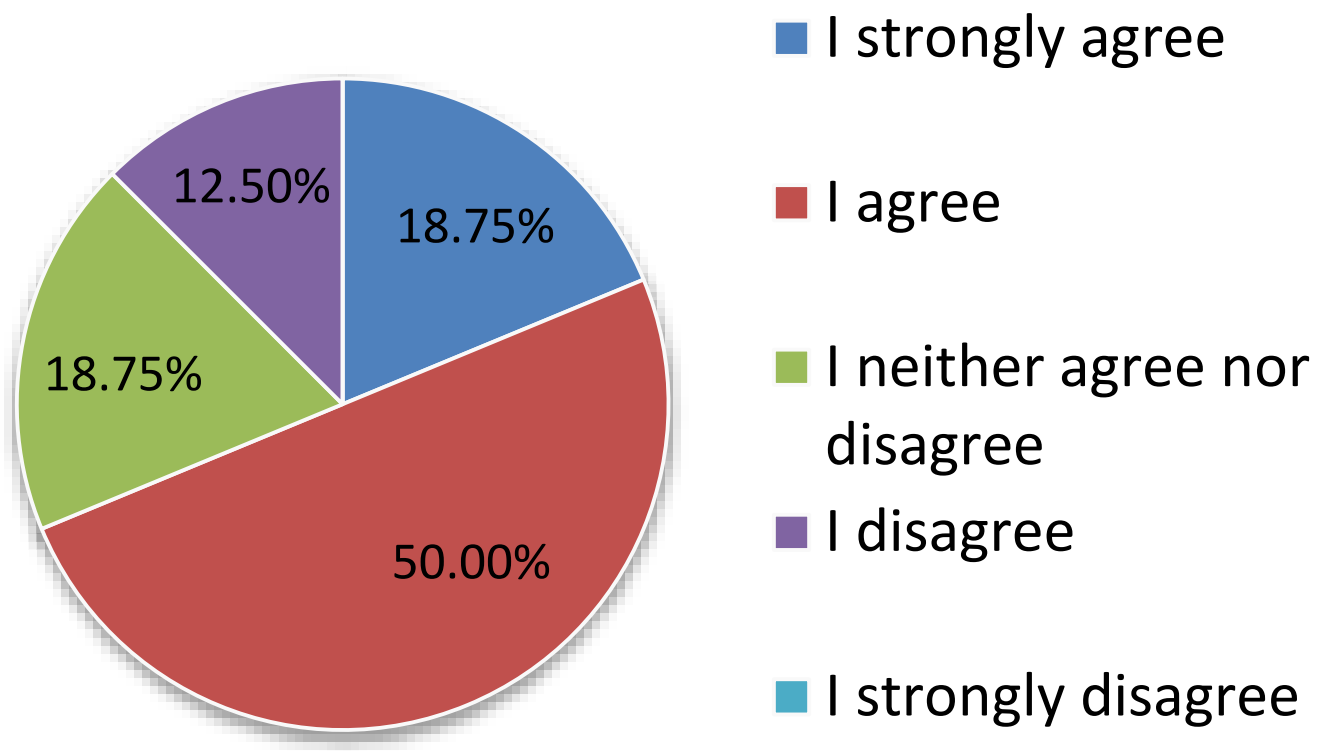

**Figure 64. The panel (3) makes it possible to generate summaries and rank related news based on the relevance of specific topics. This is a nice feature to have in a news browser.**

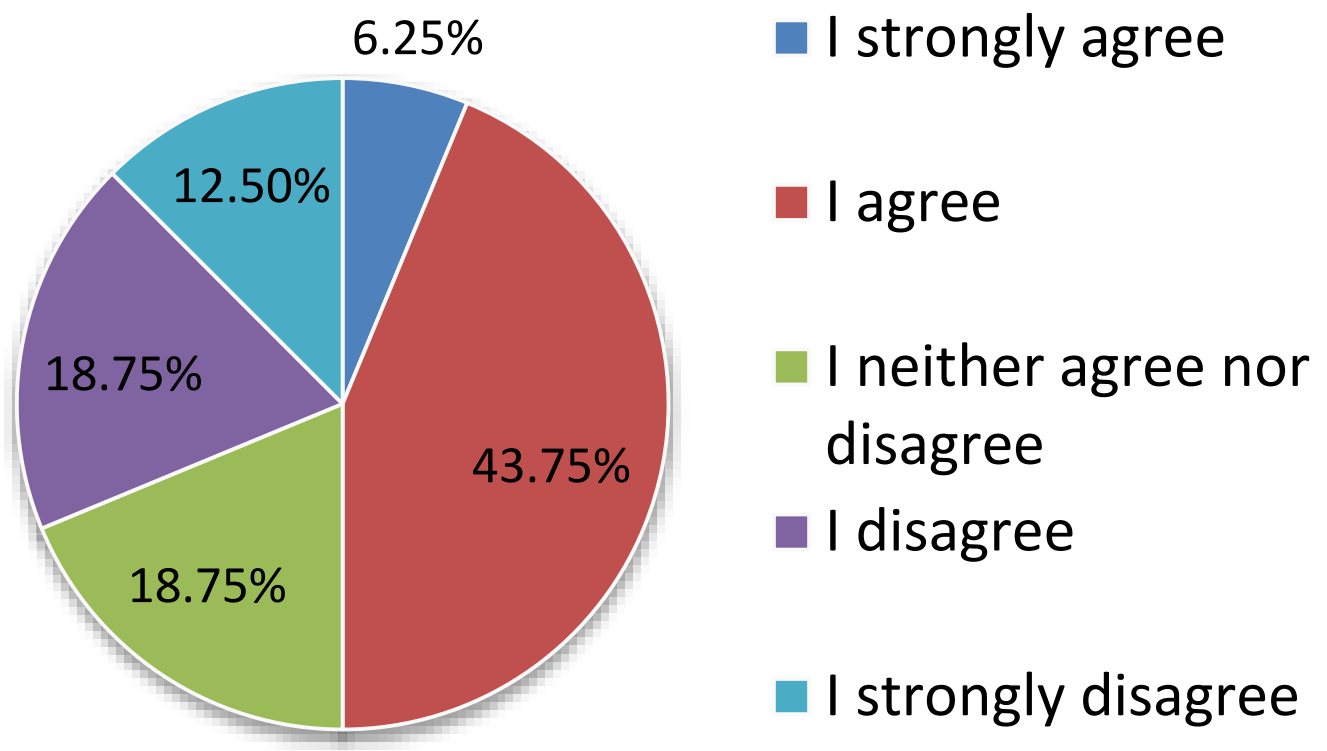

**Figure 65. The panel (3) can help highlighting different opinions and points of view in a news collection and let diversity in news stand out.**

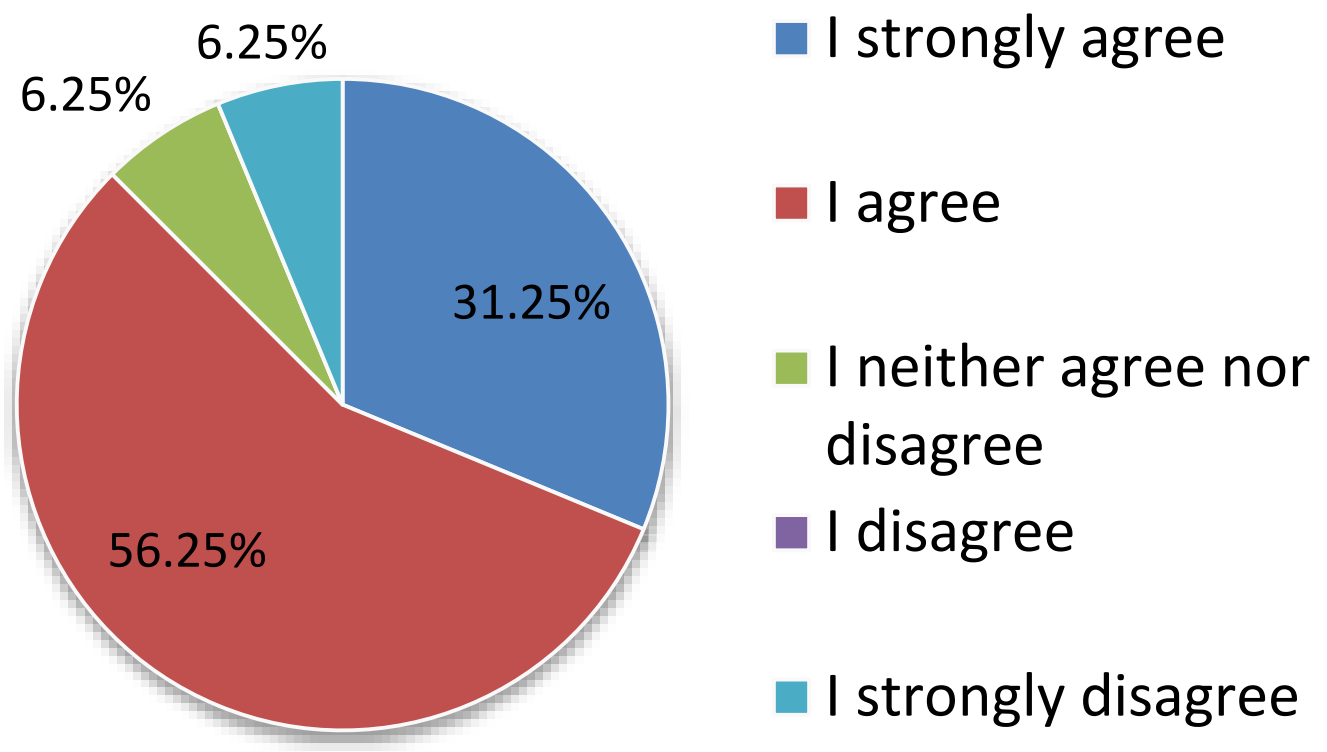

**Figure 66. The panel (4) makes it possible to generate summaries and rank related news based on the geographic source of the news. This is a nice feature to have in a news browser.**

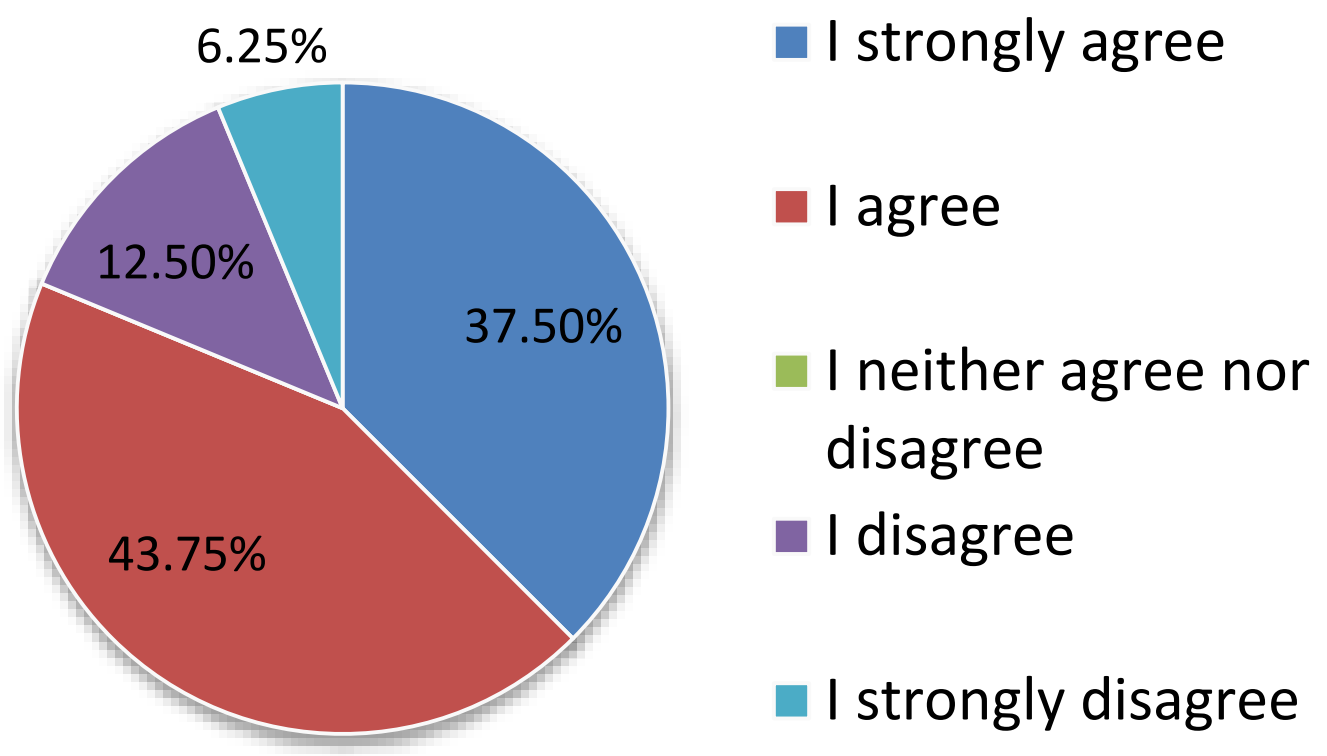

**Figure 67. The panel (4) can help highlighting different opinions and points of view in a news collection and let diversity in news stand out.**

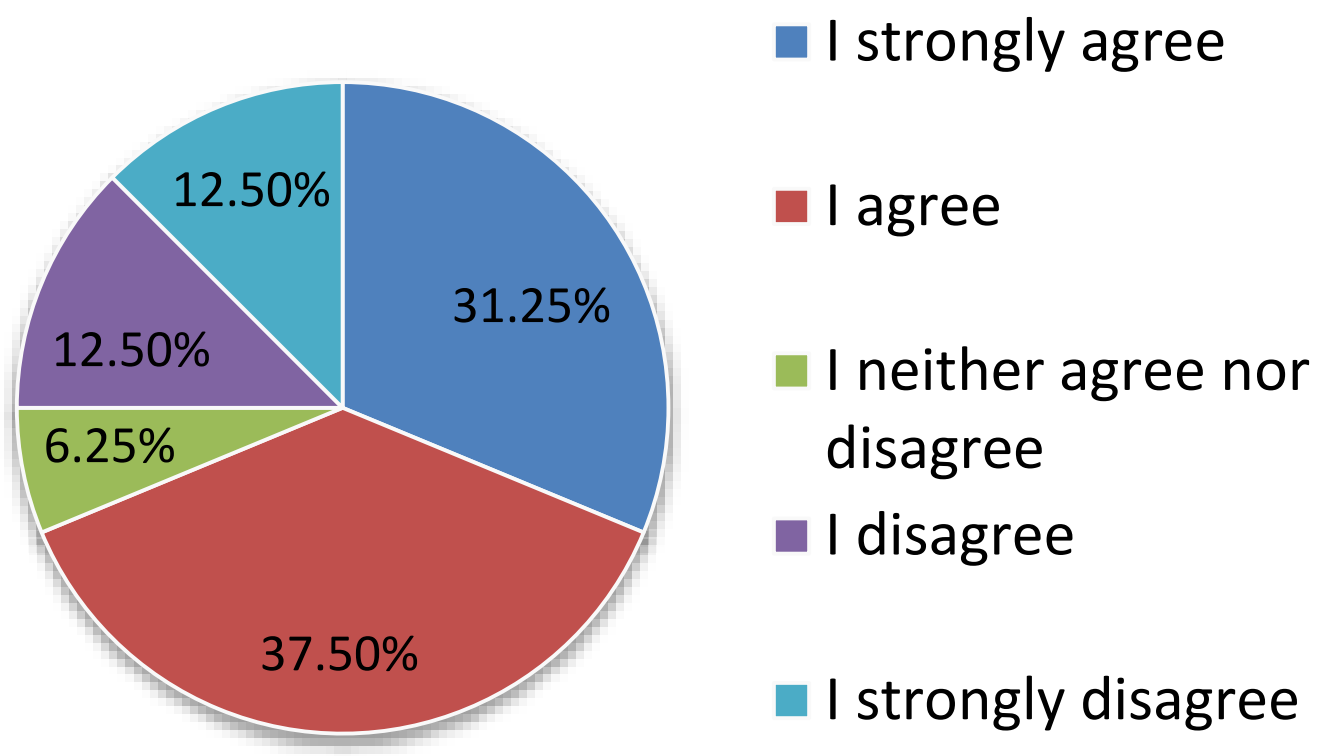

**Figure 68. The panel (5) makes it possible to generate summaries and rank related news based on their polarity. This is a nice feature to have in a news browser.**

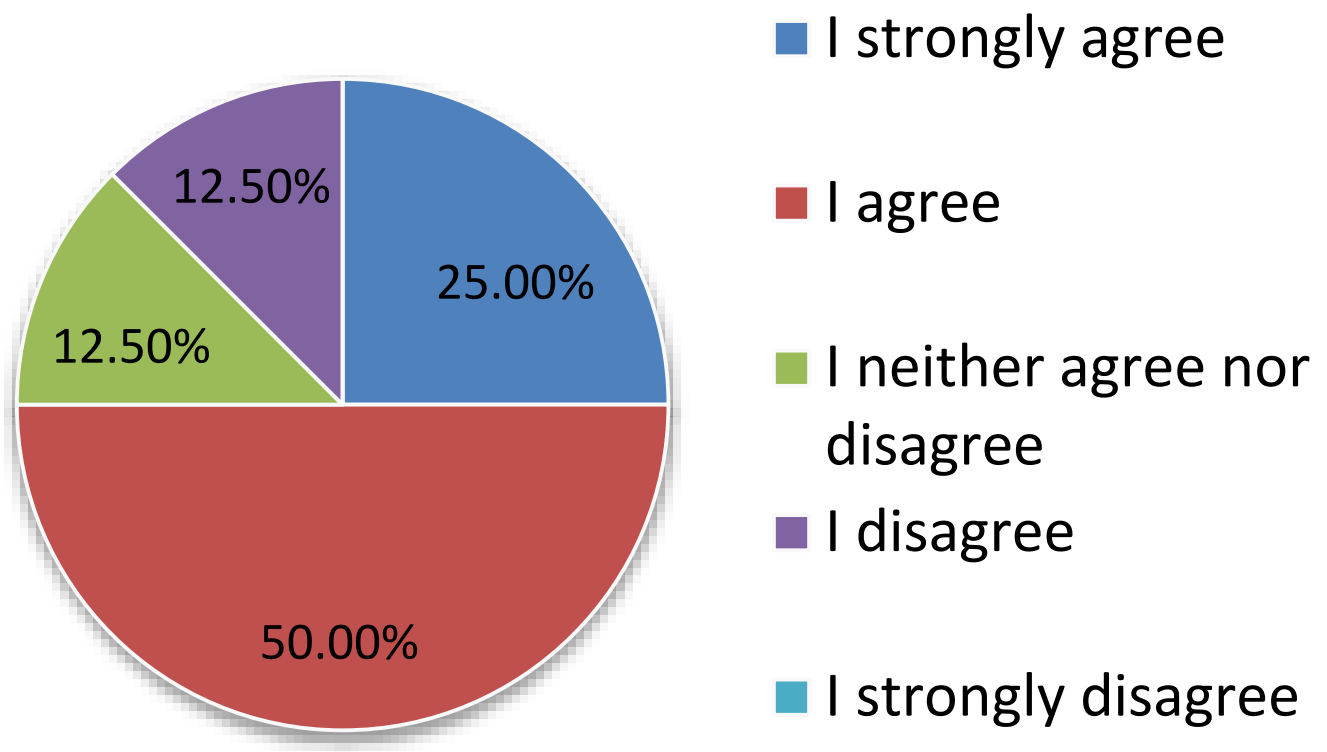

**Figure 69. The panel (5) can help highlighting different opinions and points of view in a news collection and let diversity in news stand out.**

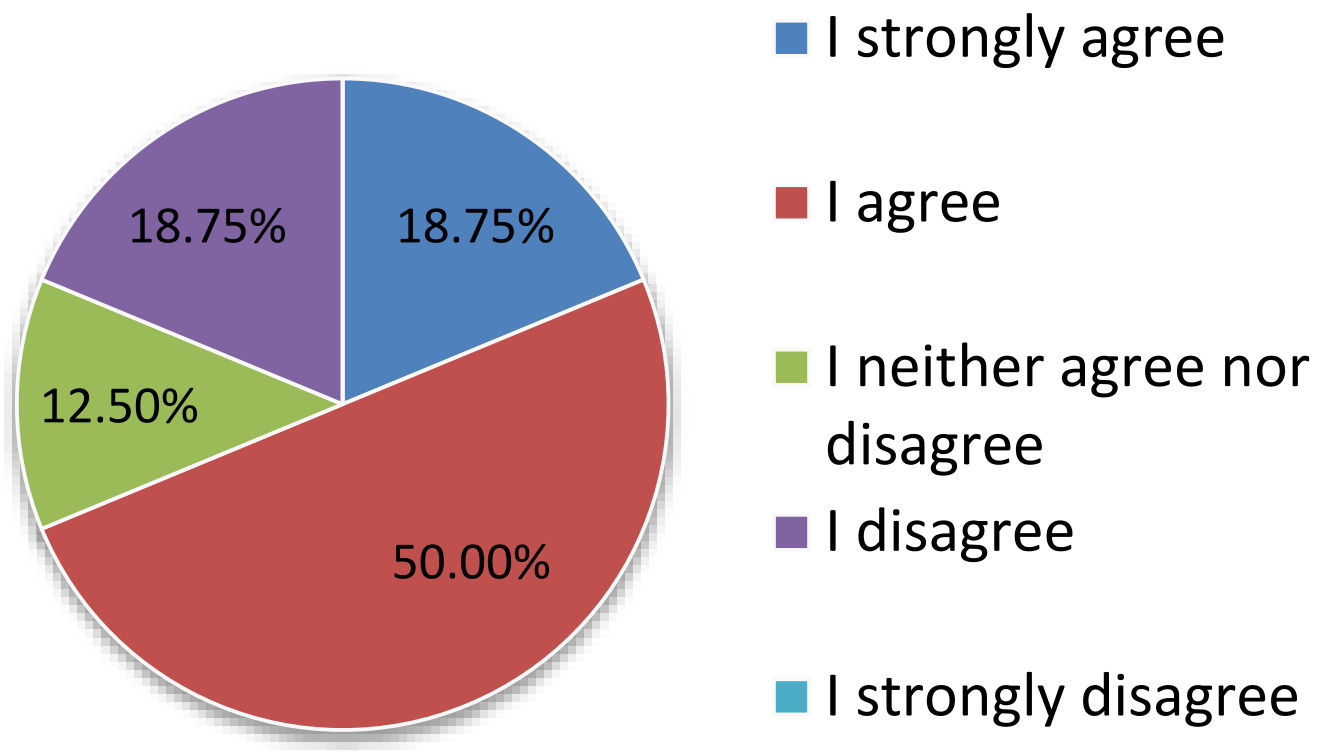

**Figure 70. The panel (6) makes it possible to discover news involving entities which are related to the currently selected news. This is a nice feature to have in a news browser.**

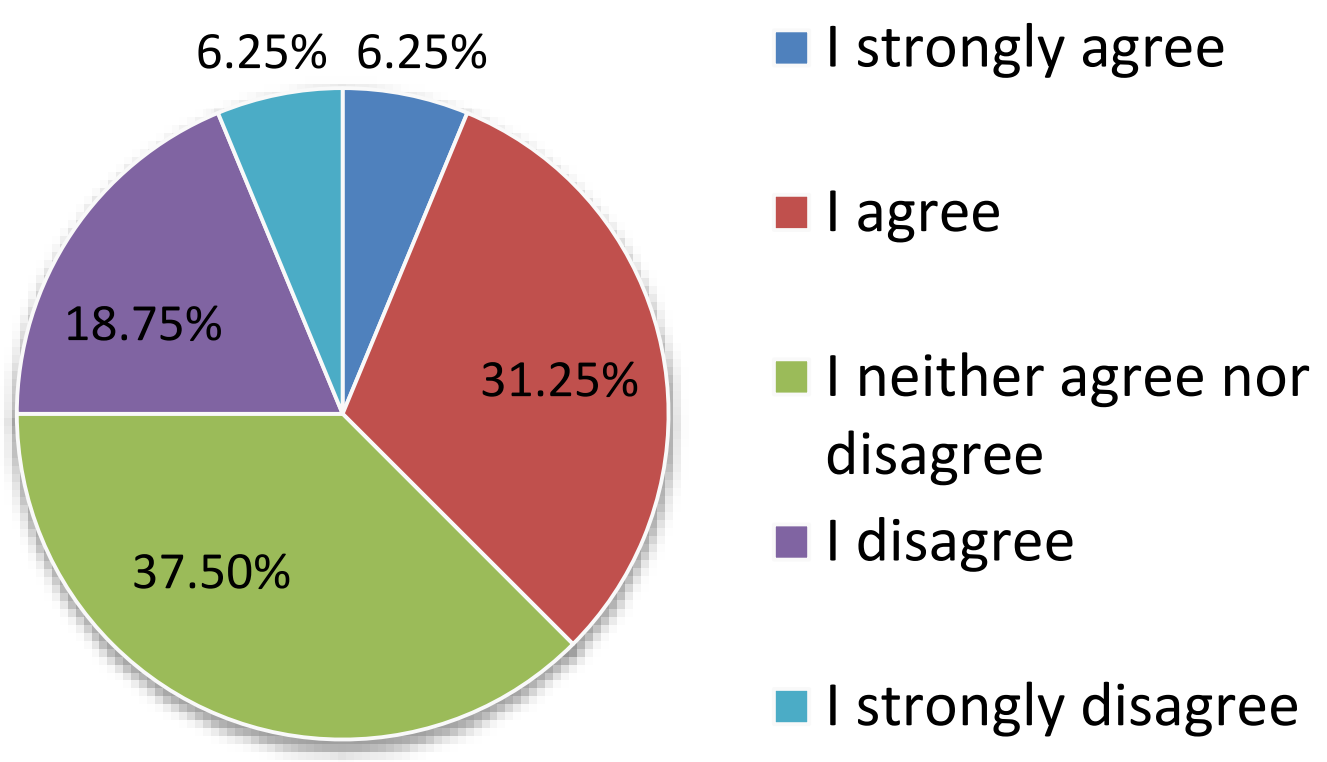

**Figure 71. The panel (6) can help highlighting different opinions and points of view in a news collection and let diversity in news stand out.**

**How would you change (3) so that it would make it easier to discover diversity of opinion in the news?**

- I would make it with more details. The problem is that we have to take some times before understand what's going on. some more groups, but with less topics in each group
- Tag cloud

- Perhaps use different colors or shading to reflect negative/positive sentiment, otherwise i think it is OK
- The relation to diversity was not clear to me. I assumed that it (slightly) switches the topic of the articles
- Make it shorter. Why would I read something if I knew that I don't need most of it? People are lazy.
- Maybe little pictures with faces or logos
- Add political/financial inclination of the newspaper news are taken from, maybe with background colors?

**How would you change (4) so that it would make it easier to discover diversity of opinion in the news?**

- I thought that by 4 you just change the location of the data and the contents are changing accordingly, you don't really have different opinions on the same subject. bigger map, hide Antarctica
- Indicate the polarity according to the geographical location of the source of the news.
- As said, the demo wasn't always producing diverse results. otherwise fine.
- Maybe also short polarity for the news articles by showing different colors for the articles on the map.including news items not from anglo-saxon world ;)
- Color dots according to inclination of the newspaper.
- The map is too small.. I would find out a way to make it more comfortable to navigate and select a region. I would add a search box to point exactly at the place I want by name. It is difficult to find the place you want on that small map.
- Maybe just on bubble per country

**How would you change (5) so that it would make it easier to discover diversity of opinion in the news?**

- Integrate it with (4). Say, the pin changes color according to how most of the articles from that region look like. I don't care in customizing manually the polarity of the news.
- It is good as it is.
- System could automatically generate additional dichotomies depending on topic, such as "pro-life" vs. "pro-choice" in abortion debate, or "pro-samsung" vs "pro-apple" sentiment in patent dispute. i know its asking much but you asked...
- The slider should only have 5 positions maximum Add another bar with "singularity", privileging original rather than main-stream opinions.

**How would you change (6) so that it would make it easier to discover diversity of opinion in the news?**

- It's ok
- Offer sentence patterns that are relevant for understanding the opinion in the news. E.g. about the re-election of Obama, a newspaper could say: "Unfortunately, Obama managed for a mild gap to win the elections again." While another could say: "Parties kick off in the streets as the first black president of the united states gets re-elected"
- I assumed that it just proposes related topics. Consequently, as I used it it was less diversity related.
- I think like that is quite ok, as we don't want the information to get too biased.
- No change.
- I'm doubtful about usefulness of (6)

**If you have any other comments we would be grateful if you could share them with us:**

- Excellent tool & project, but needs some UI adjustments to fly
- Make it so that pressing enter also gives results. Now I have to click on the search button.
- When I search for something in the search box I don't want to have to press "Search" every time. It should activate when pressing "Enter" or automatically. I really like the interface and it would be really nice to use it. I would start reading more news, as the summary part seems great.
- Great page, I like the summaries very much, the overall design could be improved